\DeclareUnicodeCharacter{3B1}{\ensuremath{\alpha}} 
\documentclass{aa}  
\usepackage{graphicx}
\usepackage[toc,page]{appendix}
\usepackage{caption}

\usepackage{txfonts}

\usepackage{array}
\newcolumntype{P}[1]{>{\centering\arraybackslash}p{#1}}
\newcolumntype{M}[1]{>{\centering\arraybackslash}m{#1}}

\usepackage{natbib}
\bibpunct{(}{)}{;}{a}{}{,}

\usepackage{color}
\usepackage{ulem}

\begin{document} 

   \title{The MUSE-Wide survey: Three-dimensional clustering analysis of Lyman-$\alpha$ emitters at $3.3<z<6$} 
   \author{Yohana Herrero Alonso,
          \inst{1}
           M. Krumpe,
          \inst{1}
          L. Wisotzki,
          \inst{1}
          T. Miyaji,
          \inst{2,1}\thanks{on sabbatical leave from IA-UNAM-E at AIP.}
          T. Garel,
          \inst{3}
          K. B. Schmidt,
          \inst{1}
          C. Diener,
          \inst{4}
          T. Urrutia,
          \inst{1}
          J. Kerutt,
          \inst{1,3}
          E. C. Herenz,
          \inst{5}
          J. Schaye,
          \inst{6}
          G. Pezzulli,
          \inst{7,8}
          M. V. Maseda,
          \inst{6}
          L. Boogaard,
          \inst{6}
          \and
          J. Richard
          \inst{9}
          }

   \institute{Leibniz-Institut f\"ur Astrophysik Potsdam (AIP), An der Sternwarte 16, D-14482 Potsdam, Germany\\
              \email{yherreroalonso@aip.de}
         \and
            Universidad Nacional Aut\'onoma de M\'exico, Instituto de Astronom\'ia (IA-UNAM-E), AP 106, Ensenada 22860, BC, M\'exico
         \and
             Observatoire de G\`eneve, Universit\`e de G\`eneve, 51 Ch. des Maillettes, CH-1290 Versoix, Switzerland
        \and
             Institute of Astronomy, University of Cambridge, Madingley Road, Cambridge CB3 0HA, England
        \and
             European Southern Observatory, Av. Alonso de C\'ordova 3107, 763 0355 Vitacura, Santiago, Chile
        \and
             Leiden Observatory, Leiden University, P.O. Box 9513, 2300 RA, Leiden, The Netherlands
        \and
             Department of Physics, ETH Zurich, Wolfgang-Pauli-Strasse 27, 8093 Zurich, Switzerland
        \and
             Kapteyn Astronomical Institute, University of Groningen, Landleven 12, 9747 AD Groningen, The Netherlands
        \and
             Univ Lyon, Univ Lyon1, Ens de Lyon, CNRS, Centre de Recherche Astrophysique de Lyon UMR5574, F-69230, Saint-Genis-Laval, France\\
             }

   \date{Received xxx/Accepted xxx} 
 
  \abstract{We present an analysis of the spatial clustering of 695 Ly$\alpha$-emitting galaxies (LAEs) in the MUSE-Wide survey. All objects have spectroscopically confirmed redshifts in the range $3.3 < z < 6$. We employ the K-estimator of \citet{adelberger}, adapted and optimized for our sample. We also explore the standard two-point correlation function (2pcf) approach, which is however less suited for a pencil-beam survey such as ours. The results from both approaches are consistent. We parametrize the clustering properties in two ways, (i) following the standard approach of modelling the clustering signal with a power law (PL), and (ii) adopting a Halo Occupation Distribution (HOD) model of the 2-halo term. Using the K-estimator and applying HOD modeling, we infer a large-scale bias of $b_{\mathrm{HOD}} = 2.80^{+0.38}_{-0.38}$ at a median redshift of the number of galaxy pairs $\langle z_{\rm pair} \rangle \simeq 3.82$, while the best-fit power-law analysis gives $b_{\mathrm{PL}} = 3.03^{+1.51}_{-0.52}$ ($r_0=3.60^{+3.10}_{-0.90}$ comoving $h^{-1}$Mpc and $\gamma=1.30^{+0.36}_{-0.45}$). The implied typical dark matter halo (DMH) mass is $\log (M_{\text{DMH}}/[h^{-1}\text{M}_\odot]) = 11.34^{+0.23}_{-0.27}$ (adopting $b = b_{\mathrm{HOD}}$ and assuming $\sigma_8 = 0.8$). We study possible dependencies of the clustering signal on object properties by bisecting the sample into disjoint subsets, considering Ly$\alpha$ luminosity, UV absolute magnitude, Ly$\alpha$ equivalent width, and redshift as variables. We find no evidence for a strong dependence on the latter three variables but detect a suggestive trend of more luminous Ly$\alpha$ emitters clustering more strongly (thus residing in more massive DMHs) than their lower Ly$\alpha$ luminosity counterparts. We also compare our results to mock LAE catalogs based on a semi-analytic model of galaxy formation and find a stronger clustering signal than in our observed sample, driven by spikes in the simulated $z$-distributions. By adopting a galaxy-conserving model we estimate that the Ly$\alpha$-bright galaxies in the MUSE-Wide survey will typically evolve into galaxies hosted by halos of $\log (M_{\text{DMH}}/[h^{-1}\text{M}_\odot]) \approx 13.5$ at redshift zero, suggesting that we observe the ancestors of present-day galaxy groups.
  }

    \keywords{large-scale structure -- high-redshift galaxies -- galaxy evolution -- observational Cosmology }

   \titlerunning{Spatial clustering of LAEs in the MUSE-Wide survey}
   \authorrunning{Yohana Herrero Alonso et al.}
 
   \maketitle
%

\section{Introduction}
\label{sec:introduction}
The distribution of galaxies in the Universe forms a well defined network, the so-called "Cosmic Web". This structure was formed when gravitational instabilities produced by primordial density fluctuations grew until they reached a critical density. Their collapse formed the gravitationally bound dark matter halos (DMH). These halos grow hierarchically through accretion and mergers with other halos. Their gravitational interaction with baryonic matter caused gas to fall into the growing potential wells, making the gas cool by radiation and collapse to form stars and galaxies. 

The evolution of the baryonic matter distribution follows the underlying dark matter (DM) but, especially in the early stages of galaxy formation, the details of this relation and how it evolved with time are still unclear. Galaxy clustering analyses constrain the masses of the typical DMHs by which these galaxies are hosted and are therefore a crucial element towards understanding the formation and evolution of galaxies \citep{coil}. 

A common way to quantify galaxy clustering is through correlation functions \citep[e.g.,][]{gawiser07,zehavi11,ouchi17}, which express the probability of finding a tuple (usually a pair) of galaxies at a certain separation \citep[e.g.][]{peebles1980}. The clustering strength (large-scale bias) and the typical DMH masses can be inferred from measured correlation functions in various ways. A widespread traditional approach is to approximate the 2-point correlation function (2pcf) as a power law \citep{davispeebles}, while more recent methods such as Halo Occupation Distribution (HOD) modelling \citep[e.g.][]{zhenghod} distinguish between the different contribution of the correlation function. In these models, pairs of galaxies either belong to the same DMH or to different DMHs.

These procedures have often been applied to galaxy surveys. At low redshifts, the major surveys cover large areas of the sky, in particular SDSS \citep[e.g.][]{sdss,ahumada} together with its successors, 2MASS \citep{2mass}, or the 2dF Galaxy Redshift survey \citep{2df}.
These samples at similar luminosities revealed a modest evolution of the clustering strength between $z=1$ and $z=0$ together with significant clustering dependencies on various galaxy properties, such as luminosity, color, morphology, galaxy type, etc.\ \citep[e.g.][]{norberg02,zehavi02,li06,zehavi11}.

At high redshifts ($z>2$), galaxy samples are however more limited. Gathering a statistically relevant number of objects and covering representative volumes of the sky is a hard task. Photometric selection techniques are often preferred because spectroscopic observations of many faint objects are observationally too expensive. The two most common techniques involve exploiting the drop in the continuum bluewards of 912\,\AA\ \citep{steidel} to search for Lyman-break galaxies (LBGs) and the use of narrow-band (NB) filters to target, for instance, the Ly$\alpha$ emission line of young, star-forming galaxies (Ly$\alpha$ emitters, LAEs). 

While each selection method provides us with its own somewhat biased view of the distribution of galaxies, LAEs are particularly interesting to probe the lower range of stellar masses ($10^8-10^9\; M_{\odot}$), highlighting a subset of galaxies heavily forming stars (star formation rates of $1-10\; M_{\odot}$yr$^{-1}$). By combining the clustering analysis of LAEs with cosmological simulations we can connect LAEs to their descendants in the local Universe.

Statistically substantial LAE samples ($> 10^2$ objects) based on narrow-band surveys were presented by  \cite{cowie,rhoads,ouchi03,gawiser07,ouchi17,sobral} and others. 
Generally, the NB selection method only provides LAE candidates implying that samples are contaminated by interlopers, which can be a problem for clustering studies. Obviously, since all objects selected by a given NB filter are assumed to be at the same redshift, their clustering can only be explored through the analysis of transverse angular correlations \citep{ouchi03,gawiser07,ouchi10, ouchi17, khostovan}. In order to study the full 3-dimensional spatial clustering behaviour of galaxies and its evolution over cosmic time, large-scale spectroscopic surveys of high-redshift galaxies with individually measured redshifts are required \citep{fevre,lilly,deep2,vipers,vimos}. It has been found that the clustering strength of high-redshift galaxies is significantly higher at similar luminosities than at intermediate and low redshifts 
\citep{durkalek}, possibly also depending on luminosity and stellar mass \citep[e.g.][]{ouchi03,ouchi17,durkalekp}.

Ideally, one would wish to perform spectroscopy of \emph{all} objects over a large area of the sky, with a wide redshift coverage. While such surveys are still beyond current means, panoramic Integral Field Units (IFUs) are currently opening an avenue for exploring this approach at least over modest areas. In particular, the Multi Unit Spectroscopic Explorer \citep[MUSE,][]{bacon} on the ESO-VLT has already produced significant samples of high-redshift galaxies with unprecedented source densities of several tens or even hundreds of objects per arcmin$^2$ \citep{inami,urrutia19}. In this paper we explore the potential of using $\approx$ 700  LAEs selected from the MUSE-Wide survey \citep{herenz17, urrutia19} for clustering studies. Our sample differs from previous studies of LAE clustering based on narrow-band imaging, but also from generic spectroscopic surveys requiring the preselection of targets from broad-band photometry. 

In a pilot study, \cite{catrina} used 238 LAEs from the first 24 MUSE-Wide fields to demonstrate that MUSE-selected LAEs do indeed reveal a significant clustering signal, even though the uncertainties were still large. Here we extend this work with a 3 times larger sample and a refined set of analysis methods and tools. The paper is structured as follows. First, we briefly describe the data used for this work and characterise the sample. In Sect.~\ref{sec:methods} we explain our methods to measure and analyse the clustering properties of our LAE sample. In Sect.~\ref{sec:results} we present the results of our measurements including a study of clustering dependencies with different galaxy parameters. In Sect.~\ref{sec:discussion} we discuss our results and compare our findings to predictions from a semi-analytic galaxy formation model. In Sect.~\ref{sec:conclusions} we give our conclusion. The Appendix of the paper is mainly discussing the LAE clustering results using the traditional 2-point correlation function.

Throughout the paper, all distances are measured in comoving coordinates and given in units of $h^{-1}$Mpc, where $h = H_0/100 = 0.70$ km s$^{-1}$ Mpc$^{-1}$. We use a $\Lambda$CDM cosmology and adopt $\Omega_M =$ 0.3, $\Omega_{\Lambda} =$ 0.7, $\sigma_8 =$ 0.8 and $H_0 =$ 70 km s$^{-1}$ Mpc$^{-1}$ \citep{constants}. All uncertainties represent a 1$\sigma$ (68.3\%) confidence interval unless otherwise stated.


\section{Data}
\label{sec:data}
\subsection{The MUSE-Wide Survey}
MUSE-Wide is an untargeted 3D spectroscopic survey \citep{herenz17,urrutia19} executed by the MUSE consortium as one of the Guaranteed Time Observations (GTO) programs. The survey covers parts of the CANDELS/GOODS-S and CANDELS/COSMOS fields and also includes 8 MUSE pointings in the so-called HUDF09 parallel fields (see \citealt{urrutia19} for details). The spectroscopic data provided by MUSE complement the rich multiwavelength data available in these fields, from which physical properties such as star formation rates or stellar masses can be derived. The full survey comprises 100 MUSE fields of 1 arcmin$^2$ each (although there is some overlap between adjacent fields), with a depth of 1~hour exposure time, each split into $4 \times 900$~s with $90\deg$ rotation between the exposures. Most fields were observed in dark time with a seeing better than $1.0\arcsec$. Spectra cover the range of 4750--9350~\AA, implying a Ly$\alpha$ redshift range of $2.9 \la z \la 6.7$. 

\begin{figure}[tb]
\centering
\includegraphics[width=\columnwidth,scale=0.5]{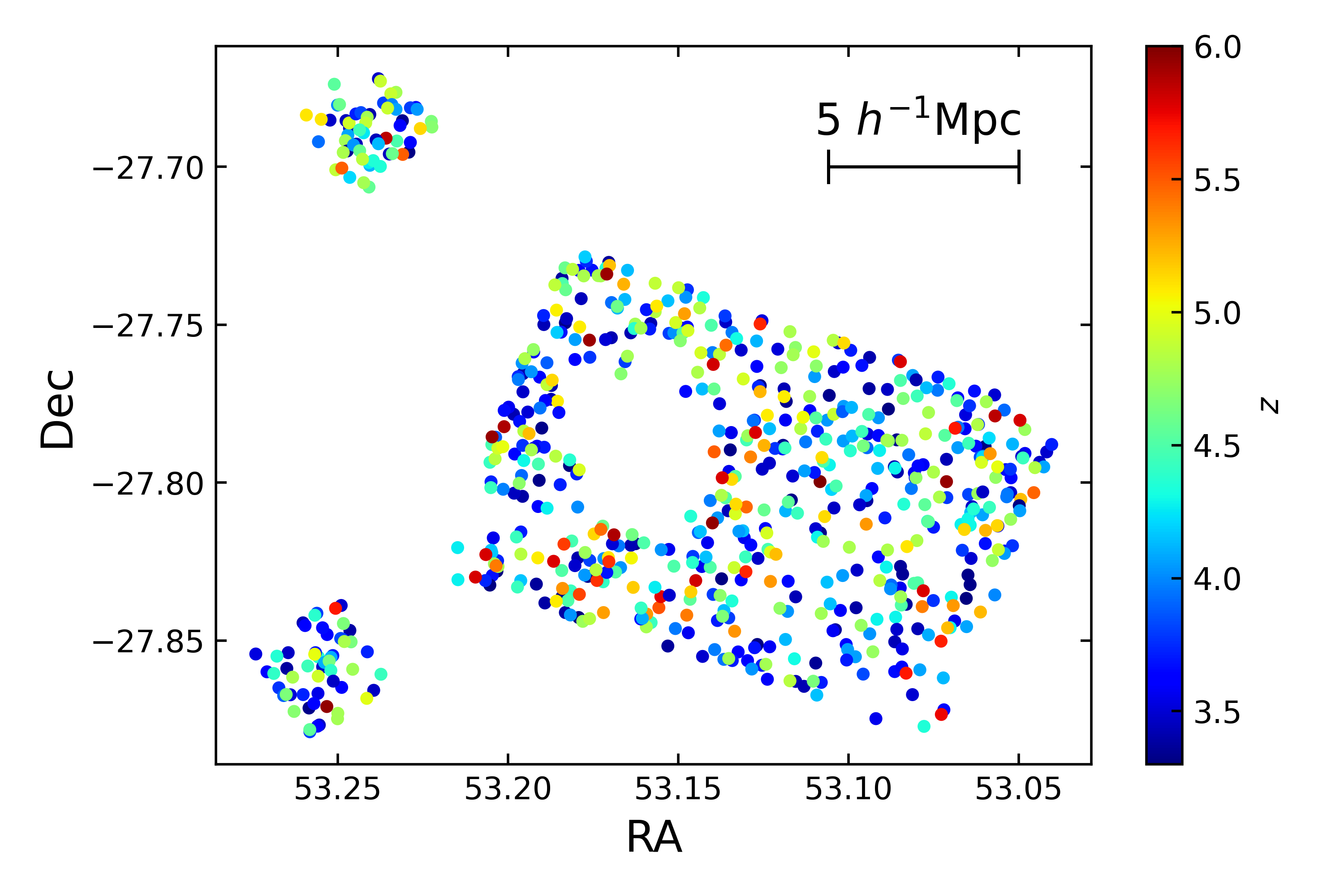}
\caption{Spatial distribution of 695 LAEs covering part of the CANDELS/GOODS-S region and the HUDF parallel fields on the left. The individual LAEs span a total of 68 fields from the MUSE-Wide survey and are color-coded by their Ly$\alpha$ spectroscopic redshift, $3.3<z<6$. The 5 $h^{-1}$Mpc bar for the mean redshift of the sample $\overline{z}$ $\approx$ 4.23 indicates the actual transverse extent of the footprint.}
\label{fig:ra-dec}
\end{figure}

The data reduction process is detailed in \cite{urrutia19}. Emission line sources were detected and their line fluxes were measured with the Line Source Detection and Cataloguing \citep[LSDCat,][]{herenzlutz17} software.   LSDCat cross-correlates a reduced and flux-calibrated data cube with a 3D source template to find emission line sources above a given significance threshold. The resulting emission line flux limit of the survey is typically around $\sim 10^{-17}$ erg s$^{-1}$ cm$^{-2}$ for LAEs, but with considerable spread between fields and also depending on the spatial extent of the Ly$\alpha$ emission \citep{herenz19}.

After the automatic detection of emission lines, a source catalog for each field was produced through visual  inspection using the QtClassify tool \citep{josie}. After an initial redshift guess of each object from LSDCat, refined redshifts of the LAEs were measured by fitting an asymmetric Gaussian profile to the Ly$\alpha$ emission line. Ly$\alpha$ fluxes were measured using the LSDCat \texttt{measure} functionality, adopting a 3D aperture of 3~Kron-like radii \citep{kron}; with the redshifts this also provides the Ly$\alpha$ luminosities.

Since our sample is based on emission lines without prior broadband selection, it includes galaxies with very faint continua but high equivalent widths -- sometimes even undetected in deep Hubble Space Telescope (HST) data \citep{maseda}. We identified the UV counterparts for our sample and measure their continuum flux densities and absolute UV magnitudes in various HST bands as described in detail by Kerutt et al.\ (in prep.). Our Ly$\alpha$ equivalent widths are based on combining the Ly$\alpha$ fluxes measured by LSDCat and continuum flux measurements from the HST counterparts. In cases when no continuum counterpart was detected, an upper limit to the continuum flux density was used to calculate lower limits to the absolute magnitudes and equivalent widths.

\subsection{LAE sample}
\label{sec:data1}
In this paper we focus on 68 fields of the MUSE-Wide survey located in the CANDELS/GOODS-S region, plus the HUDF09 parallel fields. Some of these fields are not yet included in the currently publicly available MUSE-Wide data; these will be part of the planned final data release. We decide to not take into account the 9 central fields in the MUSE-Deep area because of their different depth and selection function, in order to approach as much as possible a homogeneous sample and minimize systematic effects. Furthermore we also discard the 23 MUSE-Wide fields in the COSMOS region from this analysis because of their on average somewhat lower data quality. We keep the 8 MUSE-Wide pointings in the HUDF09 fields (see Fig.~\ref{fig:ra-dec}), since they give additional power to constrain the clustering signal at larger separations. In Appendix~\ref{appendix:k-fields}, we demonstrate that including the UDF09 parallels fields has no significant impact on our clustering results despite a minor decrease of the uncertainties.

\begin{figure}[tb]
\centering
\includegraphics[width=\columnwidth,height=6cm]{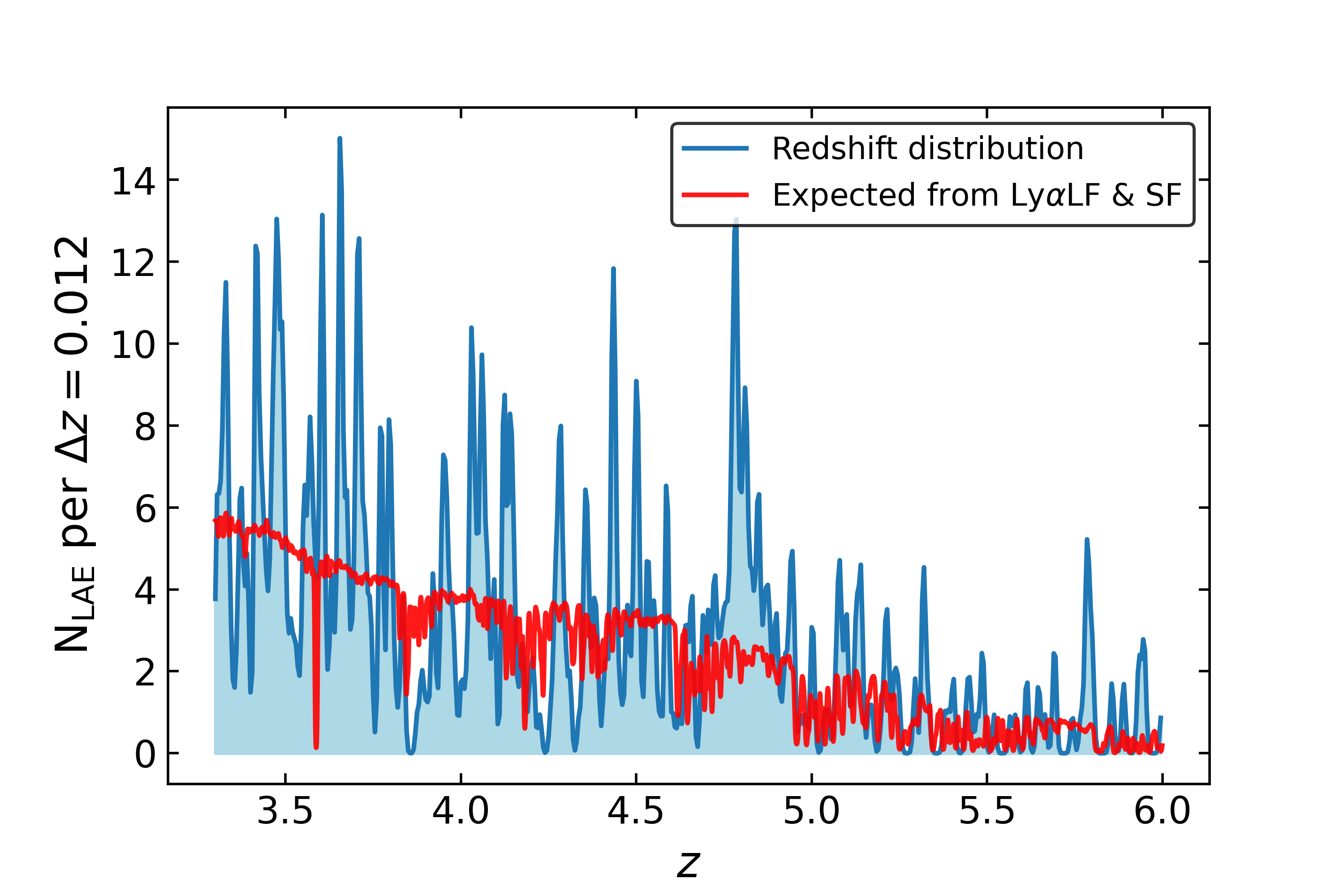}
\caption{KDE-filtered redshift distribution of the 695 LAEs of our sample, taken from 68 fields of the MUSE-Wide survey. The sample spans a redshift range of $3.3<z<6$. The kernel is chosen to be a Gaussian with standard deviation $\sigma_z$ $=$ 0.005. The expected $z$-distribution of an unclustered population following the Ly$\alpha$ luminosity function of \cite{herenz19} and the selection function of the MUSE-Wide survey is overplotted in red.} 
\label{fig:z_distribution}
\end{figure}

While MUSE is formally capable of detecting LAES with $2.91<z<6.65$, we limit the redshift range for this investigation to $3.3 < z < 6$, as the details of the selection function at redshifts close to the limits are still a matter of investigation. Thus, we arrive at a final number of 695 LAEs, distributed over 62.53 arcmin$^2$ (accounting for small field-to-field overlaps), implying an LAE density of slightly more than 11 objects per arcmin$^2$. The sample has a mean redshift of $\overline{z}$ $\approx$ 4.23, the median redshift is 4.12. The transverse extent of the footprint at $\overline{z}$ is $\sim$20 $h^{-1}$~Mpc.

\begin{table*}[ht]
   \caption[]{Properties of the LAE samples.} 
   \label{table:properties_table}
   \centering
    \begin{tabular}{l@{\qquad}cccccc}
        \hline \hline
           \noalign{\smallskip}
        LAE sample name & N$_{\rm{gal}}$  &  $\langle z \rangle$  &  $\rm{log}\langle L_{\rm{Ly}\alpha}$/[erg\; s$^{-1}]\rangle$   &  $\langle EW_{\rm{Ly}\alpha} \rangle / [\AA] $  &  $\langle M_{\rm{UV}} \rangle$  \\
             \noalign{\smallskip} 
            \hline \hline
            \noalign{\smallskip}
            Full sample  &  695   &  4.12 &   42.39 &   118.3 &   -18.4 \\
            Redshift $< 4.12$  &  348   &  3.63  &   42.31 &    109.9  &   -18.4 \\
            Redshift $> 4.12$  &  347   &  4.79 &   42.39 &    111.8 &   -18.4 \\
            log$L_{\rm{Ly}\alpha} <$ 42.36 &   349  &  4.03 &   42.14 &   110.0 &   -17.9   \\
            log$L_{\rm{Ly}\alpha} >$ 42.36   &  346   &   4.30 &   42.57  &    113.7 &   -19.0\\
            $EW_{\rm{Ly}\alpha} <$ 87.9  & 254 &   4.03  &   42.37 &    53.6 &   -19.5   \\
            $EW_{\rm{Ly}\alpha} >$ 87.9   & 255 &  4.05  &   42.45 &   167.3 &   -18.3 \\
            $M_{\rm{UV}} <$ -18.8  & 256 &   4.07  &   42.57 &   61.8   &   -19.6 \\
            $M_{\rm{UV}} >$ -18.8   & 253 &  3.92  &   42.19 &  168.6   &   -17.6       \\
            \noalign{\smallskip}
        \hline
    
         \multicolumn{6}{l}{%
          \begin{minipage}{12cm}%
          \vspace{0.2\baselineskip}
            \small \textbf{Notes}: Physical properties marked with $\langle \rangle$ represent median values for the galaxies (not pairs) in the sample.
          \end{minipage} 
          }\\
          \end{tabular}
\end{table*}

The spatial distribution of our LAEs is shown in Fig.~\ref{fig:ra-dec}, and the distribution over redshifts is presented in Fig.~\ref{fig:z_distribution}. For the latter we replaced the usual histogram counts-per-bin by a quasi-continuous kernel density estimator (KDE) to better represent the underlying probability distribution and avoid binning artefacts.
Superimposed on the KDE-filtered $z$ distribution we also show the distribution expected for an unclustered population of objects following the Ly$\alpha$ luminosity function (LF) of \cite{herenz19} and also factoring in the selection function of the MUSE-Wide survey. The curve has been normalized to the footprint size of our 68 fields.

While the formal average accuracy of our redshifts is $\Delta z\simeq 0.0007$ or $\pm 41$~km/s (limited by the accuracy of fitting the line), it is well-known that Ly$\alpha$ peak redshifts are typically offset by up to several hundreds of km/s from systemic (e.g. \citealt{hashimoto15,sowgat,kasper}), which would introduce a systematic error in the redshift-derived 3D positions of the LAEs along the line-of-sight (l.o.s.) of the order of $\sim$3~Mpc. We mitigate this systematic uncertainty by applying a correction to the Ly$\alpha$ redshifts following the two recipes described in \cite{verhamme18}: When the Ly$\alpha$ line presents two peaks with the red peak larger than the blue peak, we apply Eq. (1) in \cite{verhamme18}. When only a single peak is visible, we employ the correction given by Eq. (2) in \cite{verhamme18}. We show in Appendix~\ref{appendix:k-redshifts} that our method of measuring the clustering properties is insensitive to the details of this correction.

The range of Ly$\alpha$ luminosities ($L_{\rm{Ly}\alpha}$) of our galaxies is 40.91 $<$ log(\textit{L}$_{\rm{Ly}\alpha}/[\rm{erg\:s}^{-1}])$ $<$ 43.33, with a median Ly$\alpha$ luminosity of $\langle \log(L_{\rm{Ly}\alpha}/[\rm{erg\:s}^{-1}])\rangle  = 42.36$, the range of UV absolute magnitudes is -22.4 $<$ $M_{\rm{UV}}$ $<$ -16.8, with a median of $\langle M_{\rm{UV}} \rangle =-18.4$, and the range of rest frame equivalent widths is 10.2 $<$ $EW_{\rm{Ly}\alpha}$ $<$ 794.9 \AA, with a median of $\langle EW_{\rm{Ly}\alpha} \rangle  = 118.3$ \AA. 


\subsection{LAE subsets}
\label{sec:subsamples}

In order to explore the dependence of the clustering amplitude on physical properties of LAEs, we divide the original sample into subsamples based on different available properties. In each case we split the full sample at the median value of the LAE property under question in order to have (nearly) the same number of objects in each of the two subsets. The subsamples are summarized in Table~\ref{table:properties_table} and defined in greater detail in the following.

A first split in redshift around $\langle z \rangle  = 4.12$ leads to a low-$z$ subset of 348 LAEs with median redshift $\langle z_{\rm{low}} \rangle  = 3.56$ and a high-$z$ subset of 347 LAEs with $\langle z_{\rm{high}} \rangle  =4.59$, respectively. The median Ly$\alpha$ luminosities and equivalent widths of the two redshift subsamples are nearly the same (differences of 0.08 dex and 2 \AA, respectively). There is no difference between the median $M_{\rm{UV}}$.  

In order to explore possible clustering dependencies on Ly$\alpha$ luminosity, we generate two subsamples divided by Ly$\alpha$ luminosities. We split the full sample at $\langle \log(L_{\rm{Ly}\alpha}/[\rm{erg\:s}^{-1}]) \rangle = 42.36$. The low- and high-$L_{\rm{Ly}\alpha}$ subsamples hold 349 and 346 LAEs, respectively. Their median redshifts are $\langle z_{\rm{low L}} \rangle = 4.03$ and $\langle z_{\rm{high L}} \rangle = 4.30$. The median $\log(L_{\rm{Ly}\alpha})$ of the subsamples differs by 0.43 dex.

While at $z\simeq 3$ most of our LAEs have a photometric HST counterpart, at $z>5$ only around 50\% of the objects are detectable in the available HST images (Kerutt et al. in prep.). Hence, for those objects we can only adopt $M_{\rm{UV}}$ and $EW_{\rm{Ly}\alpha}$ lower limits, which would skew the $EW_{\rm{Ly}\alpha}$ and $M_{\rm{UV}}$ distributions for the higher redshift subset. Therefore we decide to eliminate the LAEs without HST counterparts when splitting by $EW_{\rm{Ly}\alpha}$ or $M_{\rm{UV}}$. This reduces our sample size from 695 to 509 LAEs.

We then split the HST-detected sample by equivalent width at $\langle EW_{\rm{Ly}\alpha} \rangle =87.9\,\AA$. The low- and high-$EW_{\rm{Ly}\alpha}$ subsample consists of 254 and 255 LAEs, respectively. The median redshifts and luminosities of these samples are very similar (see Table~\ref{table:properties_table}).

Finally, we divide the HST-detected LAE sample by absolute magnitude at $\langle M_{\rm{UV}} \rangle =-18.8$, leading to low- and high-$M_{\rm{UV}}$ subsets (bright and faint, respectively) of 256 and 253 LAEs. The $\langle M_{\rm{UV}} \rangle $ values differ between these two subsamples by 1.59 dex, the $\langle \rm{log}(L_{\rm{Ly}\alpha}) \rangle $ values by only 0.32~dex. 

\section{Methods}
\label{sec:methods}

\subsection{K-estimator}
\label{sec:k-estimator}

\subsubsection{Basic principles}
\label{sec:k-principles}

The specifics of MUSE as a survey instrument present a serious challenge for the commonly used 2-point correlation function (2pcf) to measure galaxy clustering. By design, a MUSE survey spans a wide redshift range but covers only small (spatial) regions in the sky. The MUSE-Wide footprint has already the largest transverse footprint of all MUSE surveys, but its nature is still that of a pencil-beam survey. While transverse scales in the MUSE-Wide survey span up to $\sim$20 $h^{-1}$Mpc, radial scales exceed the 1000 $h^{-1}$Mpc. The limitations of the transverse extent impede the application of the Jackknife technique to compute realistic uncertainties (see Sect.~\ref{sec:errorsK}) and methods such as bootstrapping fail in the 2pcf. Besides, given our spatial ranges, exploiting the redshift coverage rather than the spatial extent is strongly preferred. We thus explore possible alternatives to the 2pcf. In \cite{catrina} we applied the so-called K-estimator, introduced by \cite{adelberger} to analyse the clustering of Lyman Break Galaxies, in a subset of our pencil-beam survey. Here we build on our previous work by extending it to a larger dataset, but also paying attention to optimization aspects and comparing the method with the 2pcf.

\begin{figure}[t]
\centering
\includegraphics[width=\columnwidth]{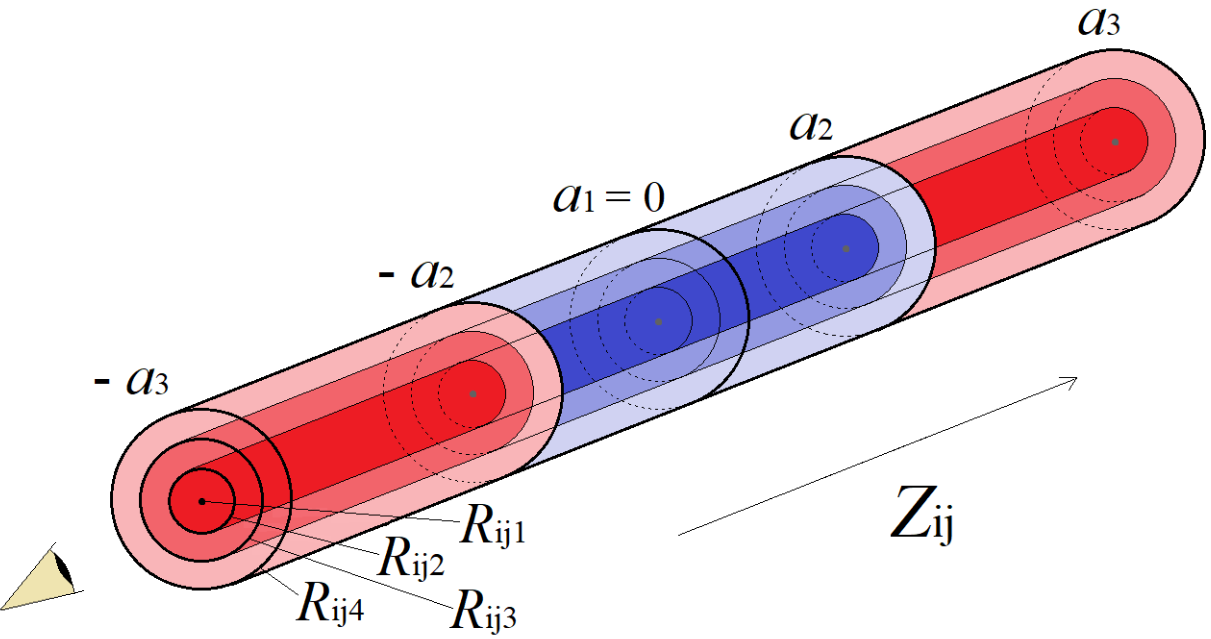}
\caption{Illustration of the K-estimator. We show three nested cylinders representing three bins of transverse separations. The number of galaxy pairs inside each blue cylindrical shell from $a_1 = 0$ to $\pm a_2$ is $N_{0, a_2}$, the number of pairs in each red cylindrical shell between $a_3 - a_2$ and $-a_2 - (-a_3)$ is $N_{a_2, a_3}$. The K-estimator for each shell is then the ratio of pair counts between the inner (blue) segment to the total (blue plus red) segment. For illustration purposes we depict linear $R_{ij}$ bins, although in practice we use a logarithmic binning scheme.}
\label{fig:Kdrawing}
\end{figure}

The K-estimator focuses on radial clustering along the l.o.s. by counting pair separations in redshift space at fixed transverse distances. In contrast to the 2pcf, no random sample is needed because the K-estimator computes the ratio between small and small+large scales. This quantity is directly related to the underlying correlation function. We adopt the following notation: Considering two galaxies with indices $i$ and $j$, their transverse distance is $R_{ij}$ (equivalent to $r_p$ in the 2pcf), and their l.o.s. redshift-space separation is $Z_{ij}$ (equivalent to $\pi$ in the 2pcf). We then count the number $N$ of pairs within a given $R_{ij}$ bin, for two different ranges of $Z_{ij}$, $\lvert a_1\rvert < Z_{ij} < \lvert a_2\rvert$ and $\lvert a_2\rvert < Z_{ij} < \lvert a_3\rvert$. The K-estimator is defined as the ratio of the numbers of galaxy pairs $N_{a_1, a_2}(R_{ij})$ and $N_{a_2, a_3}(R_{ij})$ between these two consecutive cylindrical shells, i.e.,
\begin{equation}
    K_{a_2, a_3}^{a_1, a_2}(R_{ij}) = \frac{N_{a_1, a_2}(R_{ij})}{N_{a_1, a_2}(R_{ij}) + N_{a_2, a_3}(R_{ij})}, 
    \label{Kestimator}
\end{equation}
as a function of transverse separation $R_{ij}$. We set $a_1=0$ $h^{-1}$~Mpc so that the K-estimator quantifies the excess of galaxy pairs in the range $0 < Z_{ij} < a_2$ with respect to the larger l.o.s.\ range $0 < Z_{ij} < a_3$. In other words, the K-estimator can be expressed as $K(R_{ij})= N_{0,a_2}(R_{ij}) / N_{0,a_3}(R_{ij})$. This concept is schematically illustrated in Fig.~\ref{fig:Kdrawing}. Here, ($a_2 - 0$) and ($a_3 - a_2$) are the lengths of the two cylinders within which the numbers of pairs are counted. 

The transverse distance $R_{ij}$ between LAE pairs is taken in bins of $R_{ij}$, corresponding to different cylindrical shells in Fig.~\ref{fig:Kdrawing}. These shells are defined by their radii $R_{ij}$ and their lengths $a_2,\; a_3$ in the $Z$ direction. For illustration purposes we display $R_{ij}$ in Fig.~\ref{fig:Kdrawing} using a linear scaling ($R_{ij2} - R_{ij1} = R_{ij3} - R_{ij2}$ etc.) although in practice we adopt a logarithmic spacing of subsequent transverse separations. Note that in this figure each $R_{ij}$ and $Z_{ij}$ combination corresponds to a galaxy \emph{pair} and not just a single galaxy.

$K_{a_2, a_3}^{a_1, a_2}$ is related to the 2pcf through the mean value of the correlation function $\overline{\xi}$ \citep[see][]{adelberger}
    \begin{multline}
    \label{eq:expected_k}
     \langle  K_{a_2, a_3}^{a_1, a_2}(R_{ij})  \rangle  \simeq {(a_2-a_1) \cdot \displaystyle\sum_{i>j}^{\text{pairs}}[1 + \overline{\xi}{_{a_1, a_2}}]} \; \times \\
     \left\lbrace(a_2-a_1) \cdot \displaystyle\sum_{i>j}^{\text{pairs}}[1 + \overline{\xi}{_{a_1, a_2}}] + (a_3-a_2) \cdot \displaystyle\sum_{i>j}^{\text{pairs}}[1 + \overline{\xi}{_{a_2, a_3}}]\right\rbrace^{-1},
     \end{multline}
where $\overline{\xi}{_{a_1, a_2}}$ is
\begin{equation}
\overline{\xi}{_{a_1, a_2}} = \frac{1}{a_2 - a_1}\int_{a_1}^{a_2}{\text{d}Z_{ij} \cdot \xi(R_{ij}, Z_{ij})} \label{xi_mean}
\end{equation}
and corresponds to the mean correlation function that would be theoretically measured in the blue region in Fig.~\ref{fig:Kdrawing}. The same is applied for $\overline{\xi}{_{a_2, a_3}}$ in the red region of Fig.~\ref{fig:Kdrawing}. The function $\xi(R_{ij}, Z_{ij})$ can be represented by a power law through the Limber Equation \citep{limber} in spatial coordinates $\xi(R_{ij},Z_{ij})= (\sqrt{R_{ij}^{2}+Z_{ij}^{2}}/r_0)^{-\gamma}$ or modelled with a Halo Occupation Distribution model. 

The understanding of this estimator is quite intuitive. If galaxies were randomly distributed in space ($\xi(r)=0$), the expected number of galaxy pairs at each l.o.s.\ separation would be equal. Thus, from Eq.~\eqref{eq:expected_k} and with $a_1=0$, $K_{a_2, a_3}^{0, a_2}$ is simply the ratio of volumes between the two cylindrical shell segments, $(a_2 - 0)/(a_2 - 0 + a_3 - a_2) = a_2/a_3$. Hence if $a_3 = 2a_2$, the expectation value for an unclustered galaxy population would be $K = 0.5$; if for a specific sample the value of $K$ is significantly above 0.5, we have detected a clustering signal. Note however that while this criterion (applied by both \citealt{adelberger} and \citealt{catrina}) seems natural, there is no a priori reason to keep the restriction $a_3 = 2a_2$. In fact, allowing for $a_3/a_2 > 2$ provides the analysis with a more solid statistical baseline against which the clustering signal can be evaluated. This is addressed in Sect.~\ref{sec:optimizing-k}.

\cite{adelberger} applied Eq.~\eqref{eq:expected_k} and the Limber Equation to estimate the correlation length $r_0$ while keeping the power law slope $\gamma$ of the correlation function fixed. They first measured the K-estimator in a single $R_{ij}$ bin $R_{\text{cut}}<5$ $h^{-1}$Mpc which captures the $R_{ij}$ scale for which the clustering signal is largest. They then applied Eqs.~\eqref{eq:expected_k} and \eqref{xi_mean} to predict the expectation values $\langle K \rangle $ for different assumed values of $r_0$, selecting the correlation length for which the predicted value of $K$ was closest to the measured value as their best estimate. The same procedure was adopted by \citet{catrina} in their analysis of a MUSE-Wide subset of LAEs. We refer to this approach in the following as the ``one-bin fit'' method.

Besides this simple approach to estimate $r_0$ at fixed $\gamma$ we also implemented a more elaborate procedure to fit the K-estimator with a power law correlation function with both $\gamma$ and $r_0$ as free parameters. For this we integrate $\xi(r)$ over both $Z_{ij}$ ranges as in Eq.~\eqref{xi_mean}, for each $R_{ij}$ bin and for each combination of a grid in ($r_0$, $\gamma$). Plugging the values of these integrals into Eq.~\eqref{eq:expected_k} to calculate $\langle K \rangle$ for each $R_{ij}$ bin, we obtain a global $\chi^2$ value for each grid point by summing over the squared deviations between predicted and observed values of $K$ relative to the statistical error bars (obtained by bootstrapping as explained in Sect.~\ref{sec:errorsK}). Our best-fit parameters are then finally taken as the ($r_0$, $\gamma$) grid point with the smallest $\chi^2$. For the estimation of confidence intervals, we face the complication that the $K$ values in subsequent $R_{ij}$ bins are correlated because each galaxy contributes to multiple pairs at various separations. We explain in Sect.~\ref{sec:errorsK} how we obtain realistic uncertainties for the fit parameters.

\begin{figure*}
\centering
\begin{tabular}{c c}
  \centering
  \includegraphics[width=.49\linewidth]{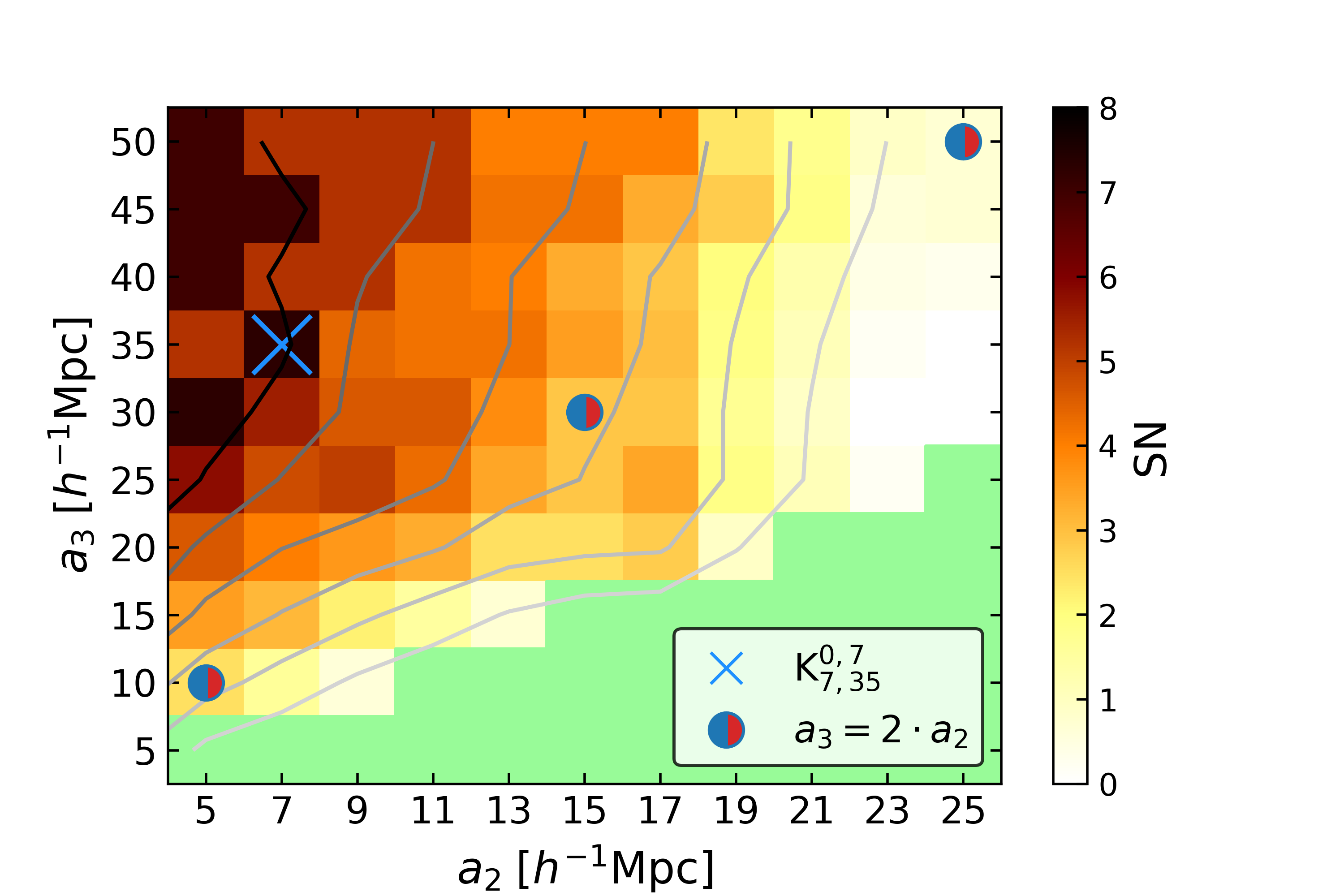}
\end{tabular}%
\begin{tabular}{c c}
  \centering
  \includegraphics[width=.49\linewidth]{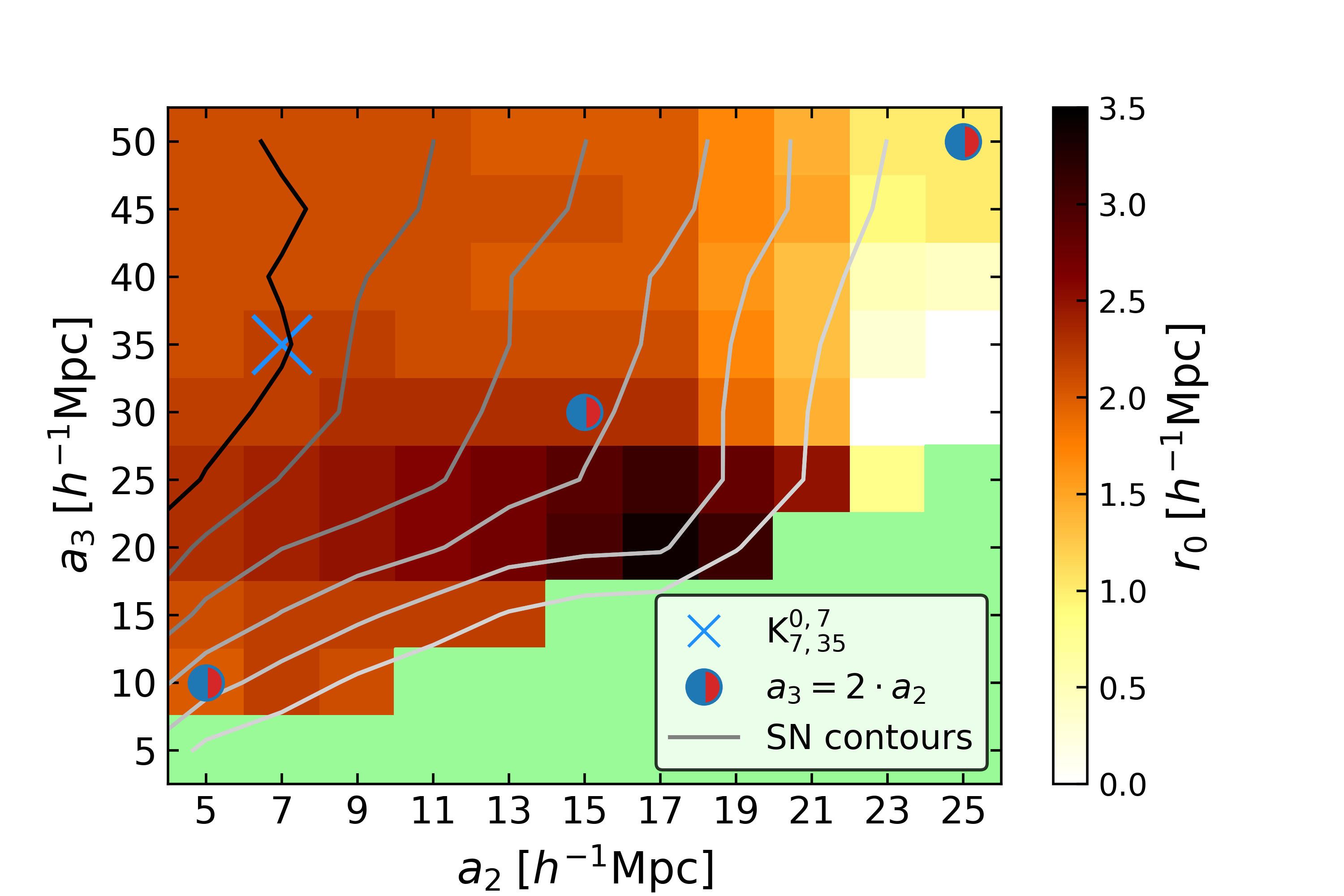}
\end{tabular}
\caption{Results of our grid study to optimize the K-estimator. Left: SN ratios obtained for each evaluated combination of ($a_2$, $a_3$), displayed as a color map. The green area indicates the ``forbidden'' range where $a_3 < a_2$. The contours trace SN increments of 2, slightly smoothed for display purposes. Right: The same but for the correlation length $r_0$, except that the contours again follow the values of SN. The blue-red colored circles represent grid points with $a_3 = 2a_2$ for which the blue-red cylinders in Fig.~\ref{fig:Kdrawing} are equally long. The blue cross indicated our adopted parameter combination for the clustering analysis, as it provides the highest SN ratio and reaches saturation in $r_0$.}
\label{fig:optimizing-k}
\end{figure*}


\subsubsection{Optimizing the K-estimator}
\label{sec:optimizing-k}

The parameters $a_2$ and $a_3$ in the definition of the K-estimator can in principle be chosen freely. We now explore for which values we obtain the best sensitivity for the clustering signal and the highest signal-to-noise ratio (SN). We compute SN from the error bars of the correlation lengths. This procedure is similar to finding the optimal $\pi_{\text{max}}$ saturation value in the case of the 2pcf, where $\pi_{\text{max}}$ is increased until most of the correlated pairs are included, while even larger values of $\pi_{\text{max}}$ only add noise to the measurement.

We perform a grid search with the full sample over the different combinations of $K^{a_1,a_2}_{a_2,a_3}$, but setting $a_1=0$ throughout. We vary $a_2$ within 5--25~$h^{-1}$Mpc in steps of 2~$h^{-1}$Mpc and $a_3$ within 5--50~$h^{-1}$Mpc in steps of 5~$h^{-1}$Mpc, with the additional restriction $a_3 \geq a_2$. We adopt 15 logarithmic bins in the range $0.6 < R_{ij} < 12.8$~$h^{-1}$Mpc, discarding $R_{ij}$ bins with less than 16 galaxy pairs. We use the one-bin fit described above with a fixed canonical $\gamma$ value of $\gamma=1.8$ (\citealt{adelberger,durkalek,ouchi17}) to calculate the correlation length $r_0$ and the SN for each combination ($a_2$, $a_3$). 

The results are shown in Fig.~\ref{fig:optimizing-k}. The left panel reveals that the SN ratio is highest for small $a_2$ and large $a_3$ values, while it decreases towards $a_2 \approx a_3$. Parameter combinations with $a_2=a_3/2$ as adopted in the two previous studies that used the K-estimator \citep{adelberger, catrina} are represented by the colored circles; it is evident that these combinations are far from optimal in bringing out the clustering signal with maximal significance.

The right panel of Fig.~\ref{fig:optimizing-k} shows that in the upper left range of the diagram where the SN is highest, the best-fit value of $r_0$ is also insensitive to the specific parameter combination. 
On the other hand, larger values of $a_2$ and smaller values of $a_3$ degrade the SN ratio. 
Comparable $a_2$ and $a_3$ values (tiny red and large blue cylinders in Fig.~\ref{fig:Kdrawing})
result in $N_{a_2, a_3}(R_{ij})<<N_{a_1, a_2}(R_{ij})$, with $N_{a_2, a_3}(R_{ij})$ strongly varying with the exact value of $a_3$. This translates into large uncertainties when computing $r_0$ from the K-estimator (see Eq.~\ref{Kestimator}). These errors are however not reflected in the right panel of Fig.~\ref{fig:optimizing-k} but are clearly visible in the low SN values on its left panel. The largest $r_0$ values correspond therefore to the most uncertain values but agree well within their uncertainties with the $r_0$ value accepted by us. We adopt the combination $a_2 = 7$~$h^{-1}$Mpc and $a_3 = 35$~$h^{-1}$Mpc, marked with a blue cross in both plots, for the rest of this paper as the grid point giving the highest SN and a $r_0$ within the saturation values. In the following we thus always refer to the specific estimator $K_{7,35}^{0,7}$ which quantifies the ratio of the number of galaxy pairs with l.o.s.\ separations between $-7 < Z_{ij}/h^{-1}\rm{Mpc} < 7$ and between $-35 < Z_{ij}/h^{-1}\rm{Mpc} < 35$ at given transverse distance $R_{ij}$. The expectation value of this estimator for an unclustered population is $(a_2 - a_1)/(a_3 - a_1) = 0.2$.

 
\subsubsection{Error estimation}
\label{sec:errorsK}

The individual data points from clustering statistics are correlated. One galaxy can contribute to galaxy pairs in more than one $R_{ij}$ bin. In order to account for this correlation one would use the jackknife resampling technique and compute a covariance matrix \citep[see e.g.][]{mirko10}. However, that method requires a division of the sky area into several independent regions, each of which must be large enough to cover the full range of scales under consideration. Due to the small sky area of our survey this approach is not feasible here. Poisson uncertainties, even if commonly used, might underestimate the real uncertainties. We therefore consider several alternatives to derive meaningful uncertainties in Appendix~\ref{sec:errors} and choose the most conservative approach.

Thus, we apply the bootstrapping technique detailed in \cite{ling} (and similar as in \citealp{durkalek}) to determine the statistical uncertainties of our data points. We create pseudo-data samples by randomly drawing 695 LAEs from our parent sample, allowing for repetitions. We generate 500 different pseudo-samples and compute the K-estimator in all of them. The standard deviations of $K$ in each $R_{ij}$ bin are adopted as error bars. We verify the robustness of our error approach in Appendix~\ref{sec:errors}.

With the bootstrapped uncertainties and the uncorrelated $\chi^2$ statistics, the uncertainties of the clustering parameters can be derived. However, we suspect that naively applying an uncorrelated $\chi^2$ analysis with the standard confidence threshold can also lead to an underestimation of the clustering uncertainties. Therefore, we test this hypothesis by investigating the behavior of the error bars when the bin size is modified. While one would expect a decrease of the individual uncertainties when the bin size is increased, we expect an increase of the error bars if the bin size is decreased.

We compute new bootstrapping error bars for five different $R_{ij}$ bin sizes (half size, double size, three times larger, four times larger and five times larger than the current binning). The error bar sizes do not vary significantly when the $R_{ij}$ bin size is modified, contrary to the expectation of the standard $\chi^2$ method.
 
We therefore recalibrate the $\chi^2$ analysis to determine realistic 68.3\% and 95.5\% confidence levels in the following way: With each of our bootstrapped samples delivering a best-fit minimal value of $\chi^2_{\text{min},i}$ corresponding to ($r_{0,i},\gamma_i$), we assume that the posterior distribution of these $\chi^2_{\text{min},i}$ approximately describes the true confidence regions. We compute the $\chi^2_i$ values using the corresponding ($r_{0,i},\gamma_i$) combinations and our real data. We sort these $\chi^2_i$ into ascending order and adopt the 68.3\% and 95.5\% parameter ranges with respect to the sorted bootstrapped $\chi^2_i$ values as marginalized single-parameter error bars. This posterior distribution is also used to provide combined confidence regions on both $r_0$ and $\gamma$. Throughout the paper, we will refer to this fitting approach as ``PL-fit''. 

\subsection{Two-point correlation function}

The 2pcf is undoubtedly the most frequently used statistic to investigate galaxy clustering. Although we argued above that it is less suited than the K-estimator for a pencil-beam survey such as MUSE-Wide, we include a 2pcf analysis of our sample for comparison in Appendix~\ref{2pcf}. We note that this is in fact the first time that such an analysis has been performed on a 100\% spectroscopically confirmed sample of LAEs. However, the challenge of estimating realistic uncertainties in the case of the 2pcf is even more problematic (due to the survey design) than for the K-estimator. We present in Appendix~\ref{2pcf} an in depth presentation and discussion of the 2pcf on our LAE sample. In summary, we show that the results from the K-estimator and 2pcf agree within their uncertainties.

\subsection{Bias and typical Dark Matter Halo masses from power-law fits}
\label{sec:bias}

The clustering strength is characterized by the large-scale bias factor $b$, which relates the distribution of galaxies to that of the underlying dark matter density. The bias factor has often been derived from the characteristic correlation length $r_0$ and the PL slope $\gamma$ by fitting a PL to the clustering signal \citep[e.g.][]{peebles1980}. 
Given $b$ one can also derive typical host DMH masses. Within the concept of linear bias, the evolution of $b$ with redshift is given by the ratio of the density variance of galaxies $\sigma_{8,\rm{gal}}(z)$ over that of dark matter $\sigma_{8,\rm{DM}}(z)$
\begin{equation}
\label{eq:bz}
  b(z) = \frac{\sigma_{8,\rm{gal}}(z)}{\sigma_{8,\rm{DM}}(z)} .
\end{equation}

 For a power-law 2pcf the relation between $\xi(r)$ and the density variance $\sigma_{8,\rm{gal}}(z)$ \citep{peebles1980, miyaji} is given by 
  \begin{align}
  \label{eq:LAES_DM_xi}
    & \xi(r,z) = \left(\frac{r}{r_0}\right)^{-\gamma} \nonumber \\
    & \sigma_{8,\rm{gal}}(z)^2 = \xi(r_8,z) \times J_2,
  \end{align}
where $\\xi(r_8,z)$ is the correlation function evaluated in spheres of comoving radius $r_8 = 8$ $h^{-1}$Mpc and $J_2 = 72/[(3-\gamma)(4-\gamma)(6-\gamma)2^\gamma]$. Simultaneously, for the DM case
\begin{equation}
\label{eq:sigma8}
  \sigma_{8,\rm{DM}}(z) = \sigma_8 \frac{D(z)}{D(0)} 
\end{equation}
with $D(z)$ as the linear growth factor.

Inserting Equations \eqref{eq:LAES_DM_xi} and \eqref{eq:sigma8} into Eq.~\eqref{eq:bz} we obtain the bias factor as a function of the growth factor
\begin{equation}
\label{eq:b}
  b(z) = \left[\frac{r_8}{r_0(z)}\right]^{-\gamma/2} \frac{J_2^{1/2}}{\sigma_8 D(z)/D(0)}.
\end{equation}

Following the bias evolution model described in \cite{sheth}, we can compute the large-scale Eulerian bias factor $b_{\text{Eul}}$ and compare it to the bias given by Eq.~\ref{eq:b} in order to estimate DMH masses. To calculate $b_{\text{Eul}}$, we consider linear overdensities in a sphere which collapses in an Einstein-de Sitter Universe at $\delta_{\text{cr}} = 1.69$. The linear root mean square fluctuations correspond to the mass at the epoch of observation $\nu = \delta_{\text{cr}}/\sigma_{8,\rm{DM}}(M_{\rm{DMH}},z)$. The theory behind the $\sigma_{8,\rm{DM}}(M_{\rm{DMH}},z)$ calculation is developed in \cite{bosch}. 
\subsection{Halo Occupation Distribution modeling}
\label{sec:hod}

It is known that bias factors and DMH masses inferred from PL fits suffer from systematic errors (e.g. \citealt{jenkins} and references therein). 
A PL correlation function treats scales in the linear and non-linear regime alike and does not differentiate between pairs of objects belonging to the same DMH and pairs residing in different halos. Even for fits performed only in the linear regime, the correlation function still deviates from the PL shape. A more appropriate treatment is achieved through HOD modeling which explicitly combines the separate contributions from the 1- and the 2-halo terms.

The HOD model we use here is an improved version of the model set presented by \citep{miyaji11,mirko12,mirko15,mirko18}. To be consistent with these studies, we use the bias-halo mass relation from \cite{tinker}, the halo mass function of \cite{sheth}, the dark matter halo profile of \citet{nfw97}, and the concentration parameter from \citet{zheng07}. We use the weakly redshift-dependent collapse overdensity $\delta_{\rm{cr}}$ \citep{nfw97,vandenbosch13}. We further include the effects of halo-halo collisions and scale-dependent bias by \citet{tinker} as well as redshift space distortions using linear theory \citep[Kaiser infall,][]{kaiser,vandenbosch13} to the 2-halo term only (see Appendix~\ref{appendix:discussion_rsd}).   

The mean occupation function is a simplified version of the five parameter model by \citet{zheng07}, where we fix the halo mass at which the satellite occupation becomes zero to $M_0=0$ and the smoothing scale of the central halo occupation lower mass cutoff to  $\sigma_{\log M}=0$.

In this simplification, the mean occupation distribution of the central galaxies can be expressed by
\begin{equation}
\label{eq:Nc}
  \langle N_{\text{c}}(M_{\text{h}})\rangle = 
  \begin{cases}
  \; 1 & (M_{\text{h}}\geq M_{\text{min}}) \\
  \; 0 & (M_{\text{h}}< M_{\text{min}})
  \end{cases}
\end{equation}
and that of the satellite galaxies $\langle N_{\text{s}}(M) \rangle$ as
\begin{equation}
\label{eq:Ns}
  \langle N_{\text{s}}(M_{\text{h}}) \rangle = \langle N_{\text{c}}(M_{\text{h}}) \rangle \cdot \left(\frac{M_{\text{h}}}{M_1}\right)^\alpha,
\end{equation}
where $M_{\rm{min}}$ is the mass scale of the central galaxy mean occupation, $M_1$ is the the mass scale of a DMH that hosts (on average) one additional satellite galaxy, and $\alpha$ is the high-mass slope of the
satellite galaxy mean occupation function.

We apply the model to obtain the $\xi(r)$ based on HOD modeling and convert the calculated $\xi(r)$ to the K-estimator using Eq.~\ref{eq:expected_k}. The minimum transverse separation of our observed $K$-estimator is $\sim$0.6~$h^{-1}$Mpc, where the 1-halo term contribution to $\xi(r)$ is typically a few to several percent. This is too low for obtaining robust constraints on the 1-halo term to perform a full HOD modeling. We therefore restrict our analysis to an estimate of the bias parameter by fitting the expected K-estimator based on only the 2-halo term to the observations. We hold $\alpha = 1$ and log$M_1/M_{\text{min}}=1$ fixed and vary only $M_{\text{min}}$ to find the best-fit model and calculate the bias parameter. This probes the typical DMH mass for the sum of central and satellite galaxy halo occupations, $N(M_{\rm h})=N_{\rm c}(M_{\rm h})+N_{\rm s}(M_{\rm h})$, without being able to distinguish between these two. 
\begin{figure*}
\centering
\begin{tabular}{c c}
  \centering
  \includegraphics[width=.49\linewidth]{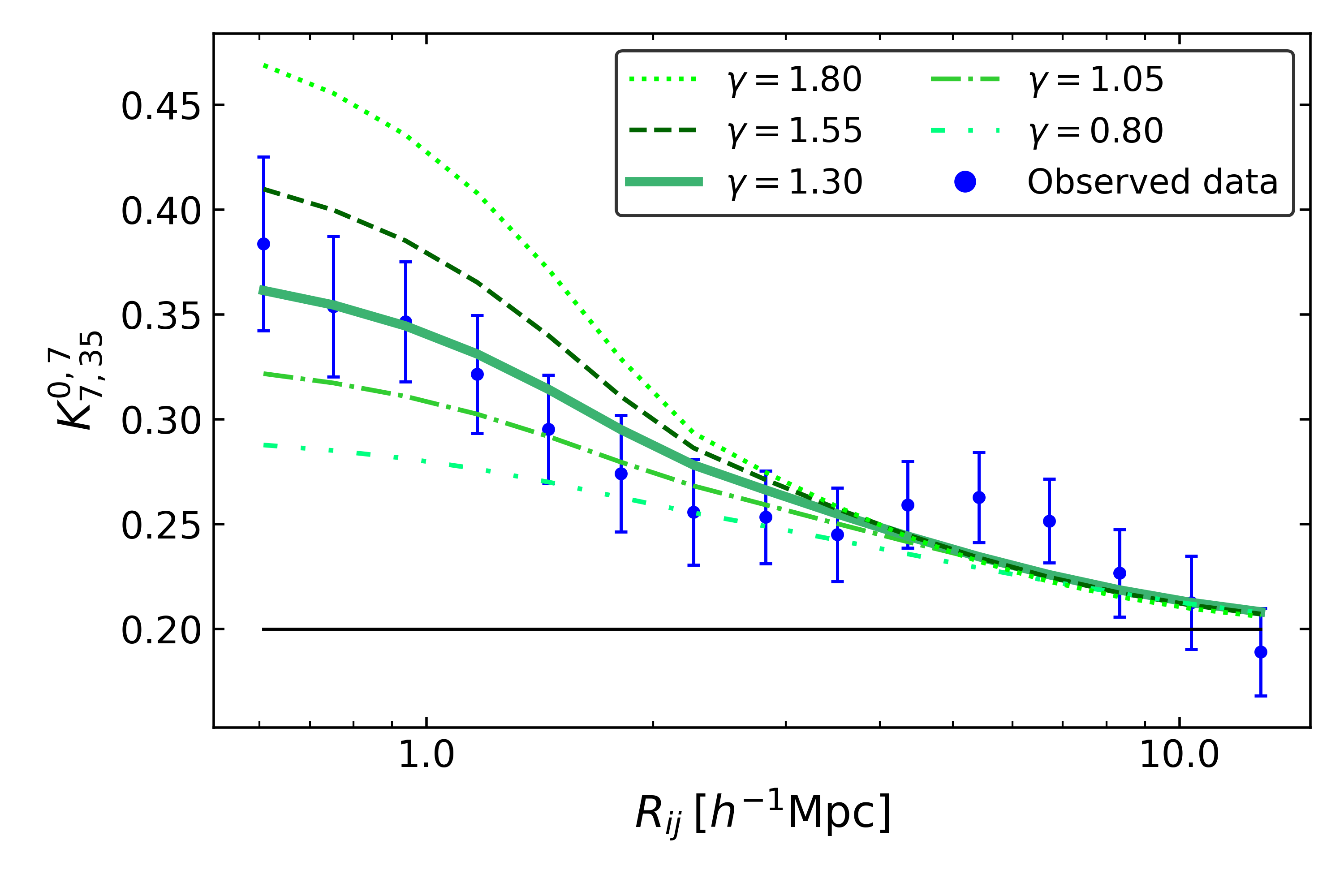}
\end{tabular}%
\begin{tabular}{c c}
  \centering
  \includegraphics[width=.49\linewidth]{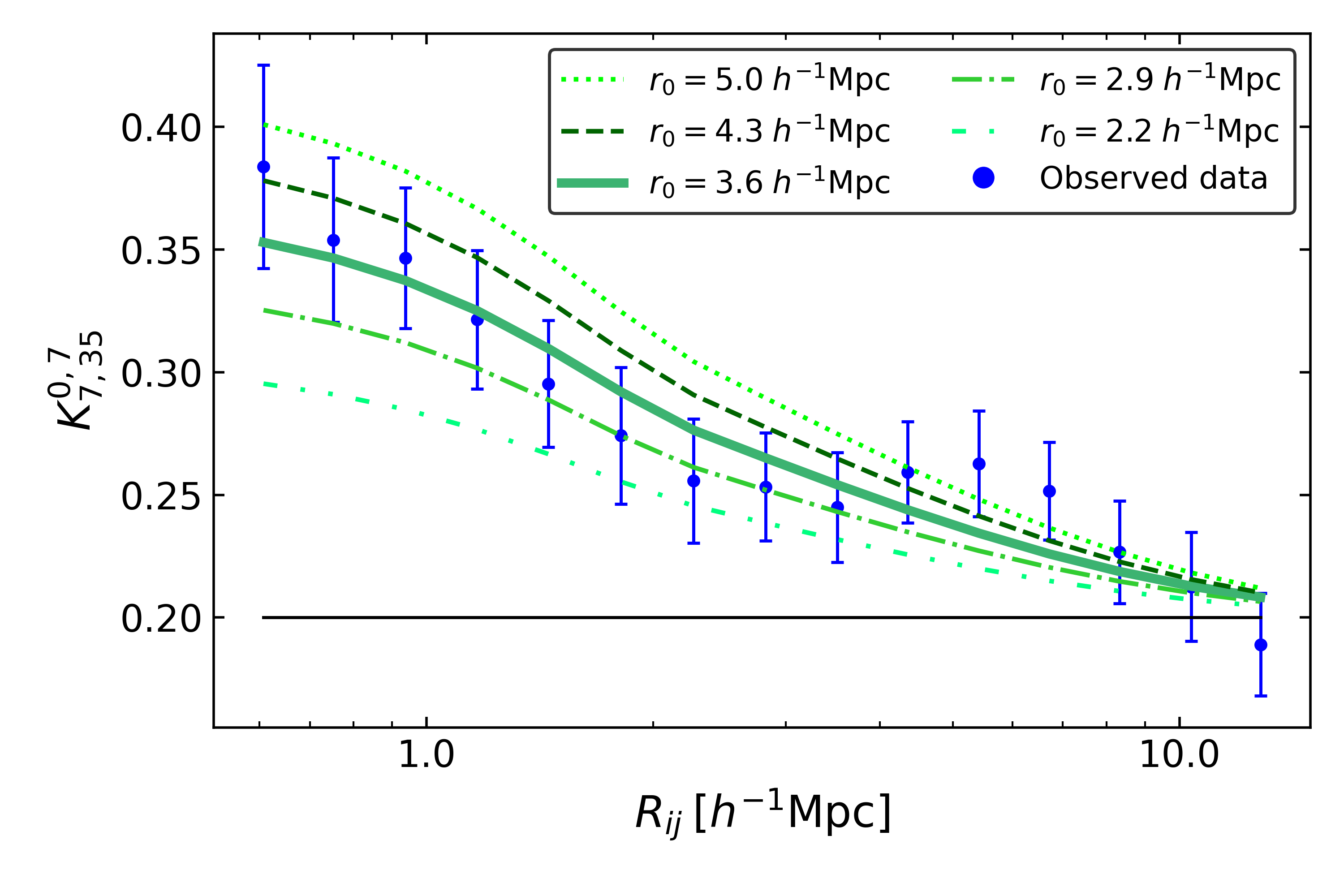}
\end{tabular}
\caption{Measured values of the K-estimator as a function of transverse distance (points with error bars) compared to the expected behaviour for a population strictly following a power-law correlation function. Left: The five curves represent different power-law indices as given in the legend, for a fixed value of $r_0=3.6\:h^{-1}$Mpc. Right: The same for five different correlation lengths at fixed $\gamma=1.3$. The central (thick solid) curves always indicate the minimum $\chi^2$ best-fit values. The horizontal straight line shows the no-clustering expectation value of $K$. The error bars are calculated with the bootstrapping technique described in Sect.~\ref{sec:errorsK}.}
\label{fig:k-gamma-r0}
\end{figure*}
The details of the HOD models (e.g. $M_1$ and $\alpha$) do not affect the  typical DMH mass estimations since we only fit the two-halo term. Some HOD modeling applications in the literature also use number density constraints \citep[e.g., Equation 18 of ][]{miyaji11}. This is however only relevant if the 1-halo term contributes significantly, which is not the case here. Thus we do not need to employ any number density constraints. The HOD model is evaluated at the median redshift of $N(z)^2$ where $N(z)$ is the redshift distribution of the sampled galaxies. For our dataset, $z_{\rm pair}=3.82$.

As above for the PL-fit parameters (Sect.~\ref{sec:errorsK}) we estimate the uncertainties of the inferred bias factor by fitting the 500 bootstrapped samples with the 2-halo term HOD modeling and obtain the 500 best bias factors from the bootstrapped samples. Those best 500 HOD models are then compared to the observed K-estimator data points to compute the bootstrapped $\chi^2$ values. We sort the bootstrapped $\chi^2_{\text{min}}$ values in ascending order and use these to recalibrate the 68.3\% (1$\sigma$) confidence interval.

\section{Results}
\label{sec:results}
\subsection{K-estimator}

Adopting the optimized K-estimator $K_{7,35}^{0,7}$ (see Sect.~\ref{sec:optimizing-k}), we measure the clustering of our LAE sample in 15 logarithmic bins of transverse separations $R_{ij}$ between 0.6 and 12.8~$h^{-1}$Mpc, with error bars calculated by bootstrapping the sample as explained in Sect.~\ref{sec:errorsK}. Fig.~\ref{fig:k-gamma-r0} shows the results for the full LAE sample. It is evident that the values of $K$ are significantly above the no-clustering expectation value of 0.2.

We verify that our clustering results are not affected by the accuracy of our redshifts (see Appendix~\ref{appendix:k-redshifts}), also taking into account our statistical corrections for the expected offset between Ly$\alpha$-based and systemic redshifts (see Sect.~\ref{sec:data1}). We emphasize that the K-estimator is insensitive to these redshift errors because of the broad ($\pm 7$~$h^{-1}$Mpc) window over which the numerator in Eq.~\eqref{Kestimator} is evaluated.

A somewhat puzzling feature, at least at first sight, is the broad hump in the $K(R_{ij})$ profile around $4 \lesssim R_{ij}/h^{-1}\rm{Mpc} \lesssim 7$, suggesting a slight excess in the clustering strength for such separations (or alternatively, a dent at $2 \lesssim R_{ij}/h^{-1}\rm{Mpc} \lesssim 4$). We test for the possibility that this feature might be introduced as an artefact of the sample footprint shape by dividing the sample into an “Eastern” and a “Western” half. Since we find the hump/dent in both subsets, as is also the case when splitting the sample by LAE properties (see Sect.~\ref{sec:discussion}), we rule out a systematic effect due to the footprint. Recalling the fact that the data points in Fig.~\ref{fig:k-gamma-r0} are strongly correlated, we underline that the significance of the feature is actually below $2\sigma$, and we consider it to most likely be due to a statistical fluctuation in the spatial distribution of the sample. The only robust test of this explanation would require an independent but statistically equivalent comparison sample, which we do not have at our disposal. We however removed the data points of the hump/dent and test the possible effect of this feature on our fits to the K-estimator. We find the same clustering parameters (within 1$\sigma$) as in the next section. For the purpose of this paper we treat the hump/dent as an insignificant statistical fluctuation not related to a true clustering excess of the MUSE-Wide LAEs.

We also checked that our clustering signal is insensitive to including or excluding the objects from the 8 HUDF09 parallel fields ($\Delta b=0.03$; see Appendix~\ref{appendix:k-fields}), again confirming the robustness of the K-estimator on the survey footprint.

\subsection{Power law fits}
\label{sec:results_traditional}

First, we apply the single-bin fit method to our clustering signal to compare our results to earlier studies which also computed the K-estimator and evaluated its strength by using the single-bin fit approach. 
We derive the best-matching correlation length $r_0$ at fixed $\gamma = 1.8$ as described in Sect.~\ref{sec:k-principles}. The calculated value of $K_{7,35}^{0,7}$ for $R_{ij,\text{max}} < 5 \; h^{-1}$Mpc corresponds to $r_0 = 2.10 \pm 0.20$ $h^{-1}$Mpc. The outcome of this single-bin fit depends somewhat on the adopted $R_{ij,\text{max}}$: Lowering the limit to 3~$h^{-1}$Mpc results in $r_0=1.90_{-0.20}^{+0.30}$ $h^{-1}$Mpc, whereas increasing $R_{ij,\text{max}}$ to 7~$h^{-1}$Mpc delivers $r_0=2.60_{-0.10}^{+0.20}$ $h^{-1}$Mpc. In principle this dependence should be included in the error bar on $r_0$. We also vary the fixed value of $\gamma$ between 1.0 and 2.0 and find that $r_0$ does not change by more than 1$\sigma$. Our single-bin fit results agree with those in \cite{catrina} but give much tighter constraints on $r_0$.

   \begin{table*}[htbp]
   \caption[]{Clustering parameters from the different fit approaches to the K-estimator in our full sample.} \label{table:main_table}
   \centering
    \begin{tabular}{l@{\qquad}cccccc}
        \hline \hline
           \noalign{\smallskip}
        Fit method & $\gamma$ &  $r_0 \; [h^{-1}$Mpc]   & \; $b_{\rm{PL}}$ &  $b_{\rm{HOD}}$ & $\log(M_{\rm{DMH}}$ / $[\it{h}^{-1}\rm{M}_{\odot}])$\\
             \noalign{\smallskip} 
            \hline \hline
            \noalign{\smallskip}
            HOD fit   &   -- &   -- &  -- & $2.80^{+0.38}_{-0.38}$ & $11.34^{+0.23}_{-0.27}$             \\
            Two-parameter PL-fit  & $1.30^{+0.36}_{-0.45}$ &   $3.60^{+3.10}_{-0.90}$ & $3.03^{+1.51}_{-0.52}$ &   --  &   -- \\
            One-parameter PL-fit  & fixed 1.8&   $2.60^{+0.72}_{-0.67}$ & $2.02^{+0.22}_{-0.24}$ &   -- &  -- \\
            Single-bin fit  &fixed 1.8&   $2.10^{+0.20}_{-0.20}$  &  $1.66^{+0.14}_{-0.14}$  &  -- &  -- \\
            \noalign{\smallskip}
        \hline 
        
        \multicolumn{6}{l}{%
          \begin{minipage}{14cm}%
          \vspace{0.2\baselineskip}
            \small \textbf{Notes}: The typical DMH masses for the full sample are derived only from our HOD results. The uncertainties in the bias factors reflect the statistical error on $r_0$ only.
          \end{minipage} 
          }\\
    \end{tabular}
\end{table*}

\begin{figure}[h]
\centering
\includegraphics[width=\columnwidth]{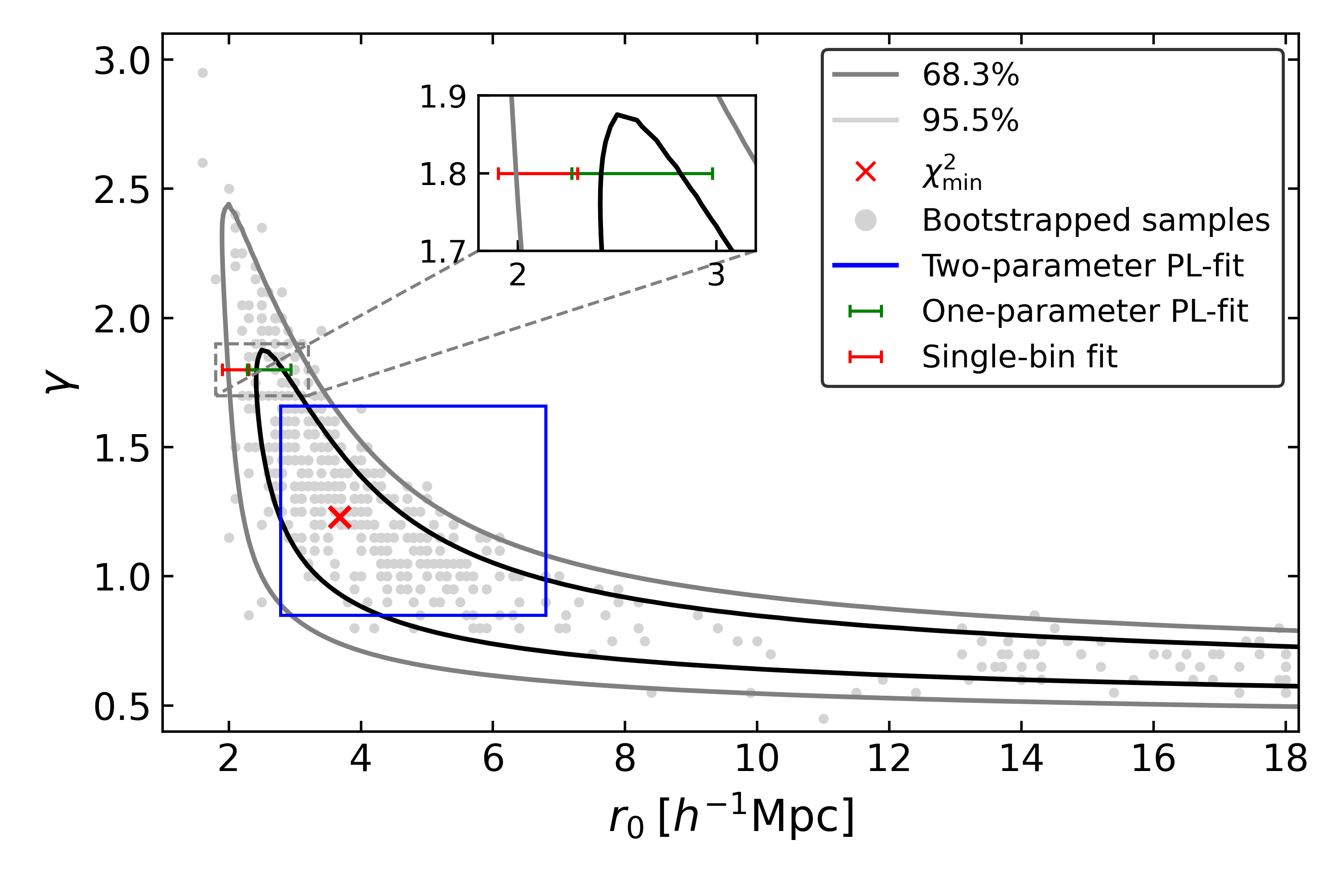}
\caption{Simultaneous fit to $r_0$ and slope $\gamma$. The black (dark grey) contour represents the 68.3\% (95.5\%) confidence. The red cross stands for the lowest  $\chi^2$ value at ($r_0 = 3.65$, $\gamma = 1.25$). The points show the 500 best-fit values from the 500 bootstrapped samples. The blue rectangle indicates the 16\% and 84\% percentiles from the marginalized single-parameter posterior distributions of the bootstrapped samples. The green (red) error bar represents the correlation length from the one-parameter PL (single-bin) fit with fixed $\gamma=1.8$. For a better visualization, we show a zoom onto the region containing these fits.}
\label{fig:k-contours}
\end{figure}

Motivated by these results we proceed to estimate both parameters simultaneously. Since in the single-bin approach the choice of $R_{ij,\text{max}}$ does affect the fit result, we now switch to fitting the K-estimator over the \textit{full} measured range of transverse separations using all bins in $0.6 < R_{ij}/h^{-1}\rm{Mpc} < 12.8$ (see Sect.~\ref{sec:k-estimator}).

To obtain a visual impression of how $K_{7,35}^{0,7}(R_{ij})$ depends on $\gamma$ and $r_0$ separately we overplot the expected curves for five different values of each quantity into Fig.~\ref{fig:k-gamma-r0}, always keeping the other parameter fixed. It can be seen that $K$ reacts in different ways to changes in the two parameters. Increasing $r_0$ leads to an elevated $K$ at all $R_{ij}$ scales, whereas increasing $\gamma$ results in changes of $K$ mainly at small transverse separations. Because the shape of $K_{7,35}^{0,7}(R_{ij})$ changes differently for $r_0$ and $\gamma$, it is in principle possible to fit both parameters simultaneously. We perform an uncorrelated $\chi^2$ analysis over a grid of $r_0$ and $\gamma$ to find the best-fit parameters as described in Sect.~\ref{sec:k-principles}.

Following the procedure laid out in Sect.~\ref{sec:errorsK} we compute confidence contours for $r_0$ and $\gamma$ by fitting the 500 bootstrapped samples in the same way. The marginalized single-parameter (1D) 16\%--84\% confidence regions are $\gamma=1.30^{+0.36}_{-0.45}$ and $r_0=3.70^{+3.10}_{-0.92}$. These and the corresponding 2-dimensional 68.3\% and 95.5\% confidence contours are displayed in Fig.~\ref{fig:k-contours}, along with the 500 best-fit parameter sets from the bootstrapped pseudo-data samples. 
\begin{figure}[h]
\centering
\includegraphics[width=\columnwidth]{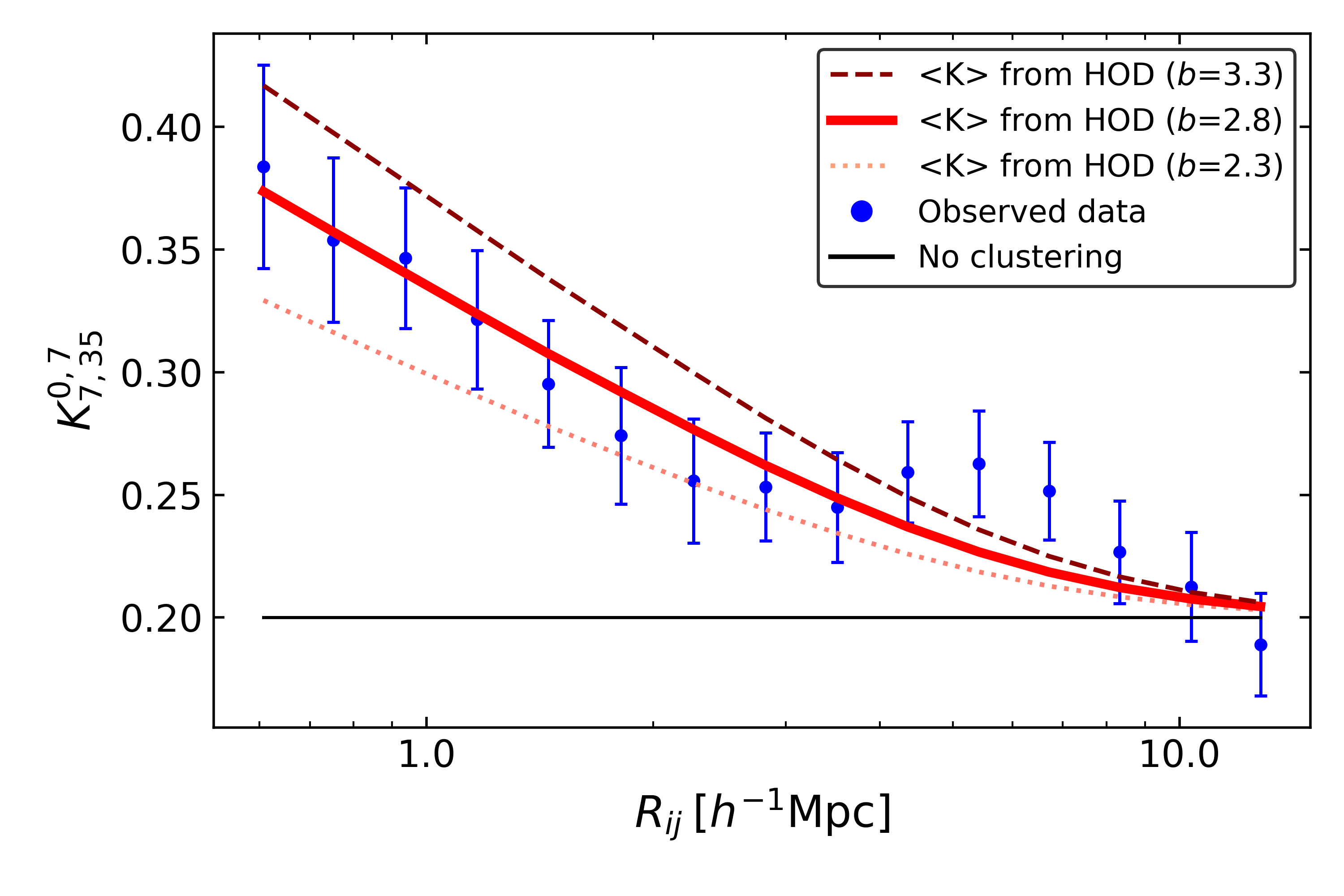}
\caption{Dependence of the HOD fits to the K-estimator on the large-scale bias factor. The dotted, solid and dashed red curves show three different bias factors $b=2.3, 2.8, 3.3$, respectively. The thicker solid red curve shows the $b$ that provides the lowest $\chi^2$ value. The $K$ values and their respective error bars are the same as in Fig.~\ref{fig:k-gamma-r0}.}
\label{fig:k-hod}
\end{figure}
For a visual comparison we also plot the estimations from the single-bin fit. We further include the results from a one-parameter PL-fit with fixed $\gamma=1.8$ for an easier comparison with the literature in Sect.~\ref{sec:literature}.

The best-fit correlation length of 2.1~$h^{-1}$Mpc obtained by the single-bin fit (at fixed $\gamma=1.8$) is lower than suggested by the one- and two-parameter fits and it is not compatible with its 68.3\% probability contour. This was expected because the single-bin fit was not optimized for $r_0$, SN and $R_{ij}$ range. We also observe a large similarity between the medians of the marginalized single-parameter posterior distributions ($r_0 = 3.60^{+3.10}_{-0.90} \;h^{-1}$Mpc, $\gamma = 1.30^{0.36}_{0.45}$) and the combination of parameters that provide the lowest $\chi^2$ value ($r_0 = 3.65$, $\gamma = 1.25$). It is also evident from Fig.~\ref{fig:k-contours} that the fit is quite degenerate between $r_0$ and $\gamma$ in the sense that parameter combinations with higher $\gamma$ and lower $r_0$ are only slightly less likely than the best-fit combination. Different $R_{ij}$ scales are affected when modifying $\gamma$ or $r_0$ (see Fig.~\ref{fig:k-gamma-r0}). This results in similarly good PL-fits when combinations of low $\gamma$ and high $r_0$ or high $\gamma$ and low $r_0$ are applied. Taking into account the sensitivity of the single-bin fit to the value of $R_{ij, \text{max}}$, the three results are in fact very similar. We therefore adopt the PL fitting approach also for our subsequent investigation of the dependence of clustering on LAE physical properties. 
This eases the comparison to the literature, where mainly PL fits are performed. The values and errors of the best-fit parameters from the different fit approaches are summarized in Table \ref{table:main_table}. 

\begin{figure}[h]
\centering
\includegraphics[width=\columnwidth]{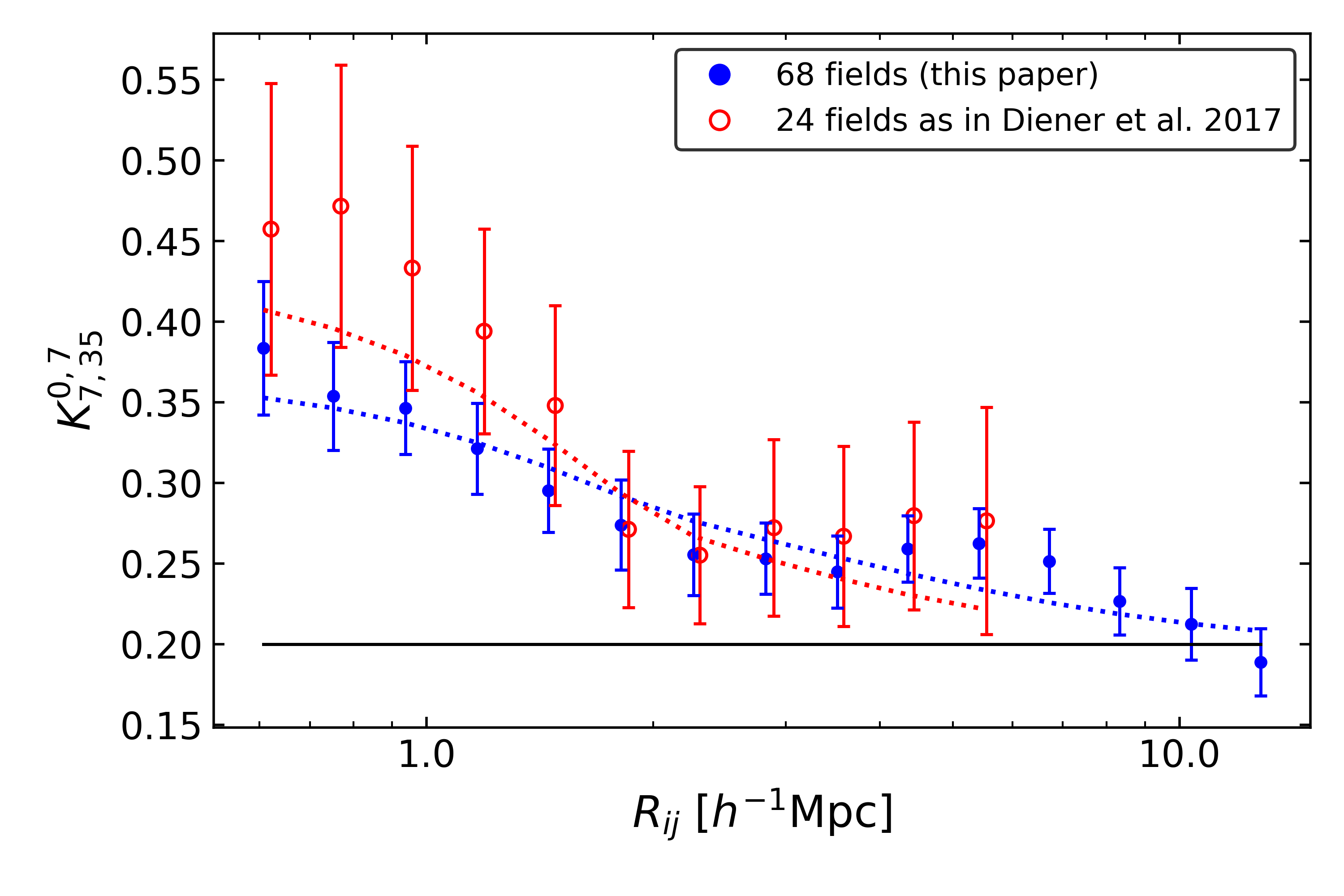}
\caption{Measured values of the K-estimator of our sample of 68 MUSE-Wide fields (blue filled circles) compared to the subset of 24 fields considered in \citet[][D17; open red circles]{catrina}. The error bars are again calculated with the bootstrapping technique described in Sect.~\ref{sec:errorsK}. The blue dotted curve represents our two-parameter best PL-fit. The red dotted curve uses the best single-bin fit results of D17 ($r_0=2.9^{+1.0}_{-1.1}$ cMpc for a fixed $\gamma=1.8$) applied to our PL-method (two-parameter PL-fit).}
\label{fig:catrina-estimator}
\end{figure}

The confidence contours of our fit are essentially open towards large $r_0$ and low $\gamma$. In fact our bootstrap sample contains a sizeable proportion of instances with best-fit combinations in the lower right corner of Fig.~\ref{fig:k-contours} (11.8\% with $r_0>10\;h^{-1}$Mpc). Upon investigation of these ``solutions'' we find that they correspond to almost constant K-estimator values with respect to $R_{ij}$, driven by the tentative hump around 5~$h^{-1}$Mpc. 
Whatever the actual origin of these extreme points, it seems clear that from the K-estimator alone without further priors we can only constrain plausible combinations of $r_0$ and $\gamma$ at one end of the distribution. 

While it appears that the best-fit power-law index for our LAEs tends to be substantially shallower than the results from other studies based on NB imaging that use the fiducial $\gamma$ value of $\gamma=1.8$ \citep[e.g.][]{ouchi10,ouchi17}, we note that they are generally compatible at the 1--2$\sigma$ level. 
The same is true for the values obtained from our own sample using the 2pcf method (see Appendix~\ref{2pcf}).

\subsection{Halo Occupation Distribution fit}

We now match the HOD model (Sect.~\ref{sec:hod}) to our measured K-estimator. Similarly to Fig.~\ref{fig:k-gamma-r0} we first visualize the basic behaviour of the HOD model for different large-scale bias factors; this is shown in Fig.~\ref{fig:k-hod}. 
Higher values of $b$ increase the expectation values of $K$ at most separation scales, but most strongly for small $R_{ij}$. Following the procedure described in Sect.~\ref{sec:hod} we recalibrate the confidence contours and obtain a best-fit large-scale bias of $b_{\rm HOD} = 2.80^{+0.38}_{-0.38}$. The corresponding typical DMH mass is $\log (M_{\text{DMH}}/[h^{-1}\text{M}_\odot]) = 11.34^{+0.23}_{-0.27}$.  

The best-fit HOD model behaves in several aspects similarly to the best-fit PL correlation function. Even if PL fits do not have a physical basis, the PL model seems to perform slightly better in terms of matching the observed $K$ values and reaching a slightly lower $\chi^2$ value, but these differences are not significant. The bias values derived from the two fits are also fully consistent as discussed in Sect.~\ref{sec:discussion_hod-pl}.

\begin{figure}[h]
\centering
\includegraphics[width=\columnwidth]{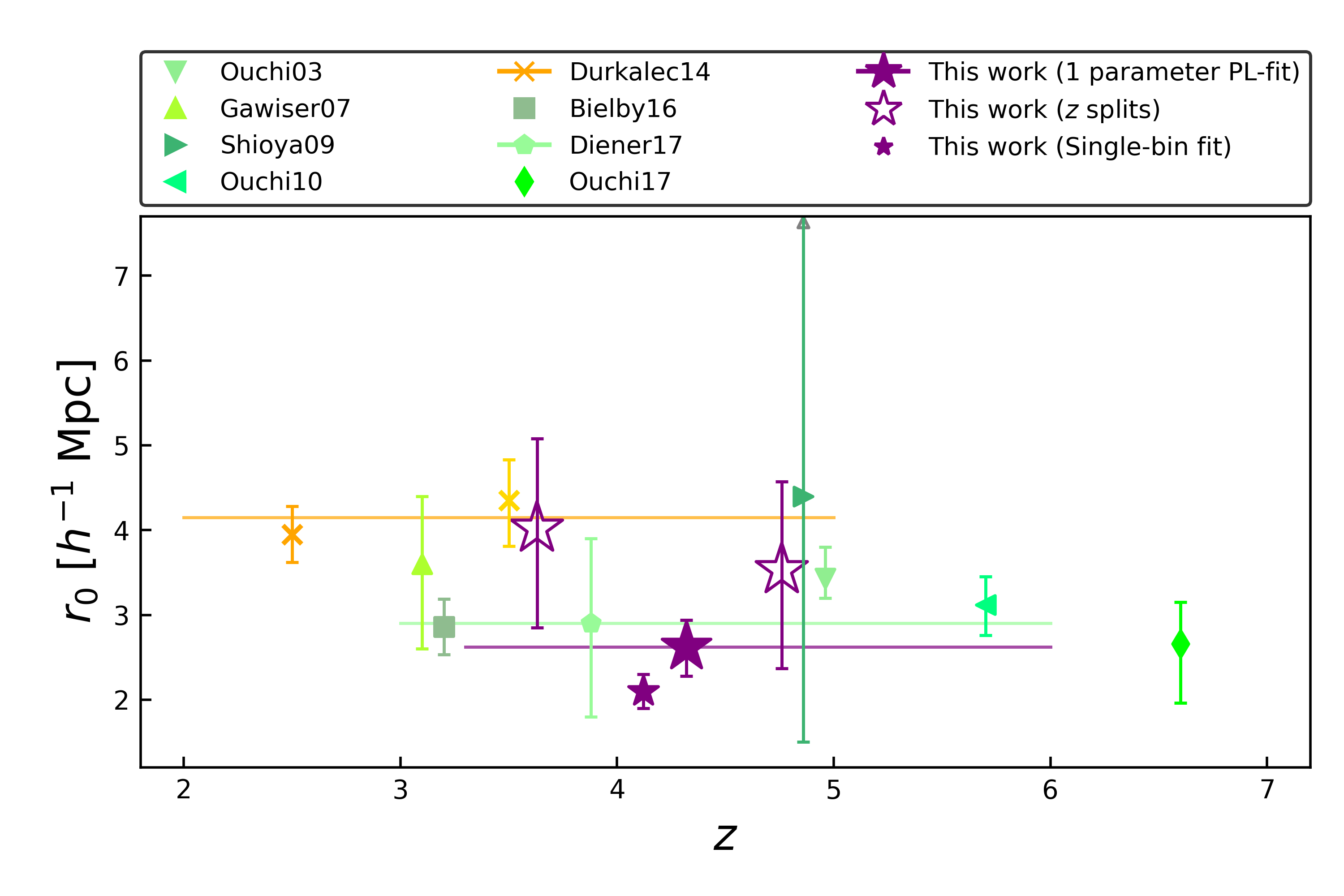}
\caption{Comparison of the derived correlation lengths to the literature. The $r_{0}$ values calculated in this study are represented with purple stars. Green symbols correspond to studies of samples based on Ly$\alpha$ selected galaxies. The samples from \citealp{durkalek} at $z\sim2.5$ and $z\sim3.5$ (dark and light yellow) are based on continuum-selected high-$z$ galaxies. The horizontal colored bars indicate the redshift ranges of the corresponding studies (spectroscopic surveys). The redshift range of the $z$-subsamples of this paper are not plotted for a better visibility. Values for $r_0$ are plotted at the median redshift of the samples. The $r_{0}$ from \cite{ouchi03} and \cite{bielby} have been shifted by +0.1 along the x-axis for visual purposes. Our one-parameter PL-fit with fixed $\gamma=1.8$ by +0.2. The upper limit of the $r_0$ from \cite{shioya} corresponds to $r_0=10.1$ Mpc.}
\label{fig:comparison-literature}
\end{figure}

We investigate the effects of the redshift space distortions (RSD) in Appendix~\ref{appendix:discussion_rsd}, where we show that the RSD do not have a significant effect on the HOD fit for K-estimator. 

\section{Discussion}
\label{sec:discussion}

\subsection{Comparison to \cite{catrina}}
\label{sec:discussion_catrina}

We first compare our results with those of our pilot study \citep[][D17]{catrina} which employed the non-optimized K-estimator $K_{25,50}^{0,25}$ for a subset of 24 fields of our current sample. In order to visualize the statistical gain of our new investigation we applied our improved $K_{7,35}^{0,7}$ estimator to the 196 LAEs at $3.3<z<6$ in the same 24 fields. The outcome of this comparison is shown in Fig.~\ref{fig:catrina-estimator}.

While the two datasets show excellent agreement given the uncertainties, as expected the error bars are much smaller in our new sample. The clustering signal of the 24 fields appears a bit higher, but the differences are at most 1$\sigma$. The smaller footprint of the 24 fields dataset limits the range of transverse separations $\la 6\; h^{-1}$Mpc. The clustering curves from the two samples are fitted with a PL correlation function, based on the results from D17 for the 24 fields and on our best PL-fit for the 68 fields. Fig.~\ref{fig:catrina-estimator} also shows that the power-law fits to the 68 fields follow much better the data points than in the 24 fields since we performed a simultaneous fit of $r_0$ and $\gamma$. Following the same procedure as in Sect.~\ref{sec:errorsK} for the 24 fields, we find $r_0=2.85^{+0.73}_{-0.76} \; h^{-1}$Mpc and $\gamma=1.62^{+1.18}_{-0.82}$. 
These results are very close to the numbers obtained in D17 ($r_0=2.9^{+1.0}_{-1.1}$ cMpc for a fixed $\gamma=1.8$), but our improved procedure substantially decreased the error bars for the same data.


\begin{figure}[h]
\centering
\includegraphics[width=\columnwidth]{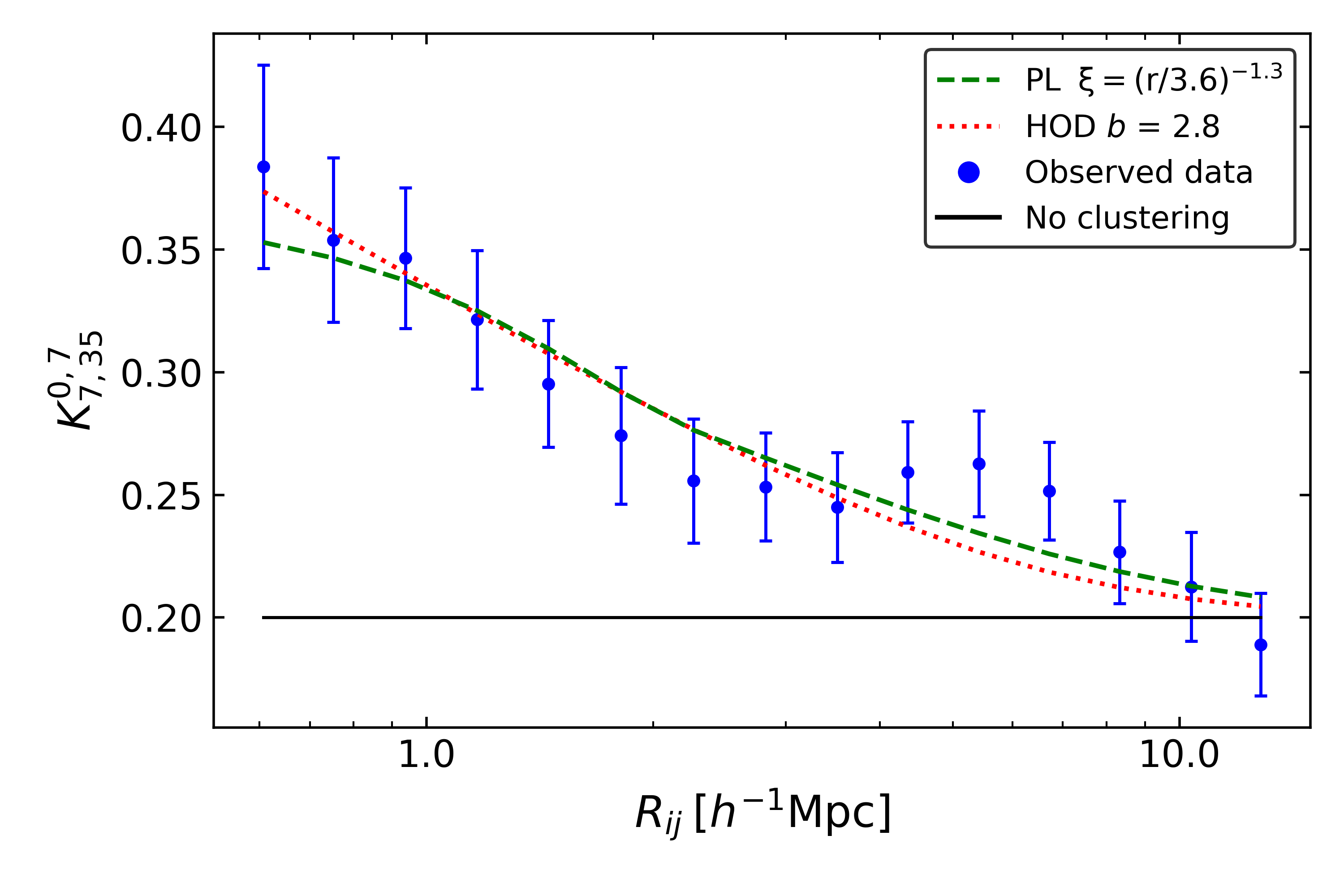}
\caption{Best PL and HOD fits to the K-estimator. The dashed green curve shows the PL-fit (same as the solid curves in Fig.~\ref{fig:k-gamma-r0}) while the dotted red curve represents the HOD fit (same as the thick curve in Fig.~\ref{fig:k-hod}). The measurements of $K_{7,35}^{0,7}$ are the same as in Figures \ref{fig:k-gamma-r0} and \ref{fig:k-hod}.}
\label{fig:pl-hod}
\end{figure}

\subsection{Comparison with the literature}
\label{sec:literature}
Most previously published work on the clustering of high-redshift galaxies is restricted to the estimation of $r_0$ at fixed power-law index $\gamma$, with the latter typically assumed to be 1.8 or thereabout. While our best-fit value for $\gamma$ based on the K-estimator is considerably lower, Fig.~\ref{fig:k-contours} shows that $\gamma$ values around 1.8 are still consistent with our data. To make a fair comparison, in Sect.~\ref{sec:results_traditional} we recompute the best-fitting power law with $\gamma$ fixed to 1.8; thus only allowing $r_0$ to vary.

Furthermore the clustering strength and thus the correlation length are predicted to evolve with cosmic time, and therefore the (average) redshifts of the
samples must also be taken into account in any comparison.

We first consider clustering measurements of LAEs selected by NB surveys. Here all objects are assumed to have the same redshift defined by the NB filter. Early studies  \citep{ouchi03,gawiser07,shioya} focused on small samples of LAEs (up to 160 objects) at $z= 3.1-4.86$ to compute angular correlations. The correlation lengths at fixed $\gamma=1.8$ (except \citealp{shioya}, who calculated $\gamma=1.90 \pm 0.22$) are consistent with our recomputed PL-fits, in particular when considering the involved uncertainties. The correlation lengths are in the range of $r_0\approx 2.5-4.5\;h^{-1}$Mpc, higher values corresponding to higher redshift samples. 
More recent studies based on NB surveys \citep{ouchi10,bielby,ouchi17} at higher redshifts ($z\approx 3-6.6$) hold much larger samples (up to 2000 objects) where they find slightly higher correlation lengths, $r_0=3-5\;h^{-1}$Mpc. Given the similarity between these and lower redshift samples, our derived correlation lengths are also in fair agreement with most recent LAE clustering studies.

We then consider clustering measurements of high-redshift galaxies selected based on photometric redshifts or magnitude and colour-colour criteria (mainly Lyman-break galaxies). \cite{durkalek,durkalekp} computed the real-space 2pcf on samples of more than 3000 objects at $2<z<5$ distributed over more than 0.8 deg$^2$. The sample is more suited for clustering studies than our MUSE-Wide survey because their large spatial coverage diminishes the effect of cosmic variance and allows the computation of the traditional 2pcf method. 
\begin{figure}[h]
\centering
\includegraphics[width=\columnwidth]{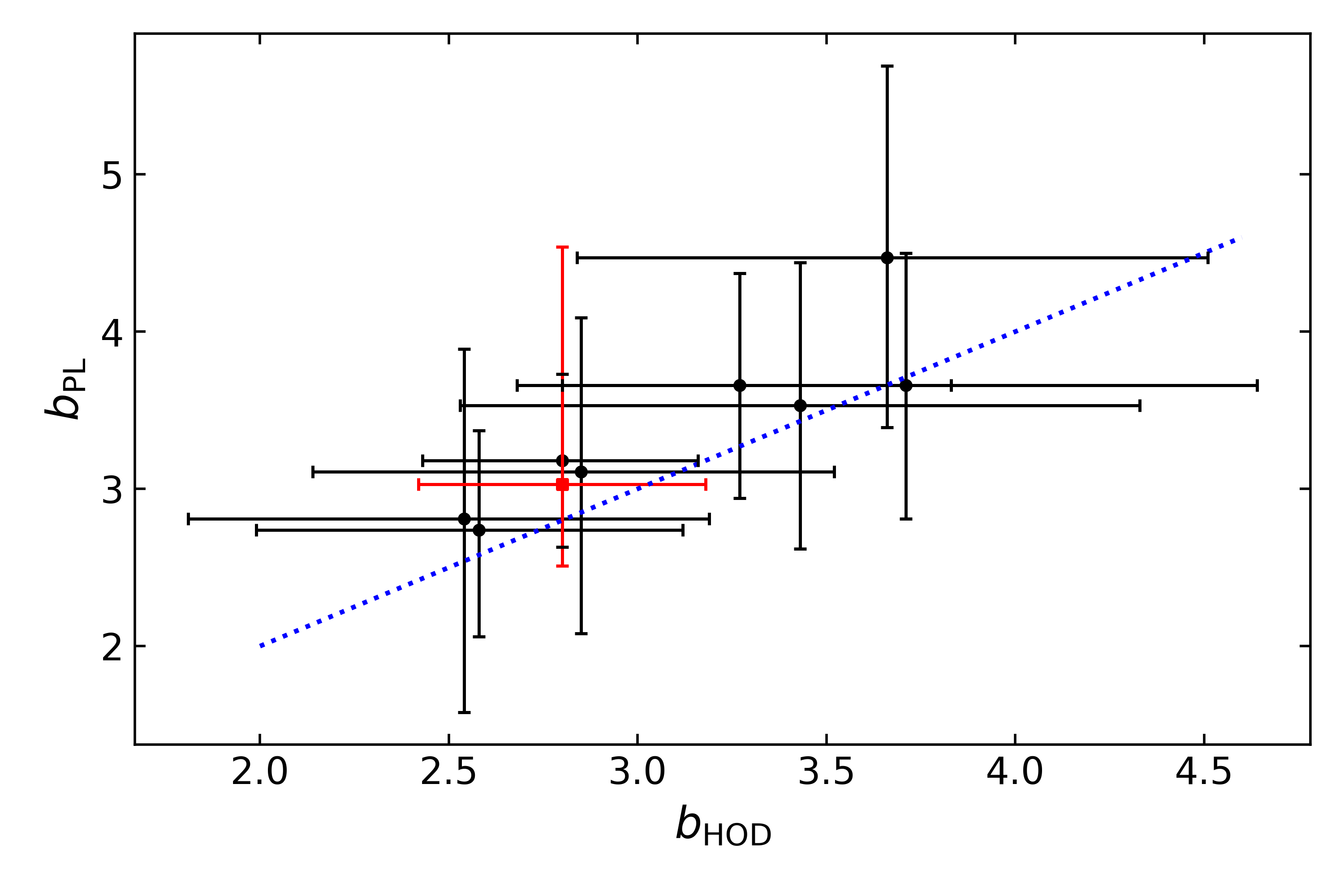}
\caption{Comparison between the bias parameters derived from the PL and HOD fits listed in Tables \ref{table:main_table} and \ref{table:subsamples_table}. We highlight the bias factor from the full sample of LAEs with a red square. The dotted blue line shows a 1:1 correspondence. }
\label{fig:b_hod_pl}
\end{figure}
Thanks to the characteristics of the survey they perform a two-parameter PL-fit and derive a correlation slope of $\gamma=1.80^{+0.02}_{-0.06}$ and a correlation length of $r_0=3.95^{+0.48}_{-0.54} \; h^{-1}$Mpc at $z\sim 2.5$. At $z\sim 3.5$ they obtain lower slopes $\gamma=1.60^{+0.12}_{-0.13}$ and higher lengths $r_0=4.35\pm 0.60$. Our results not only agree with their clustering parameters but also point toward a lower slope for higher redshift galaxies. \cite{moustaka} also reported a redshift dependence of $\gamma$ and parameterized analytically the correlation slope as a function of redshift.

Fig.~\ref{fig:comparison-literature} compiles the comparison of correlation lengths from the literature and those derived in this work with different fit approaches. We also plot the correlation lengths for our redshift subsamples (see Sect.~\ref{sec:properties}).

Most literature values are in agreement with our findings, both with the $r_0$ from the two-parameter PL-fit and from the one-parameter PL-fit with fixed $\gamma=1.8$ ($r_0=2.60^{+0.72}_{-0.67}$ $h^{-1}$Mpc). Not surprisingly, given the $r_0$ dependence on $R_{ij,\text{max}}$, the value from the single-bin fit is lower than most studies, including our robust PL-fit approach results.

A more appropriate but not so traditional comparison of the clustering strength is the bias factor, derived from $\gamma$ and $r_0$ (for PL-based correlation functions) or from HOD models. At $z\sim 3$ \cite{bielby} reported a bias factor of $b_{\rm{PL}}=2.13\pm 0.22$ and DMH masses of $M_{\text{DMH}} = 10^{11\pm 0.3} \; h^{-1}\text{M}_\odot$, whilst at the same redshift \cite{durkalek} reported a somewhat larger bias value of $b_{\rm{HOD}} = 2.82\pm 0.27$ and typical DMH masses of log$(M_{\text{DMH}}/h^{-1}\text{M}_\odot) = 11.75\pm 0.23$. 
As for \cite{ouchi17}, they obtained a bias value of $b_{\rm{HOD}} = 3.9^{+0.7}_{-1.0}$ and typical DMH masses of log$(M_{\text{DMH}}/h^{-1}\text{M}_\odot) = 11.1^{+0.2}_{-0.4}$ at $z=5.7$, whilst \cite{ouchi10} derived bias values in the range $b = 3-6$ and typical DMH masses of $M_{\text{DMH}} = 10^{11\pm 1}\; h^{-1}\text{M}_\odot$ at $z=6.6$. Our results fall between the values derived from studies at $z=3$ and $z=5.7$.

Each study however probes different luminosity and EW ranges, an effect that may have an impact on the interpretation of the clustering results from the literature. Despite these differences, it is interesting to note the general agreement in the clustering parameters from the different studies at similar $z$. Consequently, the hosting DMH mass of galaxies are also very similar. We will use our sample to test for such dependencies in the next section. 


\subsection{PL vs HOD fits}
\label{sec:discussion_hod-pl}
The various fit methods performed on the K-estimator allow us to compare the derived PL-fit results to those from HOD modeling. We have tested the performance of the different PL-fit approaches and we have developed an improved fit method.

The clustering signal provided by the K-estimator was never robustly fitted in previous studies \citep{adelberger,catrina}. The correlation length $r_0$ was obtained by measuring the K-estimator in a single $R_{ij}$ bin ($R_{\text{cut}}<5$ cMpc) and comparing the result to the expectation values $\langle K \rangle$ provided by Eq.~\ref{eq:expected_k} for different correlation lengths. In the process, a PL correlation function of fixed slope was assumed but never directly fit to the K-estimator. Instead, the correlation length that yields the closest match between $\langle K \rangle$ and $K_{\text{measured}}$ was chosen as the best correlation length. However, in Sect.~\ref{sec:results_traditional}, the result varies significantly depending on the chosen $R_{\text{cut}}$. Due to the simplicity of this approach and its dependence on $R_{\text{cut}}$, we fit the measured K-estimator as a function of $R_{ij}$ with the model predictions (Eq.~\ref{eq:expected_k}), providing a more reliable and accurate fit to the full $R_{ij}$ range covered by the K-estimator.

Taking the K-estimator one step further, we also make use of HOD models. As explained in Sect.~\ref{sec:hod}, PL-based correlation functions do not distinguish between the 1- and 2-halo term regimes. PLs are just an approximation, whereas HOD models treat galaxies residing in one DMH and in different DMHs differently, being a more advanced and physically meaningful approach. 

\begin{table*}[ht]
   \caption{Derived clustering parameters from the subsamples.}
   \label{table:subsamples_table}
   \centering
    \begin{tabular}{l@{\qquad}ccccc}
        \hline \hline
           \noalign{\smallskip}
        LAE subsample &    $r_0\;  [h^{-1}$Mpc]  &  $b_{\rm{PL}}$ &  $b_{\rm{HOD}}$ & ${\rm log}(\it{M}_{\rm{DMH}}$ / $[\it{h}^{-1}\rm{M}_{\odot}])$\\
             \noalign{\smallskip} 
            \hline \hline
            \noalign{\smallskip}
            Redshift $< 4.12$  &     $4.02^{+1.17}_{-1.06}$ & $3.18^{+0.55}_{-0.55}$ &  $2.80^{+0.36}_{-0.37}$  &  $11.39^{+0.23}_{-0.29}$ \\
            Redshift $> 4.12$  &      $3.53^{+1.16}_{-1.04}$ & $3.66^{+0.71}_{-0.72}$ & $3.27^{+0.56}_{-0.59}$ &  $11.07^{+0.31}_{-0.41}$\\
            log$L_{\rm{Ly}\alpha} <$ 42.36  &     $2.78^{+1.09}_{-1.02}$  &  $2.74^{+0.63}_{-0.68}$  & $2.58^{+0.54}_{-0.59}$ & $10.88^{+0.39}_{-0.62}$\\
            log$L_{\rm{Ly}\alpha} >$ 42.36   &      $4.08^{+1.60}_{-1.40}$ &  $3.66^{+0.84}_{-0.85}$ & $3.71^{+0.93}_{-0.91}$ & $11.57^{+0.38}_{-0.54}$\\
            $EW_{\rm{Ly}\alpha} <$ 87.9   &     $2.89^{+1.98}_{-1.74}$  &   $2.81^{+1.08}_{-1.23}$ & $2.54^{+0.65}_{-0.73}$ & $10.84^{+0.47}_{-0.83}$   \\
            $EW_{\rm{Ly}\alpha} >$ 87.9  &     $4.14^{+1.84}_{-1.57}$ &  $3.53^{+0.91}_{-0.91}$ & $3.43^{+0.90}_{-0.90}$ & $11.44^{+0.41}_{-0.62}$    
            \\
            $M_{\rm{UV}} <$ -18.8  &     $6.30^{+2.97}_{-2.26}$ &  $4.47^{+1.22}_{-1.08}$ & $3.66^{+0.85}_{-0.82}$ & $11.63^{+0.36}_{-0.48}$    \\
            $M_{\rm{UV}} >$ -18.8   &    $3.35^{+1.84}_{-1.59}$  &   $3.11^{+0.98}_{-1.03}$ & $2.85^{+0.67}_{-0.71}$ & $11.06^{+0.41}_{-0.64}$           \\
            \noalign{\smallskip}
        \hline 
        
        \multicolumn{6}{l}{%
          \begin{minipage}{12cm}%
          \vspace{0.2\baselineskip}
            \small \textbf{Notes}: Power-law derived bias values ($b_{\rm PL}$ use a fixed slope of $\gamma=1.3$; see discussion in Sect.~\ref{sec:properties}). The typical DMH masses are derived from our HOD results. The uncertainties in the bias factors and DMH masses reflect the statistical error on $r_0$ only.
          \end{minipage} 
          }\\
    \end{tabular}
\end{table*}

We measure the clustering only at $R_{ij}>0.6 \; h^{-1}$Mpc so we do not cover the one-halo term of the correlation function. Hence, we fit the 2-halo term of $\xi(r)$ from the HOD model to our $K$ values in order to obtain the large-scale bias of our sample.

In Fig.~\ref{fig:pl-hod} we show both PL and HOD best-fits to the K-estimator from the $\chi^2$ analysis described in Sect.~\ref{sec:k-estimator}. The curves perform very comparably and there are only tiny variations in the shape of the curves. Nonetheless, the PL-fit achieves the lowest $\chi^2$, indicating a modest better performance.  

Even though the curves are nearly identical at intermediate separations ($1<R_{ij}/h^{-1}\rm{Mpc}<2.5$), at smaller and larger separations the curves deviate from each other. The largest difference occurs at small separations ($R_{ij}<1 \; h^{-1}$Mpc) where the PL flattens but the HOD fit continues to increase. Less remarkable is the difference at large separations ($R_{ij}>2.5 \; h^{-1}$Mpc) where the PL-fit is somewhat higher than the HOD fit. In both cases, the differences are well within the uncertainties, 
and in the main range used to calibrate the bias factor ($R_{ij}>1 \; h^{-1}$Mpc), the variations between the fits are minute.

We show the comparison between the large-scale bias parameters calculated from the PL and HOD fits listed in Tables \ref{table:main_table} and \ref{table:subsamples_table} in Fig.~\ref{fig:b_hod_pl}. The derived bias factors from the PL fits are slightly higher than the HOD values, while the HOD uncertainties are on average smaller ($\approx$ 25\%) than those from the PL fits. Using samples of AGN and a cross-correlation function approach, \cite{mirko12} also compared PL and HOD clustering fits. They found higher bias factors and smaller uncertainties from the HOD fits because they included part of the one-halo term in the PL fit. As we have explained, strong variations between samples in the one-halo term cause a decrease in the bias factors derived from the PL fits. However, we do not include the one-halo term in any of our fits so we are not subjected to these variations.


\subsection{Clustering dependence on physical properties}
\label{sec:properties}
We search for clustering dependencies on LAE physical properties. We compute the K-estimator in the subsamples described in Sect.~\ref{sec:subsamples} but the lower number of objects in the subsets does not allow for a two-parameter PL-fit. We therefore take the prior from our full sample and assume that our subsamples present the same correlation slope as the parent sample ($\gamma=1.3$). We then perform the one-parameter PL-fit with fixed $\gamma=1.3$. We also conduct HOD fits in the same way as we did for the full sample. 


\subsubsection{Redshift}
Taking advantage of the large redshift range provided by MUSE, we investigate if LAEs occupy denser regions of the Universe at earlier epochs by measuring their clustering strength with the K-estimator.

\begin{figure*}
\centering
\begin{tabular}{c c}
  \centering
  \includegraphics[width=.49\linewidth]{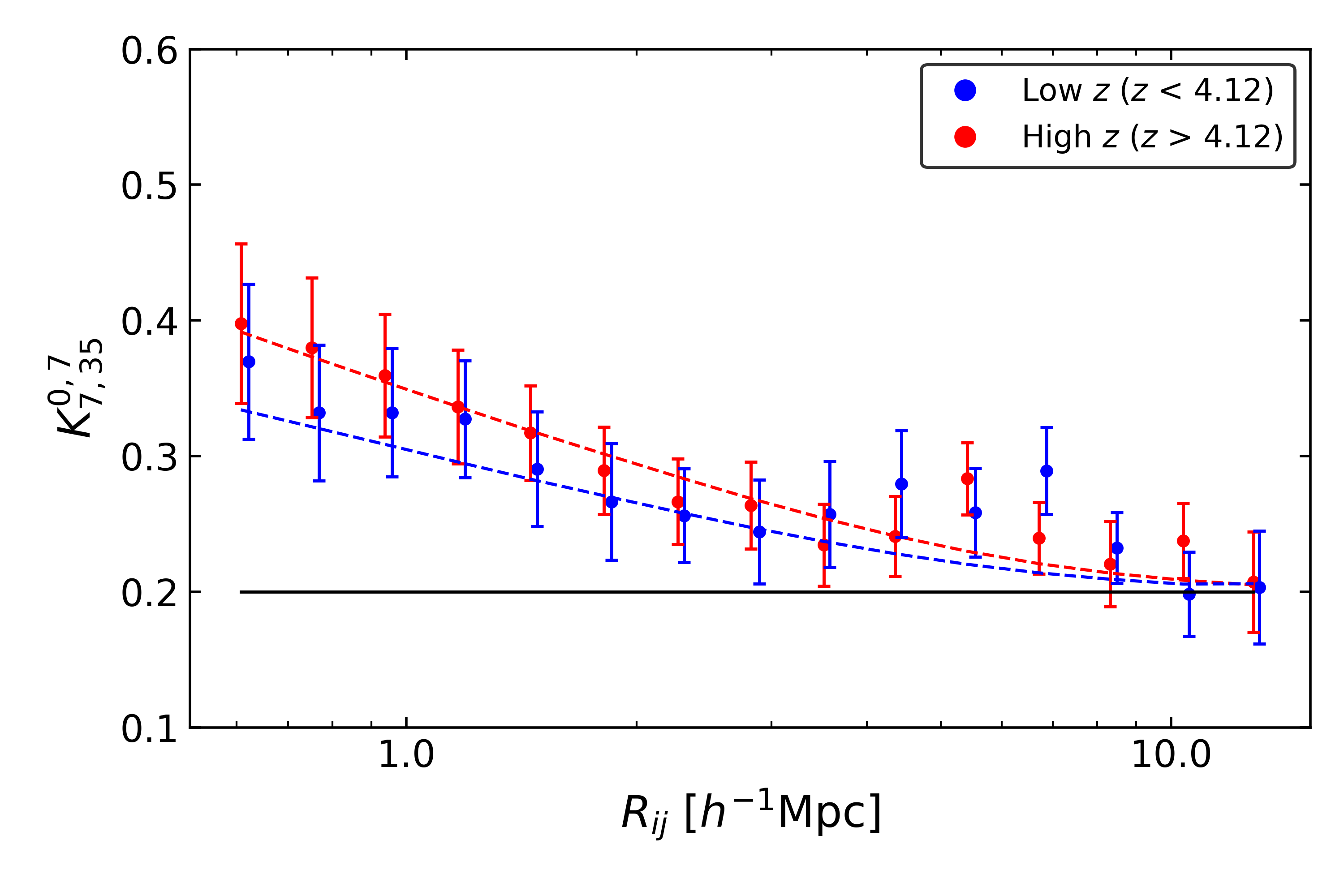}
\end{tabular}%
\begin{tabular}{c c}
  \centering
  \includegraphics[width=.49\linewidth]{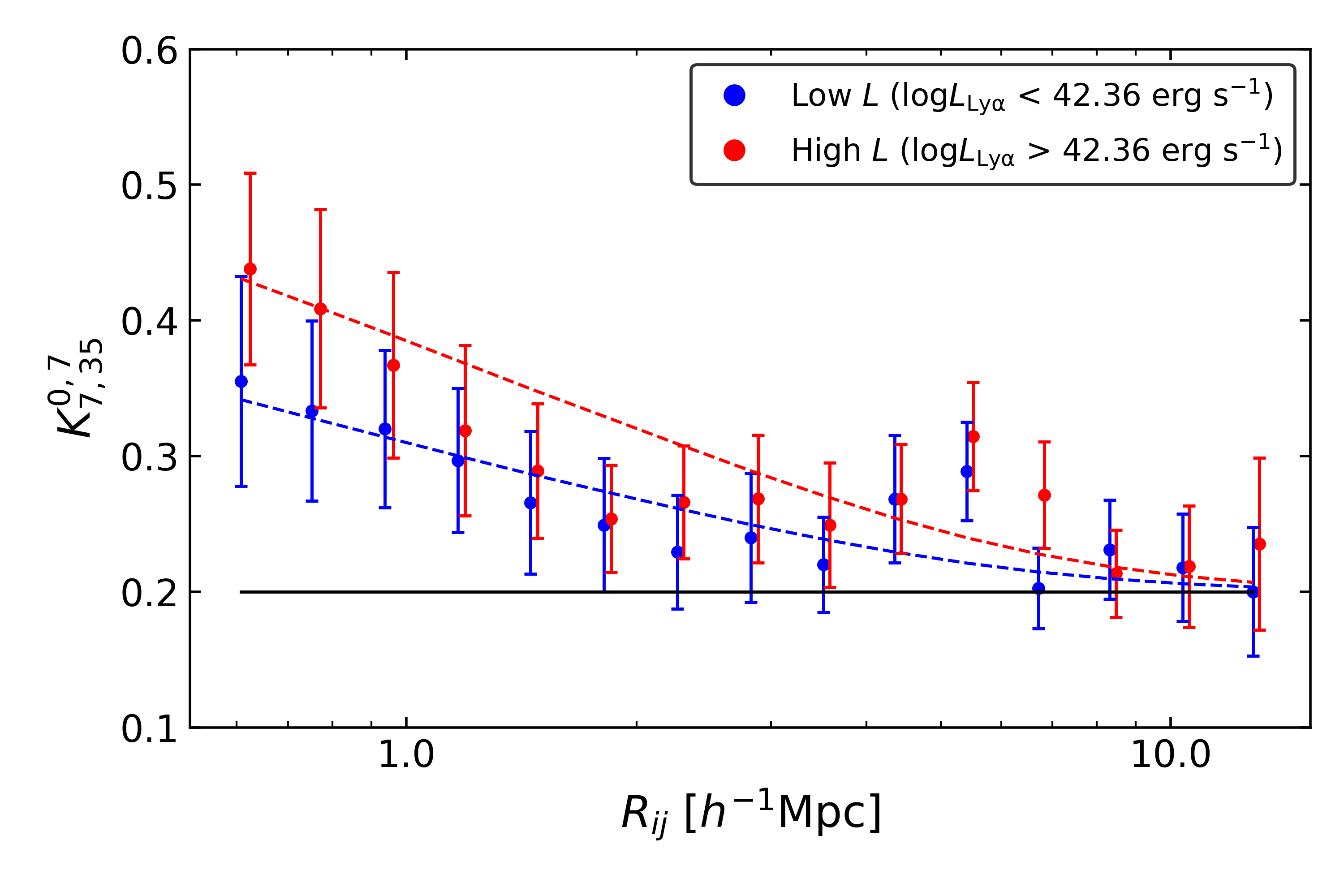}
\end{tabular}
\newline
\centering
\begin{tabular}{c c}
  \centering
  \includegraphics[width=.49\linewidth]{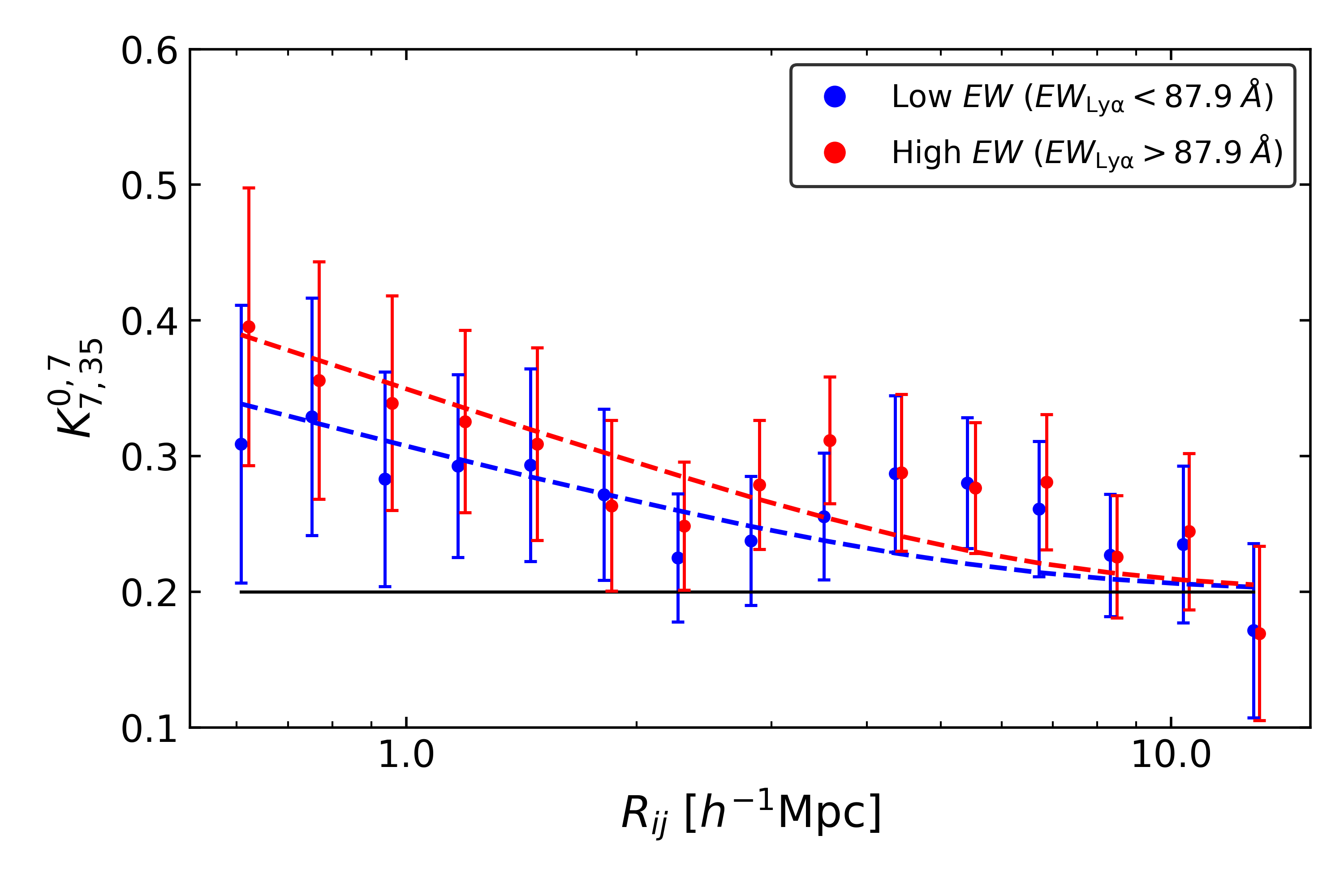}
\end{tabular}%
\begin{tabular}{c c}
  \centering
  \includegraphics[width=.49\linewidth]{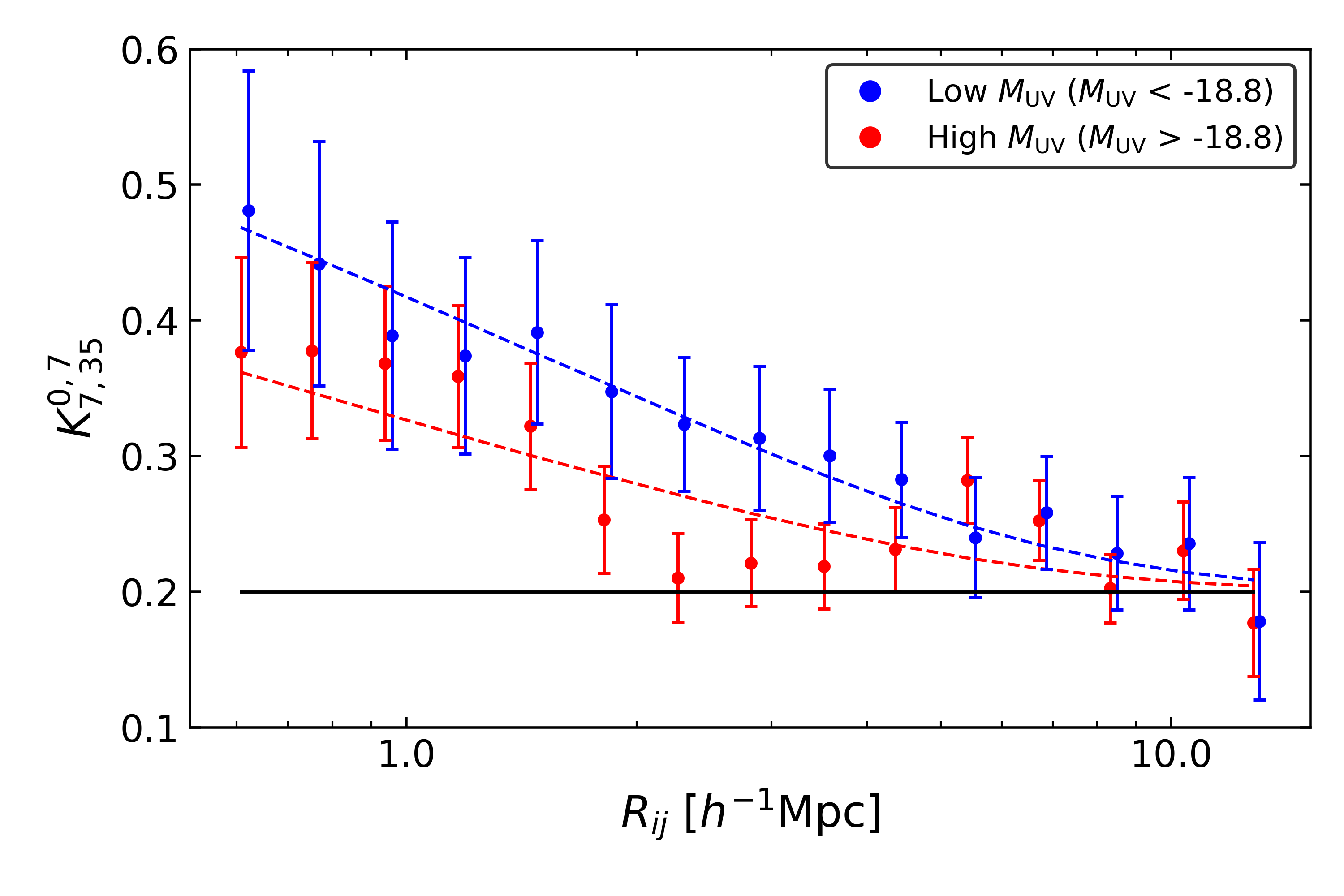}
\end{tabular}
\caption{Top left: Clustering variation in two different redshift subsamples. The blue dots show the clustering in the lower redshift bin while the red points show the higher redshift subsample. The dotted curves represent the best HOD fits. Top right: same but for two different Ly$\alpha$ luminosity subsamples. Bottom left: same but for $EW_{\text{Ly}\alpha}$. Bottom right: same but for UV absolute magnitude. The black line represents the K expectation value for an unclustered sample of galaxies and the 1$\sigma$ error bars are determined from the bootstrapping approach explained in Sect.~\ref{sec:errorsK}.}
\label{fig:properties-dependence}
\end{figure*}

At the cost of enlarging the error bars and as explained in Sect.~\ref{sec:data}, we split our sample in two bins around the median redshift, $\langle z \rangle = 4.12$. We compute the K-estimator in both subsamples and show the results in the top left panel of Fig.~\ref{fig:properties-dependence}.

The two curves are essentially indistinguishable within the error bars. Both follow the same trend and have similar shapes. Analogously to Sect.~\ref{sec:results_traditional}, we fit a PL correlation function $\xi(r)=(r/r_0)^{-\gamma}$ with fixed slope $\gamma=1.3$. For the low redshift subsample, we obtain $b_{\rm{low}} = 3.18^{+0.55}_{-0.55}$. The resulting value for the high-redshift bin is $b_{\rm{high}} = 3.66^{+0.71}_{-0.72}$. The best-fit parameters are listed in Table \ref{table:subsamples_table} along with the bias factors obtained from the HOD fit and their corresponding DMH masses.

The difference between the best-fit parameters (lower than 1$\sigma$) of the subsamples do not allow us to corroborate or contradict the general statement "LAEs reside in more massive DMHs at higher redshifts". However, other studies found higher bias factors of LAEs at higher redshifts \citep[e.g.][]{ouchi10}. Even if the study of samples at fixed luminosities is needed, this has been interpreted as evidence for downsizing, with galaxies residing in the largest DMHs going through their "LAE phase" early in the Universe, while Milky Way progenitors appear as LAEs at later times, around $z \sim 3$.

We have confirmed that our findings are not strongly affected by the selected redshift cut of the sample. Varying the cut by 10\% in $z$, from $z=4.12$ to $z=4.53$, changes the number of LAEs in each subsample by $\sim$15\% (118 objects) and results in an increase in $b$ within 1$\sigma$, equivalent considering the uncertainties. The $z$ and $L_{\text{Ly}\alpha}$ values are not independent. Therefore, the different luminosity distributions in the different redshift subsamples may bias the detection of a clustering dependence on redshift. To assure that the investigation of the clustering dependence on $z$ is not driven by the different $L_{\text{Ly}\alpha}$ distributions, we apply a "matching" technique similar to \cite{coil09} and \cite{mirko15}. To do so, we compare individual bins between the two luminosity distributions of the $z$-subsamples. In each bin, we check which subsample contains more objects. We then select the one with the higher number and randomly remove objects until we match the number counts of the other subsample in that bin. 
Once the two luminosity distributions are equivalent, we run the K-estimator in both subsamples with now matched $L_{\text{Ly}\alpha}$ distributions but still different redshift distributions. 
We find completely consistent results when comparing to our original subsample definition. The clustering difference between the "matched" and "unmatched" subsamples varies within 1$\sigma$. Therefore, we discard the possibility of a possible clustering dependence on $z$ driven by $L_{\text{Ly}\alpha}$ as well as a strong clustering dependence on $z$.

\subsubsection{Ly$\alpha$ luminosity}

In order to learn about the DMHs where LAEs of different Ly$\alpha$ luminosities reside, we study the clustering dependence on Ly$\alpha$ luminosity. We use two subsamples divided by the median Ly$\alpha$ luminosity of the full sample as explained in Sect.~\ref{sec:data}. The K-estimator is then computed for both. Details of the individual subsamples are given in Table \ref{table:properties_table} and the clustering correlations are illustrated in the top right panel of Fig.~\ref{fig:properties-dependence}.

Although the statistical uncertainties are substantial, the top right panel of Fig.~\ref{fig:properties-dependence} suggests a trend in the sense that LAEs with higher Ly$\alpha$ luminosities appear to be more strongly clustered. This trend is also seen in the correlation lengths and bias factors, see Table \ref{table:subsamples_table}. We verify that our results are not significantly altered by the chosen Ly$\alpha$ luminosity cut of the sample. Shifting the Ly$\alpha$ luminosity from log$(L_{\text{Ly}\alpha}/[\rm{erg\:s}^{-1}])=42.36$ to log$(L_{\text{Ly}\alpha}/[\rm{erg\:s}^{-1}])=43.21$ (120 objects shifted) 
does not change the results; we still find a tentative $2\sigma$ clustering dependence on Ly$\alpha$ luminosity. Furthermore, we have also investigated that the possible clustering evolution trend with $L_{\text{Ly}\alpha}$ is not caused by the different redshift distributions of the subsamples. 

As already mentioned $L_{\text{Ly}\alpha}$ and $z$ are not independent. Thus, we also create matched distributions to exclude that dependence is driven by $z$ and not $L_{\text{Ly}\alpha}$. In order to discard this possibility, we match the $z$-distributions of both subsamples such that the low- and high- $L_{\text{Ly}\alpha}$ subsamples have exactly the same $z$-distribution. We compute the K-estimator for the subsamples with the matched $z$-distributions and find a more pronounced trend than that of the top right panel of Fig.~\ref{fig:properties-dependence}. For both matched subsamples, the $K$ values vary within $\sim$7\% of the original subsamples. This translates into a difference lower than 1$\sigma$ between the "matched" bias factors and those listed in Table \ref{table:subsamples_table}. However, the "matched" bias factors between the low and high $L_{\text{Ly}\alpha}$ subsamples differ by almost 2$\sigma$, suggesting a tentative weak clustering dependence in the way that more luminous LAEs cluster more strongly than less luminous LAEs. 

The calculated bias factor for fainter LAEs is $b_{\rm{low}}=2.58^{+0.54}_{-0.59}$, while for luminous LAEs is $b_{\rm{high}}=3.71^{+0.93}_{-0.91}$. This trend is consistent with the statement that more luminous (in Ly$\alpha$) galaxies reside in more massive DMHs \citep{ouchi03}. While a more statistically significant result will require larger LAE samples, the trend we see is in agreement with \cite{ouchi03} and \cite{khostovan}, who found stronger clustering strengths for Ly$\alpha$ brighter LAEs in samples of $41.85 \leq \rm{log} (L_{\rm{Ly}\alpha}/[\rm{erg\:s}^{-1}]) \leq 42.65$ at $z\approx4.86$ and $42 \leq \rm{log} (L_{\rm{Ly}\alpha}/[\rm{erg\:s}^{-1}]) \leq 43.6$ at $2.5<z<6$, respectively.
\subsubsection{Ly$\alpha$ equivalent width}

We investigate the clustering dependence on the rest-frame Ly$\alpha$ $EW$ to explore the possibility of LAEs residing in different DMHs depending on the $EW$ of the Ly$\alpha$ emission line. We use the two subsamples described in Table \ref{table:properties_table}. The $EW$ cut is made at the median $EW_{\text{Ly}\alpha}$ of the sample of galaxies with HST counterparts as explained in Sect.~\ref{sec:data}. 
The K-estimator results are presented in the bottom left panel of Fig.~\ref{fig:properties-dependence} and the best-fit parameters from the PL- and HOD-based correlation functions are given in Table \ref{table:subsamples_table}.

There are hardly any differences between the curves shown in the bottom left panel of Fig.~\ref{fig:properties-dependence}. The low $EW_{\text{Ly}\alpha}$ subsample presents a linear bias factor of $b_{\rm{low}} = 2.54^{+0.65}_{-0.73}$, while the resulting value for the high $EW_{\text{Ly}\alpha}$ bin is $b_{\rm{high}} = 3.43^{+0.90}_{-0.90}$. Even though LAEs with higher $EW_{\text{Ly}\alpha}$ seem to reach higher $K$ values on average, the difference is smaller than 1$\sigma$. Similar results were found by \cite{ouchi03}.

We have certified that the derived correlation lengths are not affected by the selected $EW_{\text{Ly}\alpha}$ cut of the sample. Shifting the cut by 25\% in $EW_{\text{Ly}\alpha}$, from $EW_{\text{Ly}\alpha}=87.9$ to $EW_{\text{Ly}\alpha}=110 \; \AA$, changes the subsample number counts in 50 objects and results in a variation in $r_0$ within 1$\sigma$, equal within the error bars.

\subsubsection{UV absolute magnitude}

The UV absolute magnitude is related to the star formation rate which in turn is expected to scale with stellar and also the DMH mass. It is therefore interesting to explore the clustering dependence on UV absolute magnitude by dividing our full sample at the median $M_{\text{UV}}$ into two subsamples. The characteristics of both bins are listed in Table \ref{table:properties_table}. We compute the K-estimator in both subsamples and we illustrate the clustering correlations in the bottom right panel of Fig.~\ref{fig:properties-dependence}. The clustering parameters are listed in Table \ref{table:subsamples_table}.

We find large-scale bias factors in the bright- and faint-$M_{\text{UV}}$ subsamples (low and high $M_{\text{UV}}$, respectively) of $b_{\rm{low}}=4.47^{+1.22}_{-1.08}$ and $b_{\rm{high}}=3.11^{+0.98}_{-1.03}$. Given the large uncertainties, we cannot claim the detection of a clustering dependence on $M_{\text{UV}}$, even if the bottom right panel of Fig.~\ref{fig:properties-dependence} seems to indicate a stronger clustering signal for more luminous LAEs than for fainter LAEs. While \cite{durkalekp} found that high-$z$ galaxies with $M_{\text{UV}}<-20.2$ cluster stronger than those with $M_{\text{UV}}<-19.0$ ($\Delta b=0.43$),
 \cite{ouchi03} found no notable difference between their $M_{\text{UV}}$ subsamples. Since \cite{ouchi03} recognize different clustering strengths as a function of 
 Ly$\alpha$ luminosity, they claim that such dependence might dominate over a $M_{\text{UV}}$ clustering dependence.

We have checked that the derived correlation parameters are not significantly affected by the chosen $M_{\text{UV}}$ cut of the sample. Shifting the $M_{\text{UV}}$ cut, from $M_{\text{UV}}=-18.8$ to $M_{\text{UV}}=-19.18$ changes the number of counts in the subsamples by 62 objects while the correlation lengths vary within 1$\sigma$, consistent within the uncertainties.
\subsection{Cosmological simulations}
\label{sec:simulations}

We now compare our results with cosmological simulations to test if our detected clustering signal is predicted by state of the art LAE models at high redshift, and to get insight into the expected cosmic variance. 

While a plethora of cosmological simulations are now available to describe the formation and evolution of galaxies at high redshift, the complex nature of the Ly$\alpha$ line emission and propagation in the gas requires careful numerical modelling in order to make predictions for the LAE population. Various approaches based on different numerical techniques and model assumptions (i.e., cosmology, baryonic physics, etc) have incorporated Ly$\alpha$ radiation transfer effects over simple geometries in semi-analytic models \citep[e.g.][]{garel12,orsi12,garel15,gurung18} or in post-processing of hydrodynamical simulations \citep[e.g.][]{forero12,dayal12,yajima12}. Even though there is no radiative transfer (RT) model that perfectly reproduces the Ly$\alpha$ emission lines and, therefore, no cosmological simulation that succeeds in fully replicating LAE observations, here we compare our results with the GALICS semi-analytic model which includes Ly$\alpha$ radiation transfer in expanding shells \citep{garel15}.

The underlying dark matter simulation used in this model is run with GADGET \citep{springel} and features a box of 3 $\times 10^6$ Mpc$^3$ with a DMH mass resolution of $10^9$ M$_{\odot}$. As shown in \cite{garel15}, this model can reproduce the UV and Ly$\alpha$ luminosity functions at $3<z<7$ down to Ly$\alpha$ fluxes of 4$\times 10^{-17}$ erg s$^{-1}$ cm$^{-2}$. Following the procedure described in \cite{garel}, we generate 100 mock light-cones of 17 $\times$ 17 arcmin$^2$ size to obtain physical parameters such as Ly$\alpha$ fluxes or 3D positions.

\begin{figure}[h]
\centering
\includegraphics[width=\columnwidth]{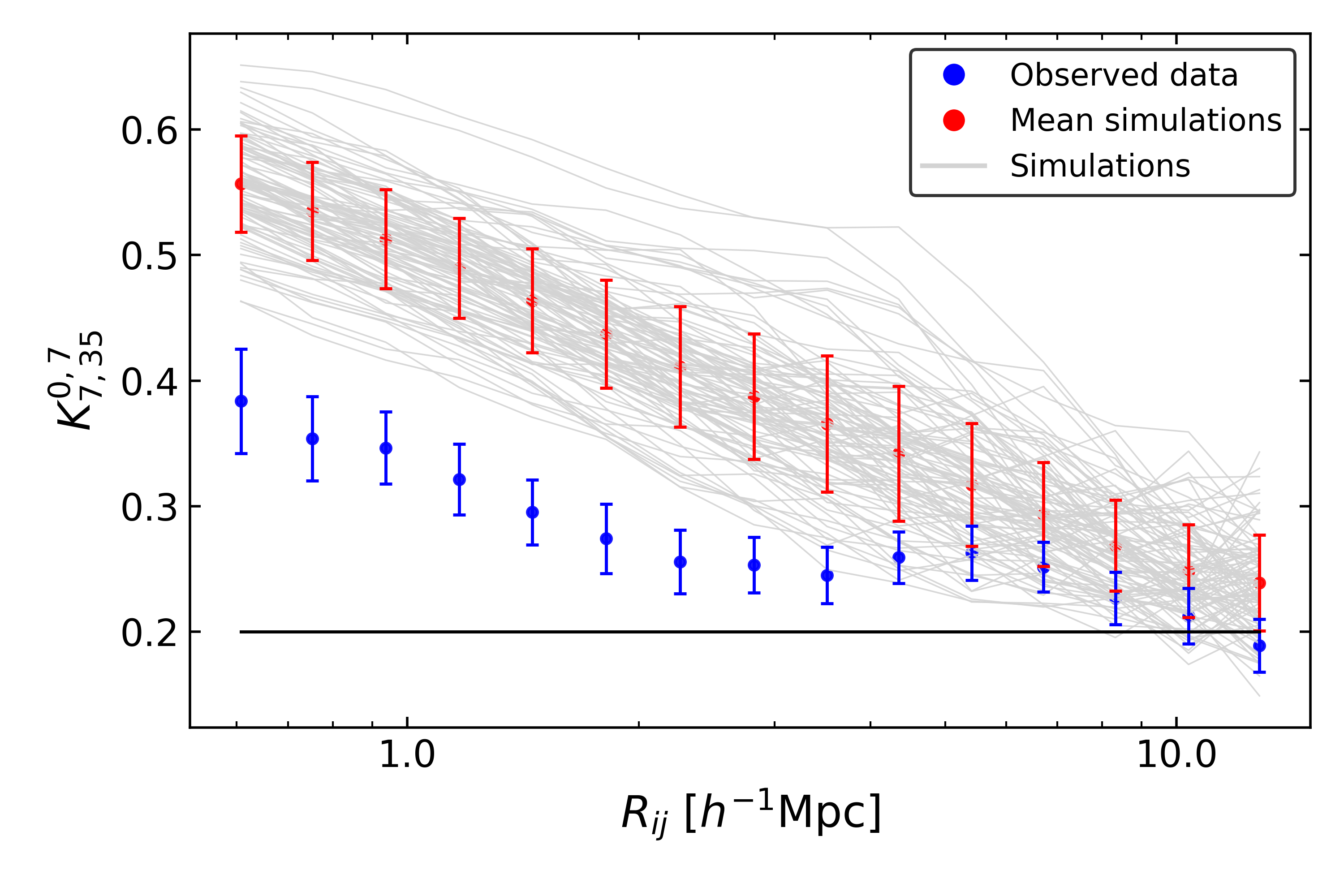}
\caption{Comparison between the clustering signal in our real survey in blue (same as in Fig.~\ref{fig:k-gamma-r0}) and in the 100 simulated samples. Each sample is drawn from
a different light-cone realization. The values of the K-estimator for each of the 100 simulated catalogues is represented in gray. The standard deviation of the 100 K-estimator values and their average values 
are shown in red.}
\label{fig:k-simulations}
\end{figure}

In order to resemble the real data, the selection function, spatial geometry, and redshift range of the 68 MUSE-Wide fields have been applied before computing the K-estimator for the 100 simulated samples. The selection function was shifted +0.28 dex in flux to recover the same total number of detections. The results given by the K-estimator in the 100 simulated samples are shown in Fig.~\ref{fig:k-simulations} along with its average and uncertainties.

We find that the clustering in the simulated samples is much stronger than the clustering in the MUSE-Wide survey. Simulated samples present much larger K-values than the observed data. 
The two-parameter PL-fit results in very large uncertainties in $r_0$ (consistent with our observed $r_0$) but also points to lower $\gamma$ values. From the one-parameter PL-fit (fixed $\gamma=1.8$), we obtain $r_0=5.80^{+1.23}_{-1.10}\; h^{-1}$Mpc and derive a bias factor of $b=4.13^{+0.78}_{-0.71}$. These are only $\approx 1.9\sigma$ away from our one-parameter best-fit results. However, from the HOD fit we compute $b=4.80^{+0.32}_{-0.32}$ and $\log (M_{\text{DMH}}/[h^{-1}\text{M}_\odot]) = 12.16
^{+0.11}_{-0.11}$. Considering the $M_{\text{DMH}}$ values, this differs by $3.2\sigma$ from our observations. We note that since the scope of this paper is not to give an extended comparison to simulations, we have restricted our comparison to one LAE model only, and we obviously cannot draw any general conclusion regarding a potential tension between the predicted and observed clustering of LAEs, and leave more detailed comparisons to future work. We briefly discuss below aspects that may plausibly explain the mismatch.

\begin{figure}[h]
\centering
\includegraphics[width=\columnwidth]{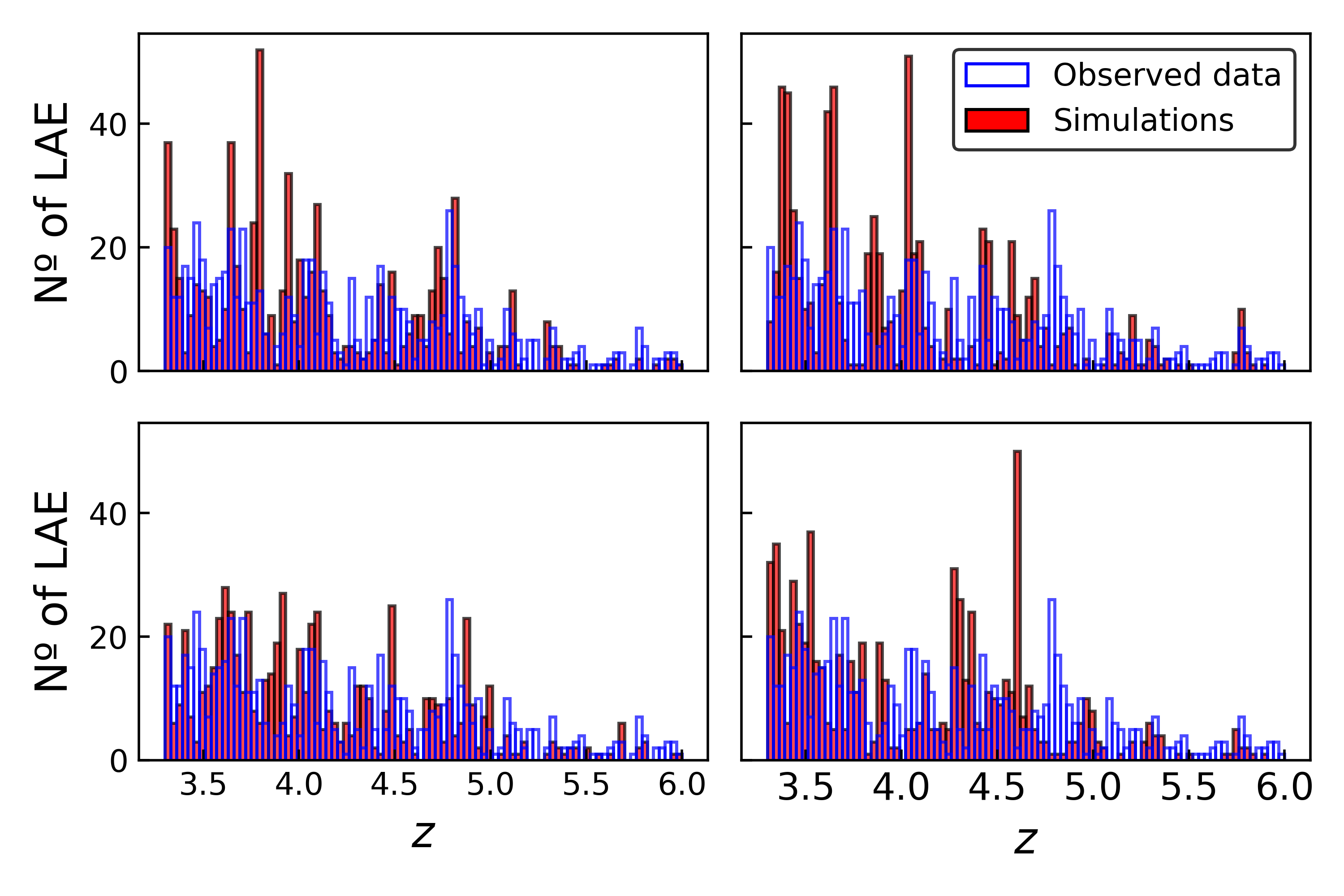}
\caption{Redshift distribution from 4 simulated catalogues chosen randomly from the full set of light-cones in red. The redshift distribution of the real LAEs from the 68 fields of the MUSE-Wide survey is shown in blue. }
\label{fig:hist-simulations}
\end{figure}

We first note that the mock light-cones show prominent redshift spikes that are not present in the real data (see Fig.~\ref{fig:hist-simulations}), a noticeable problem that was also discussed in \cite{catrina}. The super structures seem to dominate the clustering signal and be responsible for most of the disagreement between our measurements and the simulation. 
The origin of this discrepancy is unclear and could be due to several reasons that need to be addressed in future simulation work, such as inaccuracies in the Ly$\alpha$ RT modelling, the assumed cosmology, the baryonic physics modelling affecting the halo-galaxy relation, cosmic variance, intrinsic Ly$\alpha$ luminosities, or poorly-controlled LAE selection in the mocks. 

In particular, a potential cause could arise from the fact that Ly$\alpha$ luminosities in GALICS are angle-averaged whereas the Ly$\alpha$ escape fraction is supposedly highly anisotropic in real galaxies, such that observed Ly$\alpha$ luminosities strongly vary from one sight-line to another
 \citep{smith}. Since the luminosity function of our LAEs is steep, this has the same effect as photometric scattering. Uncertainties in the selection function of GALICS and of observations due to field to field variation are poorly controlled and have a similar effect. This results in the selection of more luminous galaxies at higher redshifts, which can boost the clustering signal. 

Moreover, potential deviations on the Ly$\alpha$ luminosity-halo mass relation in GALICS (e.g. star formation not sufficiently quenched in massive halos, such that $L_{\text{Ly}\alpha}=10^{42}$ erg s$^{-1}$ LAEs might reside in too massive halos) would also enhance the clustering. Besides, a duty cycle \citep[e.g.][]{ouchi10} further powers this mismatch because in GALICS most star forming galaxies emit Ly$\alpha$. Alternative modeling could lead to Ly$\alpha$-bright phases that only last a limited period of time, which could plausibly attenuate the strong spikes seen in the redshift distribution. 

Since these simulations do not closely reproduce the clustering present in the real data, it is not possible to use them for improving the error estimates on the measurements over the approaches we considered in Appendix~\ref{sec:errors}. Nevertheless, simulations can provide information on the individual contribution of the statistical uncertainty and the uncertainty due to cosmic variance. For the statistical uncertainty, we perform 500 K-estimator measurements on the same bootstrap-resampled light-cone (see Sect.~\ref{sec:errorsK} where we apply the same method to our real data). The uncertainty of the cosmic variance can be estimated by computing the standard deviation of the 100 different light-cone realizations.
Comparing the individual data points, we find that the uncertainty due to the cosmic variance (red error bars in Fig.~\ref{fig:k-simulations}) is on average $\sim$35\% larger than the error from the statistical approach only (similar to the blue error bars in Fig.~\ref{fig:k-simulations} but for one light-cone). We have verified that similar results are found if we resample different light-cones.
Therefore, our quadratically combined uncertainties (statistical and cosmic variance) in the observed clustering can be $\sim$70\% higher.  

Finally, we used the light-cones to check the effects of our special survey geometry. Even if we do not expect a strong effect on the K-estimator since we measure a contrast of galaxy pairs in two consecutive shells along l.o.s. separations (see also Appendix~\ref{appendix:k-fields}), we have compared the clustering of the 100 simulated catalogues in a 68 fields special geometry with the clustering present in the full area of the simulation without any forced geometry (just a simple square of 17 $\times$ 17 arcmin$^2$). We find that the signals agree very well. 
\subsection{The fate of LAEs through cosmic time}

Applying an HOD modelling to the K-estimator determines a typical dark matter halo mass of $\log (M_{\text{DMH}}/[h^{-1}\text{M}_\odot]) = 11.34^{+0.23}_{-0.27}$ for our LAEs. It is expected that these high redshift DMHs significantly grow all the way down to $z=0$. We here explore the evolution of those DMHs from $\langle z_{\rm pair} \rangle \simeq 3.82$ to $z=0$ to find the typical descendants of our LAE sample.

Considering the galaxy-conserving evolution model of \cite{fry}, which assumes the absence of mergers and that the motion of galaxies are driven by gravity only, the large-scale bias factor evolves as
\begin{equation}
    b(z) = 1+(b_0-1)/D(z), \label{eq:galaxy-conserving}
\end{equation}
where $b_0$ is the bias at $z=0$ and $D(z)$ is the linear growth factor \citep[e.g.][]{hamilton}. 

Using this model, we infer that the halos of LAEs with a median redshift of the number of pairs $\langle z_{\rm pair} \rangle \simeq 3.82$ evolve into halos with $b_0 \approx 1.4$ by $z=0$, which translates into typical DMH masses of log$(M_{\text{DMH}}/[h^{-1}\text{M}_\odot]) \sim 13.5$. This is $\approx$15 times more massive than the Milky Way. Based on \cite{ouchi10} calculations, in a more realistic Press-Schechter formalism \citep[e.g.][]{lacey}, the bias evolution curve is slightly lower, meaning a bias closer to $b_0 \approx 1.2$ rather than $b_0=1.4$. 

Similar results are also derived from simulations. The information stored in the DM halo merger trees used in GALICS \citep{garel} allowed a similar study. They found median descendant halo masses of $M_{\text{DMH}}/[h^{-1}\text{M}_\odot] \approx 2\cdot 10^{12}$ for $z=3$ LAEs, corresponding to the upper limit estimate of the Milky-Way halo mass. For $z=6$ LAEs they found median descendant halo masses of $M_{\text{DMH}}/[h^{-1}\text{M}_\odot] \approx 5\cdot 10^{13}$, corresponding to group/cluster galaxy halos. These assessments are well in agreement with our estimations and reinforce that our LAEs actually contain a diverse population of objects (as expected, since MUSE-Wide is a Ly$\alpha$-flux limited survey over a wide $z$ range).

Our work and the various studies presented in the literature cover wide 
$z$ and Ly$\alpha$ luminosity ranges. 
If there are indeed clustering dependencies with one or both parameters this would lead to different typical DMH masses depending on the details of the sample selection. 
It is thus necessary to discuss the descendants with respect to different redshift and luminosity progenitors.
The combination of clustering measurements of NB-selected LAEs and the galaxy-conserving model considered in this work, leads to the conclusion that $z=5.7-6.6$ LAEs will evolve into DMH with bias values of $b_0=1.5-2$ at $z=0$ \citep{ouchi10}. This is in agreement with our findings. These higher values, in comparison to those of \cite{gawiser07} at $z=3.1$, indicate that descendants of LAEs at different redshifts differ. While most LAEs at $4<z<7$ are probably today's large galaxies, LAEs at $z=3$ are more likely ancestors of Milky Way type galaxies.

\cite{khostovan} considered narrowband-selected LAEs
with typical $L_{\text{Ly$\alpha$}} \sim 10^{42-43}$ erg s$^{-1}$ and intermediate-band-selected LAEs with typical $L_{\text{Ly$\alpha$}} \sim 10^{43-43.6}$ erg s$^{-1}$ at $2.5<z<6$. Assuming halo mass accretion models, they found that the former evolved into galaxies residing in halos of typically log$(M_{\text{DMH}}/[h^{-1}\text{M}_\odot]) = 12-13$ (Milky Way-like) while the later evolved into galaxies residing in halos of log$(M_{\text{DMH}}/[h^{-1}\text{M}_\odot]) > 13$ (cluster-like) in the local Universe.  

Since our LAEs are in the redshift range $3.3<z<6$ and present typical Ly$\alpha$ luminosities in the range $40.9 < \text{log}(L_{\text{Ly$\alpha$}}/[\rm{erg\:s}^{-1}]) < 43.3$, the derived descendant masses (log$(M_{\text{DMH}}/[h^{-1}\text{M}_\odot]) \sim 13.5$) are well in agreement with the cluster-like descendants found by \cite{ouchi10} and \cite{khostovan}. 
These results, along with the literature, reveal that LAEs cover a wide range of present-day descendants depending on their luminosity and redshift, from Milky Way-type galaxies all the way to clusters of galaxies.

\section{Conclusions}
\label{sec:conclusions}
We have examined the galaxy clustering properties of a sample of 695 LAEs from the MUSE-Wide survey in the redshift range $3.3 < z < 6$. We applied an optimized version of the K-estimator and supported our results with the traditional two-point correlation function, measuring for the first time the spatial clustering as a function of distance in a spectroscopic sample of Ly$\alpha$-selected galaxies. 

Due to the characteristics of the survey (special geometry, large redshift range and limited angular coverage), we focus on the more appropriate clustering method, the K-estimator. We then relied on the radial clustering and quantified the clustering signal following different approaches. We first obtained $r_0 = 3.60^{+3.10}_{-0.90} \; h^{-1}$Mpc and $\gamma = 1.30^{+0.36}_{-0.45}$ by fitting the clustering signal with a power-law-based correlation function. We derived a bias parameter of $b=3.03^{+1.51}_{-0.52}$ and compared it to that derived from the second fit approach, $b=2.80^{+0.38}_{-0.38}$, by scaling a halo occupation distribution model to the measured signal. The large-scale bias corresponds to typical dark matter halo masses of log$(M_{\text{DMH}}/[h^{-1}\text{M}_\odot]) = 11.34^{+0.23}_{-0.27}$. In order to support the less known K-estimator method, we also computed the traditional 2pcf, whose results are consistent with those obtained with the K-estimator.

The results are also in general agreement with the last available measurements at similar redshifts, with bias factors slightly higher than those in the literature. Nevertheless, most of the previous studies have been carried out in surveys with somewhat different flux limits than that of the MUSE-Wide survey. This could play an important role since we may probe disparate stellar masses. The chosen cosmology and the redshift evolution of these parameters through different epochs also contribute to those slight differences.

We also explore possible clustering dependencies on physical properties. We exclude the possibility of a strong clustering dependence on Ly$\alpha$ equivalent width, UV absolute magnitude and redshift. However, we see a tentative weak trend when we split the sample at the median Ly$\alpha$ luminosity i.e., more luminous LAEs cluster more strongly than less luminous LAEs.

We compare the clustering in the MUSE-Wide survey with the clustering in 100 light-cones from a GADGET dark matter only cosmological simulation coupled to the GALICS semi-analytical modeling of LAEs. We find that even though the simulation mimics the flux/luminosity of the LAEs, it is far away from reproducing the observed clustering. Simulated data show a stronger clustering than measured in our sample. In order to better imitate the clustering of LAEs, determine the reliable 2pcf scales, compute more realistic uncertainties for our methods and constrain a physically robust model for LAEs, future simulation work needs to address this challenge.

Assuming galaxy-conserving evolution models we have inferred that our DMHs should evolve into halos of log$(M_{\text{DMH}}/[h^{-1}\text{M}_\odot]) \sim 13.5$ in the local Universe. Since these models assume that the motion of galaxies is driven entirely by gravity, and that mergers do not occur, our evolved DMH masses would be slightly higher in a (more realistic) Press-Schechter formalism. We deduce that the LAEs observed at $3.3 < z < 6$ with $40.9<\rm{log}(L_{\rm{Ly}\alpha}/[\rm{erg\:s}^{-1}])<43.3$ have mainly evolved into halos hosting galaxies or groups $\approx$15 times more massive than the halo hosting the Milky Way. 

A radial extension of the MUSE-Wide survey would benefit the development of LAE clustering studies. Larger areas of the sky would be covered (i.e., larger clustering scales), a larger sample of LAEs would be detected and the higher SN ratio would further decrease the uncertainties in the measurements. This will be the case for HETDEX \citep{hill}, which will provide a higher SN ratio and a much larger coverage of the sky, assisting the understanding of the cosmology behind LAEs.

\begin{acknowledgements}
      The authors give thanks to the staff at ESO for extensive support during the visitor-mode campaigns at Paranal Observatory. We thank the eScience group at AIP for help with the functionality of the MUSE-Wide data release webpage. M.K. acknowledges support by DLR grant 50OR1904 and DFG grant KR 3338/4-1. L.W. and T.U. acknowledge funding by the Competitive Fund of the Leibniz Association through grants SAW-2013-AIP-4 and SAW-2015-AIP-2. T.M. thanks
      for financial support by CONACyT Grant Cient\'ifica B\'asica \#252531 and by UNAM-DGAPA (PASPA and PAPIIT IN111319). T.G. is supported by the ERC Starting Grant 757258 "TRIPLE". The data were obtained with the European Southern Observatory Very Large Telescope, Paranal, Chile, under Large Program 185.A-0791. This research made use of Astropy, a community-developed core Python package for Astronomy \citep{astropy}. We also thank the referee for an useful and constructive report.
\end{acknowledgements}


\begin{appendices} 

\renewcommand\thefigure{\thesection.\arabic{figure}}
\counterwithin{figure}{section}

\renewcommand\thetable{\thesection.\arabic{table}}
\counterwithin{table}{section}

\section{Effect of the HUDF parallel fields on the K-estimator}
\label{appendix:k-fields}

In this research we have focused on 68 fields of the MUSE-Wide survey, including part of the CANDELS/GOODS-S region and the 8 parallel HUDF fields. In order to assess a homogeneous sample, cover a larger area of the sky, and maximize the number of detected galaxies the 8 HUDF parallel fields have been included. 
In this section we explore the possible effects on the clustering measurements of including the parallel fields. We study the clustering in the 68 fields (same as throughout the paper) and that present in the 60 fields (without including the 8 parallel HUDF fields). The characteristics of the first sample are described in Sect.~\ref{sec:data}, while the second sample covers a total of 54.74 arcmin$^2$ and has 581 LAEs. 

\begin{figure}[h]
\centering
\includegraphics[width=\columnwidth]{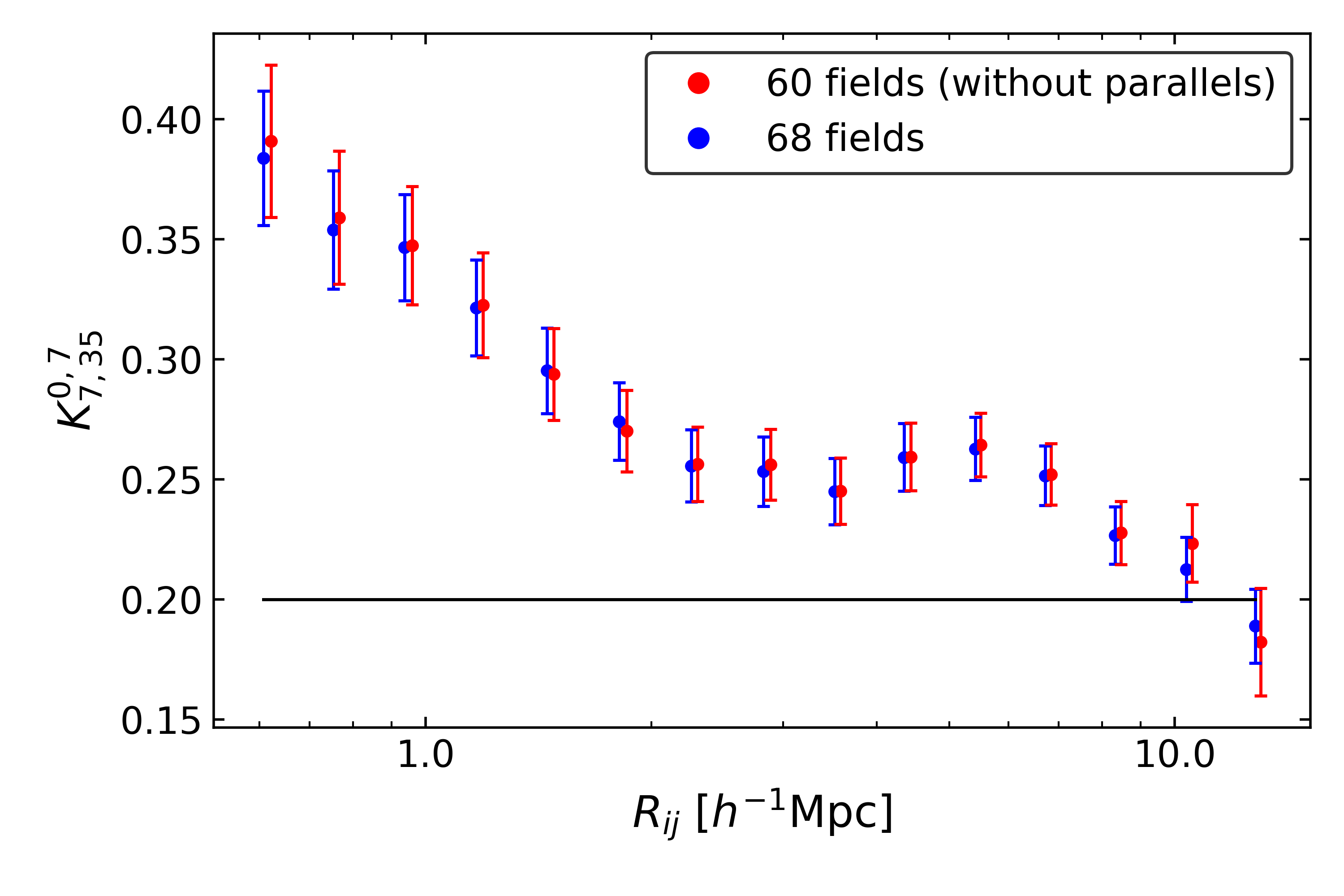}
\caption{$K_{7,35}^{0,7}$ estimator for the LAEs in 60 and 68 fields of the MUSE-Wide survey in red and blue, respectively. The black straight line shows the expected $K$ value of an unclustered sample. All error bars are Poissonian. The red dots have been shifted along the x-axis for visual purposes. }
\label{fig:k-fields}
\end{figure}

The K-estimator is then run in both samples and shown in Fig.~\ref{fig:k-fields}. We demonstrate the insignificant effect on our main results due to the inclusion or exclusion of the parallel fields. The two curves are indistinguishable within the approximated Poissonian uncertainties $\sqrt{N_{a1,a2}}/(N_{a1,a2}+N_{a2,a3})$ (see Sect. 4.2 in \citealt{adelberger}) but due to the lower number of LAEs in the 60 fields the uncertainties are somewhat larger (8\%) than those of the 68 fields.

The minimal effect on the clustering signal, the larger area of the sky covered, which makes it more representative in terms of cosmic variance, and the larger number of LAEs in the sample, which reduce the uncertainties, lead us to include the 8 parallel HUDF fields in our main sample of analysis.

\section{Effect of Ly$\alpha$ derived redshifts on the K-estimator}
\label{appendix:k-redshifts}

Inferring the redshift of galaxies from their Ly$\alpha$ lines introduces an offset (i.e., a few hundreds of km/s) with respect to their systemic redshift \citep{hashimoto15}. This offset translates into small uncertainties in the derived positions of the galaxies (i.e., $\sim$3 Mpc) that would affect the clustering measurements when scrutinized through traditional methods \citep[see e.g.][]{gurung2020}. The K-estimator compares galaxy pair counts in two consecutive shells along l.o.s. separations. Thus, the offset introduced in the separations affects both shells equally, being simultaneously compensated. Besides, we work with much larger scales than $\sim$3 Mpc so even from a theoretical point of view the K-estimator should not be sensitive to these redshift offsets.

\begin{figure}[h]
\centering
\includegraphics[width=\columnwidth]{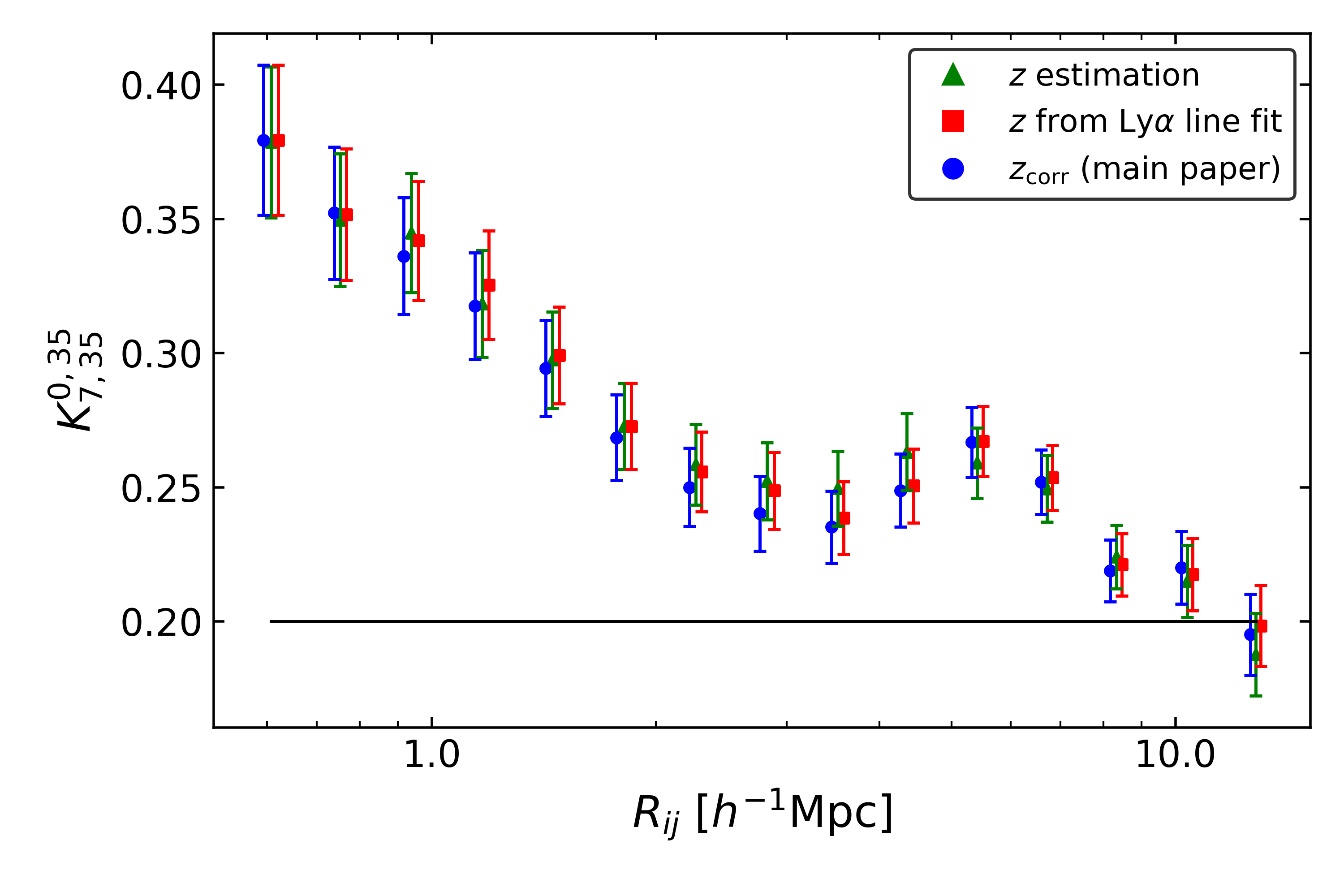}
\caption{$K_{7,35}^{0,7}$ estimator for the LAEs in the MUSE-Wide survey. The green triangles represent the K-estimator values of the sample of LAEs with redshift estimations from QtClassify, the red squares show the $K$ values when the redshifts are obtained from the Ly$\alpha$ line fit with asymmetric Gaussians and the blue circles show the same previous redshifts but including the correction for the offset between the Ly$\alpha$ and the systemic redshift (same as in all plots in the main paper where the K-estimator results are shown). The black straight line shows the expected $K$ value of an unclustered sample. All sets of data points are plotted along with Poisson errors. The blue circle and red square values have been shifted along the x-axis for visual purposes. }
\label{fig:zerrors}
\end{figure}

We prove this fact in Fig.~\ref{fig:zerrors}. We consider the same sample of LAEs but we obtain the redshift of the galaxies in different manners. We first use the redshift estimates from QtClassify (see Sect.~\ref{sec:data}) and then we use more precise redshifts obtained by fitting asymmetric Gaussian profiles to the Ly$\alpha$ emission lines. Finally, from those precise redshifts we correct the redshift following \cite{verhamme18} as described in Sect.~\ref{sec:data1}. We run the K-estimator for these three samples, which only differ by their source redshift estimates. In Fig.~\ref{fig:zerrors} we show the minimal impact of the redshift uncertainties in our results, showing that the three different redshift samples provide negligible variations in the K-estimator values. 
The K-estimator on the sample with redshift estimations provides large-scale bias factors of $b = 3.00^{+1.73}_{-0.56}$ from PL fits, while using the sample with the corrected redshifts (main paper) $b = 3.03^{+1.51}_{-0.52}$.


\section{Error estimates in the K-estimator}
\label{sec:errors}

When computing the clustering signal with the K-estimator, one has to recognize that the individual data points are correlated. Various galaxy pairs can be part of more than one $R_{ij}$ bin and the same galaxy may be counted in more than one galaxy pair. The extent of how bin $i$ correlates with bin $j$ is usually expressed with the covariance matrix. However, the small area covered by our survey does not allow us to calculate a covariance matrix. Therefore, we have investigated several error approaches for our K-estimator measurements.

We apply the bootstrapping technique described in \cite{ling}, which creates pseudo-data sets by sampling $N$ sources with replacement from the real sample of $N$ galaxies. In other words, we randomly draw objects from the real sample, allowing multiple selections of the same object, to generate a pseudo-sample with the same number of objects $N$ as the real sample. We repeat the process 500 times, obtaining a large set of pseudo-samples, which vary moderately from the original data. We compute the K-estimator in the 500 pseudo-samples, $N_{\text{boots}}=500$. The scatter from all the measurements is used as our uncertainty estimations.

We consider a second technique, in which we generate random samples. Ideally, 500 different realizations from cosmological simulations should be applied but we showed in Sect.~\ref{sec:simulations} that the simulated data cannot directly be compared to our clustering measurements. Therefore, we use the selection and luminosity functions of the survey \citep{herenz19} to obtain the real $z$-distribution of our LAEs (see Sect.~\ref{sec:data} and red curve in Fig.~\ref{fig:z_distribution}). 
\begin{figure}[h]
\centering
\includegraphics[width=\columnwidth]{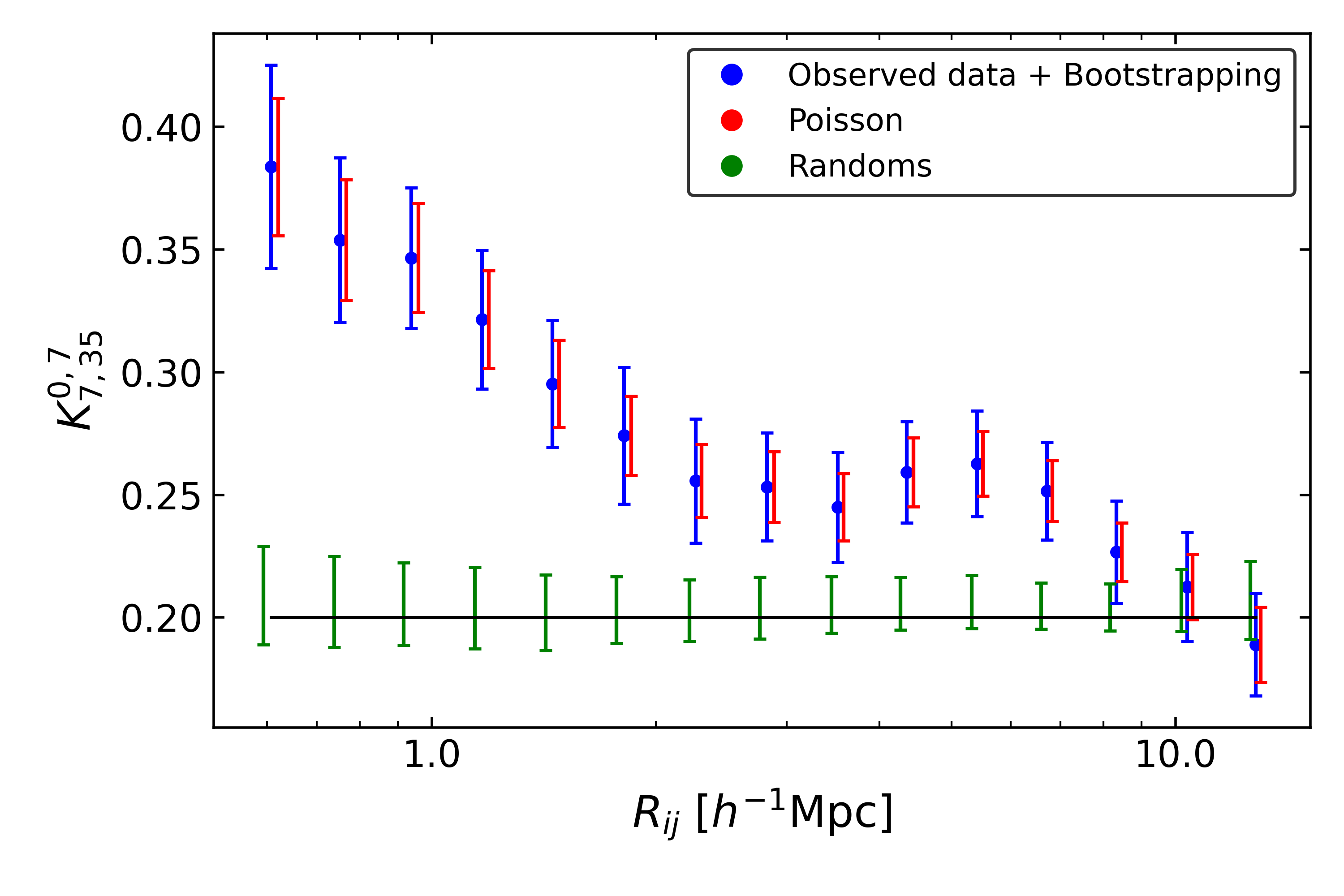}
\caption{$K_{7,35}^{0,7}$ estimator of the LAEs in the 68 fields of the MUSE-Wide survey with the different error approaches. Approximated Poisson errors (i.e., $\sqrt{N_{a1,a2}}/(N_{a1,a2}+N_{a2,a3})$) are shown in red, while the blue and green uncertainties are obtained from the bootstrapping and the random sample generation methods, respectively. Both sort of error bars have been shifted in the $R_{ij}$ direction for illustration purposes. As usual, the black line
represents an unclustered sample of galaxies.}
\label{fig:errors}
\end{figure}
Randomly clustered samples with the same number of objects as the real sample are created from the real redshift distribution. Each new galaxy is located at a random position within the MUSE-Wide survey footprint, with a specific redshift provided by the combination of LF and SF. Following this procedure, we generate 500 different random samples and compute $K^{0,a_2}_{a_2,a_3}$ in each of the samples. For each bin, the uncertainty has been obtained as the standard deviation from the resulting $K_{a2,a3}^{a1,a2}$ of the different 500 random samples.

We show the two main uncertainty approaches for the K-estimator in Fig.~\ref{fig:errors}. We also include the approximated Poisson errors $\sqrt{N_{a1,a2}}/(N_{a1,a2}+N_{a2,a3})$ calculated by the error propagation of Eq.~\ref{Kestimator}.

Whereas the random errors are 20\% smaller than Poisson errors, we find that the uncertainties from bootstrapping are moderately larger ($\sim$35\%) than those of Poisson. In order to be conservative and even if all uncertainties are comparable, we decide to compute the error bars in our K-estimator analyses following the classical bootstrapping approach. Hence, uncertainties due to cosmic variance are not represented yet in our error estimates of the real data. In other words, repeating the same LAE clustering studies in different regions of the sky can lead to clustering signals outside our expected uncertainty range. The cosmic variance contribution to the total error budget was explored in Sect.~\ref{sec:simulations}. Including this contribution results in $\sim70$\% larger uncertainties. However, this additional uncertainty is not impacting our results based on comparing the subsamples because they are obtained in the same sky field.

\section{Two-point correlation function analysis}
\label{2pcf}
\setcounter{equation}{0}
\renewcommand\theequation{D.\arabic{equation}}
\subsection{2pcf method}

The 2pcf is the most commonly used statistical approach to explore the clustering in a sample of objects. Traditionally those samples cover a broad spatial coverage that accounts for cosmic variance and allows the computation of a covariance matrix to estimate clustering uncertainties. With the MUSE-Wide survey, we are facing the opposite scenario: small spatial coverage, wide redshift range.

In reconstructing the 2pcf we follow standard recipes \citep{landy1993}. To recall, $\xi(r)$ quantifies the excess probability $P$ over a random Poisson distribution of finding a pair of galaxies separated by a distance $r$ \citep{peebles1980}
\begin{equation}
    dP = n[1 + \xi(r)]\text{d}V, \label{peebles}
\end{equation}
where $\text{d}V$ is the infinitesimal volume occupied by the pair and $n$ is the average number density of galaxies.

Galaxy distances along the light of sight (l.o.s.) cannot be measured directly. Instead the redshift information of the galaxies is used, which is affected by their peculiar velocities. In order to eliminate this effect, the so-called redshift space distortions (RSD), we compute the correlation function in a 2D grid. We measure the separation of pairs in the perpendicular distance to the l.o.s., $r_p$, and parallel to the l.o.s., $\pi$. We then count pairs of LAEs within given separations and compare them to those in a random sample of galaxies by means of the Landy-Szalay estimator \citep{landy1993} 
\begin{equation}
\xi(r_p, \pi) = \frac{DD(r_p, \pi) - 2DR(r_p, \pi) + RR(r_p, \pi)}{RR(r_p, \pi)}, \label{eq:landy-szalay} 
\end{equation}
where $DD$, $RR$ and $DR$ are the normalized data-data, random-random and data-random pairs. In other words, expressing the actual number of pairs as $n_{\text{pair},DD}(r_p, \pi)$, $n_{\text{pair},DR}(r_p, \pi)$ and $n_{\text{pair},RR}(r_p,\pi)$
\begin{align}
  & DD = n_{\text{pair},DD}(r_p, \pi)/[N_D(N_D-1)] \nonumber \\
  & DR = \frac{1}{2} n_{\text{pair},DR}(r_p, \pi)/(N_D N_D) \nonumber \\
  & RR = n_{\text{pair},RR}(r_p, \pi)/[N_R(N_R-1)].
\end{align}
$N_D$ and $N_R$ are the total number of galaxies in the real and random sample, respectively.

We then calculate the projected correlation function $\omega(r_p)$ by integrating $\xi(r_p, \pi)$ along the $\pi$-direction \citep{davispeebles}
\begin{equation}
\label{eq:wp}
  \omega_p(r_p) \approx 2 \int_{0}^{\pi_{\text{max}}}  \xi(r_p, \pi) d\pi,
\end{equation}
where $\omega_p$ is the projected 2pcf and $\pi_{\text{max}}$ is the maximum allowed l.o.s. distance between pairs of galaxies to be considered as a pair. $\pi_{\text{max}}$ is chosen such that it accounts for most correlated pairs and the amplitude of $\omega_p(r_p)$ is able to converge. Very large values would mainly increase the noise since there are not many correlated pairs at large l.o.s. distances. Contrarily, very low values would not include most correlated pairs and would underestimate $\omega_p(r_p)$.

 We compute $\pi$ within 10 $-$ 300 $h^{-1}$Mpc in steps of 10 $h^{-1}$Mpc and $r_p$ in the range of 0.375 \textless $\; r_p/h^{-1}\rm{Mpc}$ \textless $\;$13.155 in 9 logarithmic bins. We then calculate the projected correlation function $\omega_p(r_p)$ for each $\pi$ value and fit the analytical solution 
\begin{equation}
\label{eq:fit}
  \omega_p(r_p) = r_p \; \left(\frac{r_0}{r_p}\right)^\gamma \; \frac{\Gamma(1/2) \;\Gamma((\gamma-1)/2)}{\Gamma(\gamma/2)},
\end{equation}
where $\Gamma(x)$ is the Gamma function. 
The fittings are performed in the range 0.584 \textless $\; r_p/h^{-1}\rm{Mpc}$ \textless $\;$13.155 (two-halo term only) for each of the $\omega_p(r_p)$ curves. 

We measure the correlation length of the curves using Eq.~\ref{eq:fit} with a fixed slope of $\gamma=1.8$. In order to be conservative, we use $\pi_{\text{max}} = 60$  $h^{-1}$Mpc. Similar $\pi_{\text{max}}$ values are obtained with the simulation described in Sect.~\ref{sec:simulations}. Literature $\pi_{\text{max}}$ values in similar LAE clustering studies used less conservative values \citep[15--20 $h^{-1}$Mpc; ][]{durkalek,durkalekp}.

\begin{figure}[h]
\centering
\includegraphics[width=\columnwidth]{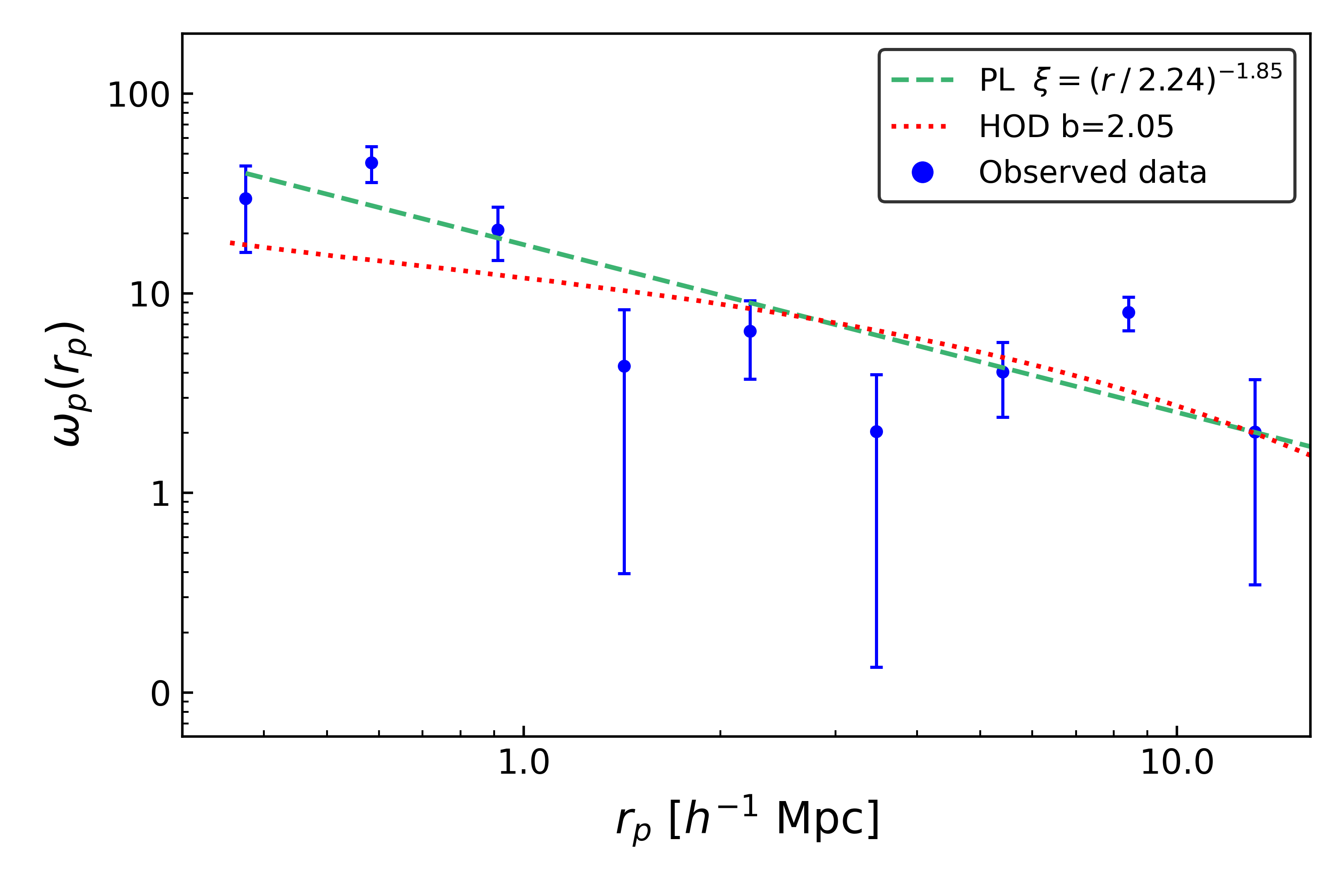} 
\caption{Best PL and HOD fits to the projected 2pcf $\omega_p(r_p)$ for $\pi_{\text{max}} = 60$ $h^{-1}$Mpc. The dashed green curve shows the PL-fit while the dotted red curve represents the HOD fit. The error bars are determined from the random approach explained in Appendix~\ref{appendix:error}.} 
\label{fig:2pcf-fit}
\end{figure}

   \begin{table*}[ht]
   \caption{Best-fit clustering parameters from 2pcf measurements.}
   \label{table:2pcf1}
   \centering
    \begin{tabular}{l@{\qquad}ccccc}
        \hline \hline
           \noalign{\smallskip}
        $r_0 \; [h^{-1}$Mpc] & $\gamma$ &  $b_{\rm{PL}}$ &  $b_{\rm{HOD}}$  & log$(M_{\rm{DMH}}/ [h^{-1}$M$_{\odot}$])  \\
             \noalign{\smallskip} 
            \hline \hline
            \noalign{\smallskip}
            $2.24^{+0.25}_{-0.35}$ & 
            $1.85^{+0.25}_{-0.25}$ & 
            $1.66^{+0.36}_{-0.42}$ & 
            $2.05^{+0.14}_{-0.14}$ & 
            $10.51^{+0.16}_{-0.17}$    \\
            \noalign{\smallskip}
            \noalign{\smallskip}
        \hline 
        \multicolumn{5}{l}{%
          \begin{minipage}{10.5cm}%
          \vspace{0.2\baselineskip}
            \small \textbf{Notes}: The correlation length and slope, the linear bias factor assuming a PL correlation function, the linear bias factor and typical dark matter halo masses from the HOD model are indicated. The uncertainties in the bias factors and DMH masses reflect the statistical error on $r_0$ only.
          \end{minipage} 
          }\\
    \end{tabular}
\end{table*}

Besides the $\pi_{\text{max}}$ determination, the measurement of the 2pcf also demands the modeling of a random sample of galaxies with the same geometry, selection effects and observational conditions as the real sample. Hence, we constructed a random sample from the real $z$-distribution of the sample (red curve in Fig.~\ref{fig:z_distribution}) obtained from the SF and the LF of the MUSE-Wide survey \citep{herenz19}. The number of random objects is chosen to be 100 times the number of galaxies in the real sample. This makes the variance of $DR$ and $RR$ in Eq.~\ref{eq:landy-szalay} negligible. We have verified that increasing the number of random galaxies or using different random samples have an insignificant effect on our estimates. Each random galaxy is then located at a random position of the sky within the MUSE-Wide survey footprint, with a redshift taken from the real $z$-distribution. Given that our  $\pi_{\text{max}}$ is much smaller than the radial comoving distance corresponding to our sample redshift range, over which the random sample is constructed, the effects for the integral constraint are negligible for our $\omega_p(r_p)$.

\subsection{Error estimates}
\label{appendix:error}

\begin{figure}[h]
\centering
\includegraphics[width=\columnwidth]{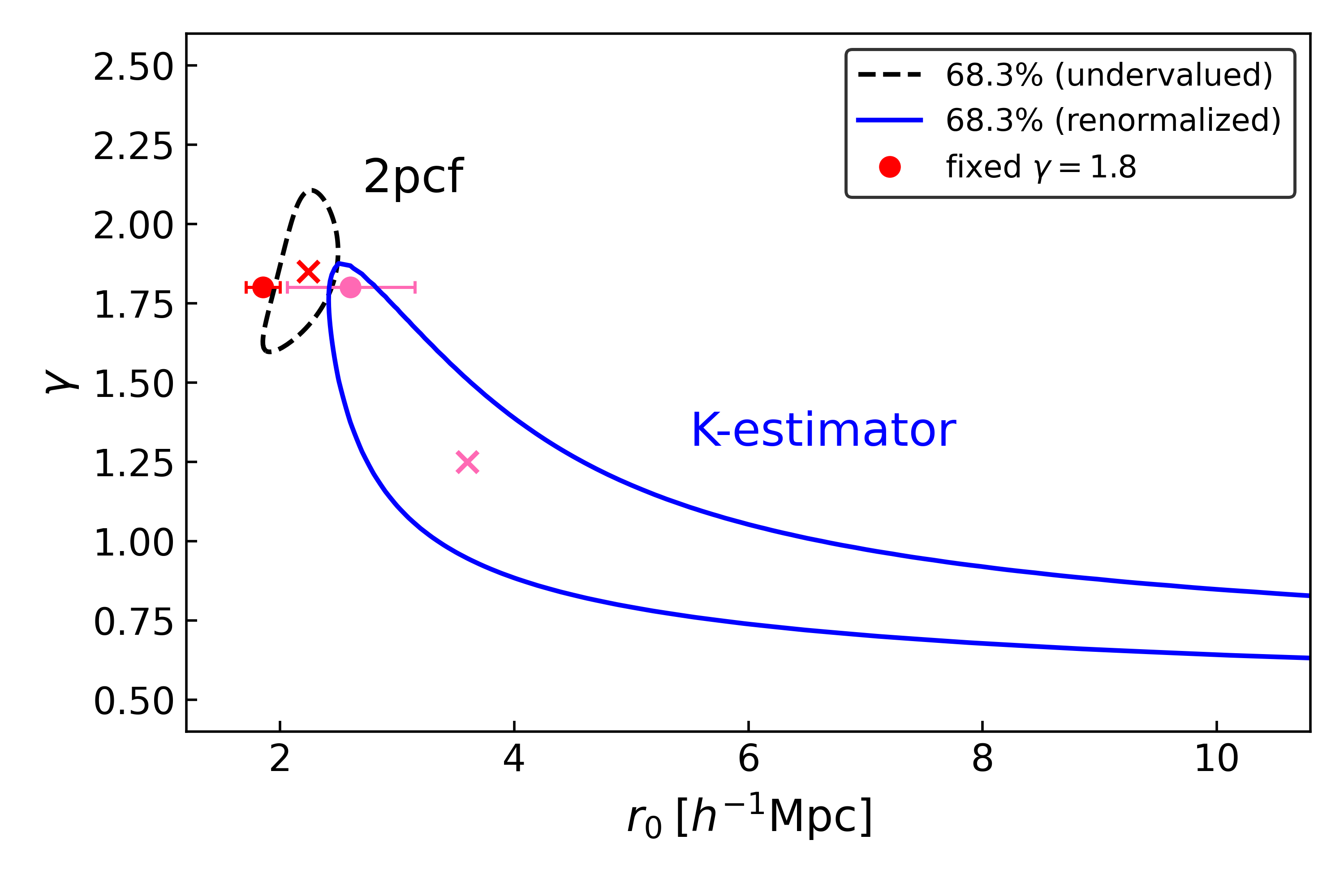}
\caption{Analogous to Fig.~\ref{fig:k-contours}, here showing the contours from the 2pcf method in black-gray. For a direct comparison to the K-estimator, its contours have been represented in blue, as well as the $r_0$ computed from both methods when fixing the slope of the PL to the standard value, shown with dots. Please note that we consider the contours of the 2pcf to be likely underestimated (see text for further details).}
\label{fig:contours-2pcf}
\end{figure}

The bootstrapping approach applied in the K-estimator method allows for replacement and galaxy repetitions. This produces an overlap of galaxies, which introduces an unrealistic clustering excess in the 2pcf. Whereas this does not affect the K-estimator because it measures a contrast of galaxy pairs between two l.o.s. regions and overlaps of galaxies cancel out in both areas, the 2pcf is severely affected. We therefore do not consider the bootstrapping approach as a possible error determination technique for the 2pcf. 

We consider an alternative approach by generating random samples. Analogous to the error estimates in the K-estimator, 100 different light-cones from cosmological simulations should be used instead but it was shown in Sect.~\ref{sec:simulations} that the simulated data should not be directly compared to our measurements. Thus, we create random samples from the real redshift distribution of our LAEs. The random samples have the same number of objects as the MUSE-Wide survey. 
The real redshift distribution is calculated from the luminosity and the selection function of the sample \citep{herenz19} as described in Sect.~\ref{sec:data}. We denote the newly created random sample by $R'$ to distinguish from the random sample $R$ in Eq.~\ref{eq:landy-szalay}. The 2pcf in the random samples is calculated by replacing $D$ by $R'$ in Eq.~\ref{eq:landy-szalay} for 100 different newly generated random samples $R'$ (i.e., $N_{\text{ran}} = 100$). The scatter of the 100 runs is used as our uncertainty estimation. 

This approach is compared to Poissonian errors calculated by error propagation in Eq.~\ref{eq:landy-szalay}, $\sqrt{ (\delta_{DD}/RR)^2+4\cdot(\delta_{DR}/RR)^2+((2DR-DD)\cdot\delta_{RR}/RR^2)^2})$. We find that the Poissonian uncertainties underestimate the true clustering errors compared to the random-sample approach (uncertainties from the random error approach are $\sim$65\% larger than Poisson errors). However, even the uncertainties from the random error approach should be understood only as a first guess. The combination of the 2pcf and the special design of our MUSE-Wide survey leave this as the only option to estimate the extent of the uncertainties on the 2pcf.


\subsection{Results}
\label{sec:results-2pcf}
      We present the projected correlation function $\omega_p(r_p)$ for $\pi_{\text{max}}=60 \; h^{-1}$Mpc over the range 0.375 \textless $\; r_p/h^{-1}\rm{Mpc}$ \textless $\;$13.155  in Fig.~\ref{fig:2pcf-fit}. The error bars in $\omega_p(r_p)$ have been computed following the random-sample approach described in Appendix~\ref{appendix:error}.  

Despite the small area covered by the MUSE-Wide survey, the $\omega_p(r_p)$ curve shows a clear clustering signal. The large number of galaxies in the MUSE-Wide survey allow us to fit the clustering signal with a PL-based correlation function, where both correlation length and slope are constrained simultaneously. 
Thus, we set $r_0$ and $\gamma$ as the free parameters to be determined from the fit. 

We then use Eq.~\ref{eq:fit} to fit $\omega_p(r_p)$ in the 2-halo term, $0.584<r_p/h^{-1}\rm{Mpc}<13.155$. We also use this $r_p$ range to fit the curve with the HOD model described in Sect.~\ref{sec:hod}, same as for the K-estimator. The measured best-fit parameters are listed in Table \ref{table:2pcf1} and shown in Fig. \ref{fig:2pcf-fit}. The probability contours from the $r_0$-$\gamma$ grid of the PL-fit are shown in Fig.~\ref{fig:contours-2pcf}, along with those from the K-estimator to allow a direct comparison. See Appendix \ref{kvs2pcf} for a discussion of the different clustering methods and results.

With the PL-fit we find $r_0 = 2.24^{+0.25}_{-0.35} \; h^{-1}$Mpc with a correlation slope $\gamma = 1.85 \pm 0.25$. These parameters correspond to $b=1.66^{+0.36}_{-0.42}$. These results are in agreement with the derived correlation lengths from the one-bin fit (fixed $\gamma=1.8$) and from the PL-fit (free $r_0$ and $\gamma$) to the K-estimator (contours in Fig.~\ref{fig:contours-2pcf}). However, for the 2pcf case, $\gamma$ is much closer to the canonical value than the K-estimator. When considering HOD fits, we obtain $b=2.05\pm 0.14$, somewhat lower ($1.3\sigma$) than the $b=2.80\pm 0.38$ obtained with the K-estimator.  The differences in the derived linear bias factors from PL and HOD fits are discussed in detail in Sect.~\ref{sec:discussion_hod-pl}.

In order to better constrain the correlation parameters, we fix the correlation slope and determine only the correlation length. Following this procedure and in order to be consistent with the literature, we fix the slope to $\gamma=1.8$ and derive from the one-parameter fit $ r_0= 1.85 \pm 0.15 \; h^{-1}$Mpc. This value agrees at a one sigma level with the one derived from the K-estimator ($ r_0= 2.60^{+0.72}_{-0.67} \; h^{-1}$Mpc; fixed $\gamma=1.8$).
\subsection{K-estimator vs 2pcf}
\label{kvs2pcf}

The various clustering methods studied in this paper allow us to compare not only their respective results but also their success as methods themselves in these sort of surveys. Whereas the 2pcf is very well known, the K-estimator is still relatively unexplored. 
However, for galaxy surveys that cover small areas of the sky but span wide redshift ranges the K-estimator seems to be a more suitable clustering method than the commonly used 2pcf. Both methods have important similarities but also present critical differences. First of all, the concept of measuring clustering itself differs. While the 2pcf measures the spatial clustering by comparing pairs of galaxies to those in random samples, the K-estimator compares the contrast of galaxy pairs in two consecutive shells along l.o.s. distances, without introducing any random sample and focusing on redshift clustering rather than on spatial clustering. Second, choosing the most suitable K-estimator or the upper integration limit $\pi_{\text{max}}$ in the 2pcf share some concepts. The $\pi_{\text{max}}$ value where the 2pcf saturates collects the maximum number of galaxy pairs and tries to discard noise from distant uncorrelated pairs. 
The $a_2$ and $a_3$ values of the K-estimator boost the clustering signal by finding the two shells along l.o.s. separations where the highest contrast of galaxy pairs is encountered. 
Therefore, $\pi_{\text{max}}$ just represents an upper integration limit in l.o.s. distances and $a_2$ and $a_3$ are the length of the shells with the highest difference in pair counts. Typically, $a_3$ should be below the upper integration limit $\pi_{\text{max}}$. Finally, both methods quantify the clustering of a sample of galaxies by counting galaxy pairs in 3D space.

Although the two methods present a clustering signal over equal transverse distances, $R_{ij}$ and $r_{p}$, and the $a_i$ values are within the $\pi_{\text{max}}$ limit, the fitting parameters are somewhat distinct. This was expected because, first, the 2pcf in this type of surveys has issues. Its performance is affected by the small spatial coverage of the data and the survey geometry.
 We do not see a clear saturation point of the $\omega(r_p)$ curves for the different $\pi_{\text{max}}$ values and error estimations such as the jackknife method fail. Hence, we believe that we are exploring the limit of the method. Second, the fits for both methods are carried out in different manners (see Sect.~\ref{sec:k-estimator} and Appendix \ref{sec:results-2pcf}). Whereas we fit $\omega_p(r_p)$ with its analytical solution (Eq.~\ref{eq:fit}), the K-estimator is either compared to the expectation value of $K_{a_2, a_3}^{a_1, a_2}$ (Eq.~\ref{eq:expected_k}) through the one-fit approach or fitted with a PL with the PL-fit approach. If we allow $r_0$ and $\gamma$ to freely vary in the fit for both methods, the correlation lengths, bias factors and DMH masses obtained from the K-estimator and the 2pcf are $r_0=3.60^{+3.10}_{-0.90}$ $h^{-1}$Mpc, $\gamma=1.30^{+0.36}_{-0.45}$ $b_{\text{HOD}}=2.80^{+0.38}_{-0.38}$, log$(M_{\text{DMH}}/[h^{-1}\text{M}_\odot])=11.34^{+0.23}_{-0.27}$ and $r_0=2.24^{+0.25}_{-0.35}$ $h^{-1}$Mpc, $\gamma=1.85\pm 0.25$, $b_{\text{HOD}}=2.05^{+0.14}_{-0.14}$ and log$(M_{\text{DMH}}/[h^{-1}\text{M}_\odot])=10.51^{+0.16}_{-0.17}$, respectively. Even if the HOD fits from the K-estimator are higher ($1.3\sigma$) than those computed with the 2pcf, the 68.3\% confidence intervals of the PL-fits agree. 
 
A noteworthy addition to our discussion is the uncertainty dissimilarities between the methods, where the difference in the probability contour sizes of Fig.~\ref{fig:contours-2pcf} come from. It is important to notice that the uncertainties in the K-estimator were obtained through the bootstrapping approach. However, bootstrapping causes overlaps of galaxies, to which the K-estimator is insensitive but in the 2pcf case, this causes a significant boost in the clustering signal. Thus, we have to consider the random sample approach as the only way to give some educated guess on the 2pcf uncertainties, even if it most likely still underestimates the real uncertainties. Thus, the 2pcf contour shown in Fig.~\ref{fig:contours-2pcf} is also most likely underestimated.

For the 2pcf we had to apply the standard (uncorrelated) $\chi^2$ analysis with likely underestimated uncertainties, while for the K-estimator we were able to renormalize the $\chi^2$ analysis and used conservative uncertainty estimates as described in Sect.~\ref{sec:errorsK}.
Despite these fundamental differences the clustering results derived from both methods still agree within their combined $2\sigma$ uncertainties.

The K-estimator is a more suitable clustering statistic than the 2pcf in these kind of surveys for the following reasons: (i) The K-estimator exploits the large redshift coverage rather than the spatial extent (more than 1000 $h^{-1}$Mpc along the l.o.s. direction vs only 20 $h^{-1}$Mpc over transverse separations), (ii) it does not require a random sample so integral constraint issues do not take place, (iii) we can use bootstrapping uncertainties when the Jackknife technique is not an option, and (iv) with the K-estimator we provide a straightforward recipe to obtain rough $a_2$ and $a_3$ values (unlike $\pi_{\rm{max}}$ in the 2pcf).
\section{Effect of redshift space distortions on our measurements}
\label{appendix:discussion_rsd}

Both PL and HOD fit approaches show a high performance on the K-estimator. Nevertheless, none of them accounts for the redshift space distortions present in the observations of the high-redshift Universe. Galaxy structures are observed "falling into" large-scale overdensities, which varies the projected velocity along the l.o.s. from that linked to its cosmological redshift. As explained in Sect.~\ref{sec:hod}, we include the effects of the redshift distortion in the HOD model using the linear theory to the 2-halo term only. Therefore, the large-scale streaming motion towards overdense regions \citep[i.e., Kaiser infall, ][]{kaiser} is corrected in the linear regime.

\begin{figure}[h]
\centering
\includegraphics[width=\columnwidth]{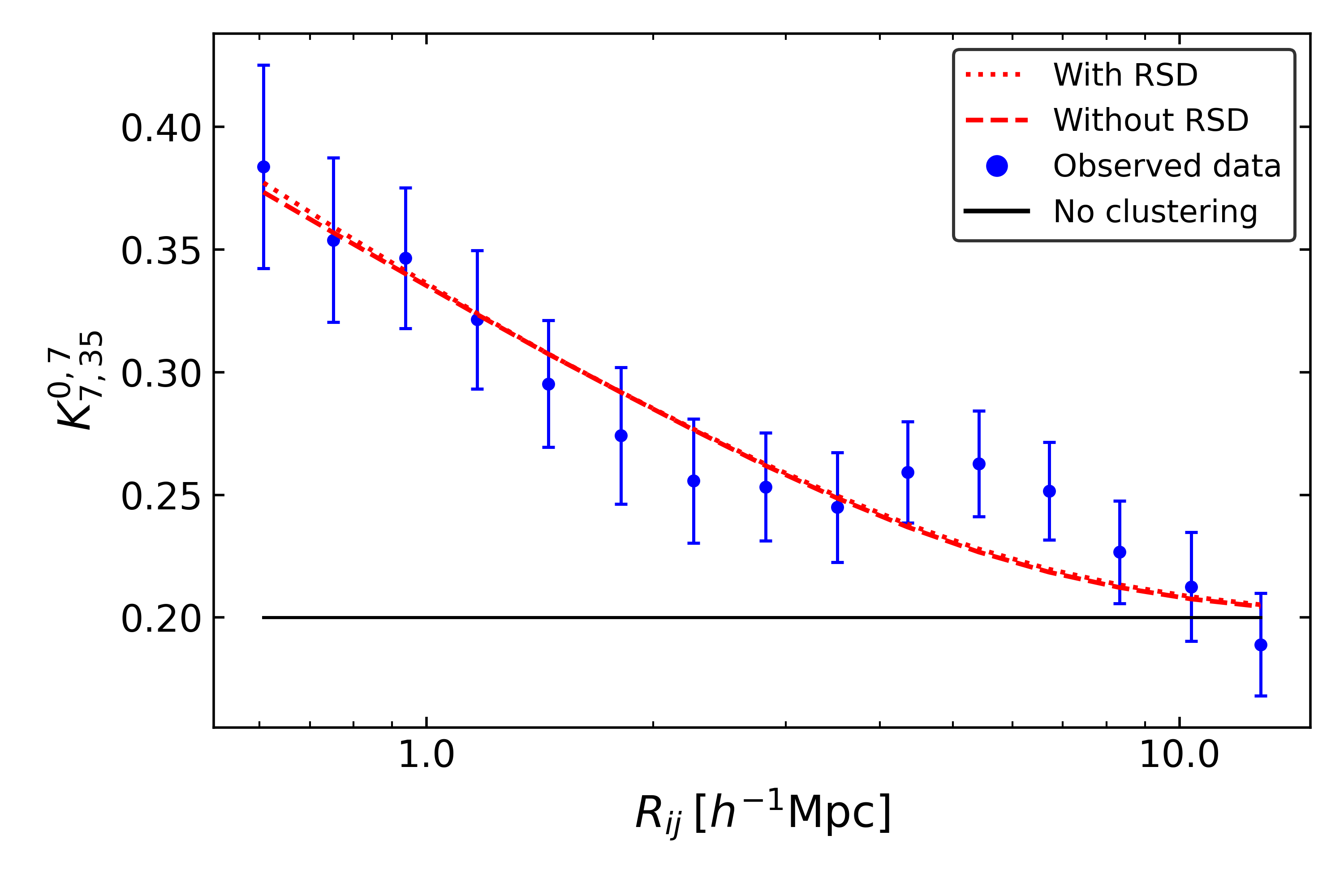}
\caption{Best HOD fits to the K-estimator from $\xi(r)$ (same as the thick curve in Fig.~\ref{fig:k-hod}) and $\xi(s)$. The dotted red curve represents the HOD that takes into account the effect of the RSD, while the dashed red curve shows the HOD fit without RSD. The black line represents an unclustered sample of galaxies.}
\label{fig:k-rsd}
\end{figure}

In order to account for these effects, we implement the HOD correlation function in redshift space, $\xi(s)$, accounting thus for RSD. We represent the HOD fit to the K-estimator both from the spatial space, $\xi(r)$, (same as in Sect.~\ref{sec:results}) and the redshift space, $\xi(s)$, in Fig.~\ref{fig:k-rsd}.

The HOD fit from the redshift space correlation function (i.e., with RSD) is slightly higher than that from the real space correlation function (i.e., without RSD) at small separations ($R_{ij} < 1 \; h^{-1}$Mpc), showing the minimal RSD effect on the K-estimator. This small rise translates into an increase in the derived bias factor, from $b=2.80^{+0.38}_{-0.38}$ to $b=2.84^{+0.33}_{-0.34}$, indistinguishable within the 1$\sigma$ error bars.

\end{appendices}


\begin{thebibliography}{}

    \bibitem[Adelberger et al.(2005)]{adelberger} 
    Adelberger, K. L., Steidel, C. C., Pettini, M., Shapley, A. E., Reddy, N. A., \& Erb, D. K. 2005, ApJ, 619, 697-713
    
    \bibitem[Ahumada et al.(2020)]{ahumada} 
    Ahumada, R., Prieto, C. A., Almeida, A., et al. 2020, ApJSS, 249, 3
    
    \bibitem[Astropy Collaboration et al.(2013)]{astropy} Adelberger, Astropy Collaboration, Robitaille, T. P., Tollerud, E. J., et al. 2013, A\&A, 558, A33
    
    \bibitem[Bacon et al.(2010)]{bacon} 
    Bacon, R., Accardo, M., Adjali, L., et al. 2010,  Proc. SPIE, 7735, 773508
    
    \bibitem[Bielby et al.(2016)]{bielby} 
    Bielby, R. M., Tummuangpak, P., Shanks, T., et al. 2001, MNRAS, 456, 4061
    
    \bibitem[Coil et al.(2009)]{coil09} 
    Coil, A. L., Georgakakis, A., Newman, J. A., et al. 2009, ApJ, 701, 1484
    
    \bibitem[Coil (2012)]{coil} 
     Coil, A. L. 2012, Planets, Stars and Stellar Systems (Springer, Dordrecht), Volume 6
     
    \bibitem[Colless et al.(2012)]{2df} 
    Colless, M., Dalton, G., Maddox, S., et al. 2012, MNRAS, 328, 1039-1063
    
    \bibitem[Cowie \& Hu(1998)]{cowie} 
    Cowie, L. L. \& Hu E. M. 1998, ApJ, 115, 1319
    
    \bibitem[Davis \& Peebles(1983)]{davispeebles} 
     Davis, M. \& Peebles, P. J. E. 1983, ApJ, 267, 465
     
     \bibitem[Dayal et al.(2012)]{dayal12} 
     Dayal, P. \& Libeskind, N. I. 2012, MNRAS, 419, L9

    \bibitem[Diener et al.(2017)]{catrina} 
    Diener, C., Wisotzki, L., Schmidt, K. B., et al. 2017, MNRAS, 471, 3186-3192
    
    \bibitem[Durkalec et al.(2014)]{durkalek} 
    Durkalec, A., Le F\`evre, O., Pollo, A., et al. 2014, A\&A, 583, A128
    
    \bibitem[Durkalec et al.(2018)]{durkalekp} 
    Durkalec, A., Le F\`evre, O., Pollo, A., et al. 2018, A\&A, 612, A42
    
    \bibitem[Forero-Romero et al.(2012)]{forero12} 
   Forero-Romero, J. E., Yepes, G., Gottl\"ober, S., et al. 2011, MNRAS, 415, 3666
 
    \bibitem[Fry(1996)]{fry} Fry, J. N. 1996, ApJ, 461, L65
    
    \bibitem[Garel et al.(2012)]{garel12} 
    Garel, T., Blaizot, J., Guiderdoni, B., et al. 2012, MNRAS, 422, 310
    
    \bibitem[Garel et al.(2015)]{garel15} 
    Garel, T., Blaizot, J., Guiderdoni, B., et al. 2015, MNRAS, 450, 1279
    
    \bibitem[Garel et al.(2016)]{garel} 
    Garel, T., Guiderdoni, B. \& Blaizot, J. 2016, MNRAS, 455, 3436
    
    \bibitem[Gawiser et al.(2007)]{gawiser07} 
    Gawiser, E., Francke, H., Lai, K., et al. 2007, ApJ, 671, 278 
    
    \bibitem[Gurung-L\'opez et al.(2018)]{gurung18} Gurung-L\'opez, S., Orsi, A., Bonoli, S., et al. 2018, MNRAS, 486, 1882-1906
    
     \bibitem[Gurung-L\'opez et al.(2021)]{gurung2020} Gurung-L\'opez, S., Saito, S., Baugh, C. M., et al. 2021, MNRAS, 500, 603--626
    
    \bibitem[Guzzo et al.(2014)]{vipers} 
    Guzzo, L., Scodeggio, M., Garilli, B., et al. 2014, A\&A, 566, A108
    
    \bibitem[Hamilton et al.(2001)]{hamilton} 
     Hamilton, A.J.S., et al. 2001, MNRAS, 322, 419
    
    \bibitem[Hashimoto et al.(2015)]{hashimoto15} 
     Hashimoto, T., Verhamme, A., Ouchi, M., et al. 2015, ApJ, 812, 157
    
    \bibitem[Herenz et al.(2017)]{herenz17} 
     Herenz, E. C., Urrutia, T., Wisotzki, L., et al. 2017, A\&A, 606, A12
     
     \bibitem[Herenz \& Wisotzki(2017)]{herenzlutz17} 
     Herenz, E. C. \& Wisotzki, L. 2017, A\&A, 602, A111
     
     \bibitem[Herenz et al.(2019)]{herenz19} 
     Herenz, E. C., Wisotzki, L., Saust, R., et al. 2019, A\&A, 621, A107
     
    \bibitem[Hill et al.(2008)]{hill} 
    Hill, G. J., MacQueen, P. J., Smith, M. P., et al. 2008, 7014, 701470
    
    \bibitem[Hinshaw et al.(2013)]{constants} 
     Hinshaw, G., Larson, D., Komatsu, E., et al. 2013, AJSS, 208, 19
     
    \bibitem[Inami et al.(2017)]{inami} 
     Inami, H., Bacon, R., Brinchmann, J., et al. 2017, A\&A, 608, 26
     
     \bibitem[Jenkins et al.(1998)]{jenkins} 
     Jenkins, A., Frenk, C. S., Pearce, F. R., et al. 1998, ApJ, 499, 20
    
    \bibitem[Kaiser(1987)]{kaiser} 
     Kaiser, N. 1987, MNRAS, 227, 1-21
     
     \bibitem[Kerutt(2017)]{josie} 
     Kerutt, J. 2017, QtClassify: IFS data emission line candidates classifier, Astrophysics Source Code Library
     
     \bibitem[Khostovan et al.(2019)]{khostovan} 
     Khostovan, A. A., Sobral, D., Mobasher, B., et al. 2019, MNRAS, 489, 555-573
     
     \bibitem[Kron(1980)]{kron} 
     Kron, R. 1980, ApJS, 43, 305
     
     \bibitem[Krumpe et al.(2010)]{mirko10} 
     Krumpe, M., Miyaji, T. \& Coil, A. L. 2010, ApJ, 713, 558
     
     \bibitem[Krumpe et al.(2012)]{mirko12} 
     Krumpe, M., Miyaji, T., Coil, A. L. \& Aceves H. 2012, ApJ, 746, 1
     
     \bibitem[Krumpe et al.(2015)]{mirko15} 
     Krumpe, M., Miyaji, T. Husemann, B., et al. 2015, ApJ, 815, 21
     
     \bibitem[Krumpe et al.(2018)]{mirko18} 
     Krumpe, M., Miyaji, T. Coil, A. L. \& Aceves, H. 2018, MNRAS, 474, 1773
     
     \bibitem[Lacey \& Cole(1993)]{lacey} 
     Lacey, C., \& Cole, S. 1993, MNRAS, 262, 627
     
     \bibitem[Landy \& Szalay(1993)]{landy1993} 
     Landy, S. D. \& Szalay, A. S. 1993, ApJ, 412, 64
     
    \bibitem[Le F\`evre et al.(2005)]{fevre} 
    Le F\`evre, O., Guzzo, L., Meneux, B., et al. 2005, A\&A, 439, 877
    
    \bibitem[Le F\`evre et al.(2015)]{vimos} 
    Le F\`evre, O., Tasca, LAM., Cassata, P., et al. 2015, A\&A, 576, A79

     \bibitem[Li et al.(2006)]{li06} 
     Li, C., Kaummann, G., Jing, Y., et al. 2006,  MNRAS, 368, 21-36

     \bibitem[Lilly et al.(2007)]{lilly} 
     Lilly, S. J., Le F\`evre, O., Renzini, A., et al. 2007,  ApJS, 172, 70
    
     \bibitem[Limber(1953)]{limber} 
     Limber, D. N. 1953, ApJ, 117, 134
     
     \bibitem[Ling et al.(1986)]{ling} 
     Ling, E. N., Frenk, C. S. \& Barrow, J. D. 1986, MNRAS, 223, 21P
     
     \bibitem[Maseda et al.(2018)]{maseda} 
     Maseda, M. V., van der Wel, A., Rix, H.-W., et al. 2018, ApJ, 854, 29
     
     \bibitem[Miyaji et al.(2007)]{miyaji} 
     Miyaji, T., Zamorani, G., Cappelluti, N., et al. 2007, ApJS, 172, 396
     
     \bibitem[Miyaji et al.(2011)]{miyaji11} 
     Miyaji, T., Krumpe, M., Coil, A. \& Aceves, H. 2011, The X-ray Universe 2011

     \bibitem[Moustakas \& Somerville(2002)]{moustaka} 
     Moustakas, Leonidas A. \& Somerville, Rachel S. 2002, ApJ, 577, 1
     
     \bibitem[Muzahid et al.(2020)]{sowgat} 
     Muzahid, S., Schaye, J., Marino, R. A., et al. 2013, MNRAS, 496, 1013-1022
     
     \bibitem[Navarro, Frenk \& White(1997)]{nfw97} 
     Navarro, J. F., Frenk, C. S. \& White, S.D.M 1997, ApJ, 490, 493
     
     \bibitem[Newman et al.(2013)]{deep2} 
     Newman, J. A., Cooper, M. C., Davis, M., et al. 2013, ApJSS, 208, 5
     
     \bibitem[Norberg et al.(2002)]{norberg02} 
     Norberg, P., Baugh, C. M., Hawkins, E., et al. 2002, MNRAS, 332, 827-838
     
     \bibitem[Ouchi et al.(2003)]{ouchi03} 
     Ouchi, M., Shimasaku, K., Furusawa, H., et al. 2003, ApJ, 582, 60 
   
     \bibitem[Ouchi et al.(2010)]{ouchi10} 
     Ouchi, M., Shimasaku, K., Furusawa, H., et al. 2010, ApJ, 723, 869
 
     \bibitem[Ouchi et al.(2017)]{ouchi17} 
     Ouchi, M., Harikane, Y., Shibuya, T., et al. 2017, PASJ, 70, S13
     
     \bibitem[Orsi et al.(2012)]{orsi12} 
     Orsi, A., Lacey, C. G., \& Baugh C. M., 2012, MNRAS, 425, 87
     
     \bibitem[Peebles(1980)]{peebles1980} 
     Peebles, P. J. E. 1980, The Large-Scale Structure of the Universe (Princeton, N.J., Princeton Univ. Press)
    
    \bibitem[Rhoads et al.(2000)]{rhoads} 
    Rhoads, J. E., Malhotra, S., Dey, A., et al. 2000, ApJ, 545, L85
    
    \bibitem[Schmidt et al.(2021)]{kasper} 
    Schmidt, B. K., Kerutt, J., Wisotzki, L., et al. 2021, submitted to A\& A
    
    \bibitem[Sheth et al.(2001)]{sheth} 
    Sheth, R., Mo, H. J. \& Tormen, G. 2001, 323, 1-12
    
    \bibitem[Shioya et al.(2009)]{shioya} 
    Shioya, Y., Taniguchi, Y., Sasaki, S. S., Nagao, T., Murayama, T., \& Saito, T. 2009, ApJ, 696, 546 
     
     \bibitem[Skrutskie et al.(2006)]{2mass} 
     Skrutskie, M. F., Cutri, R. M., Stiening, R., et al. 2006, AJ, 131, 1163-1183
     
     \bibitem[Smith et al.(2019)]{smith} 
     Smith, A., Ma, X., Bromm, V., et al. 2019, MNRAS, 484, 39-59
     
     \bibitem[Sobral et al.(2017b)]{sobral} 
     Sobral, D., Matthee, J., Best, P., et al. 2017, MNRAS, 466, 1242–1258

    \bibitem[Springel et al.(2005)]{springel} 
    Springel, V., White, S. D. M., Jenkins, A., et al. 2005, Nature, 435, 629
    
    \bibitem[Steidel \& Hamilton(1992)]{steidel} 
    Steidel, C. C., \& Hamilton, D. 1992, AJ, 104, 941
    
    \bibitem[Strauss et al.(2002)]{sdss} 
    Strauss, M. A., Weinberg, D. H., Lupton, R. H., et al. 2002, AJ, 124, 1810-1824
    
    \bibitem[Tinker et al.(2005)]{tinker} 
    Tinker, J. L., Weinberg, D. H. \& and Zheng, Z. 2005, MNRAS, 368, 85
    
    \bibitem[Urrutia et al.(2019)]{urrutia19} 
    Urrutia, T., Wisotzki, L., Kerutt, J., et al. 2019, A\&A, 624, 24
    
    \bibitem[van den Bosch(2002)]{bosch} 
    Van Den Bosch, F. C. 2002, MNRAS, 331, 98
    
    \bibitem[van den Bosch et al.(2013)]{vandenbosch13} 
    Van Den Bosch, F. C., More, S., Cacciato, M., et al. 2013, MNRAS, 430, 725

    \bibitem[Verhamme et al.(2018)]{verhamme18} 
     Verhamme, A., Garel, T., Ventou, E., et al. 2018, MNRAS, 478, 60-65
     
     \bibitem[Yajima et al.(2012)]{yajima12} 
     Yajima, H., Li, Y., \& Zhu, Q. 2012, AASMA, 219, 129-06

    \bibitem[Zehavi et al.(2002)]{zehavi02} 
    Zehavi, I., Blanton, M. R., Frieman, J. A., et al. 2002, ApJ, 571, 172
    
    \bibitem[Zehavi et al.(2011)]{zehavi11} 
    Zehavi, I., Zheng, Z., Weinberg, D., et al. 2011, ApJ, 736, 59
    
    \bibitem[Zheng \& Weinberg(2007)]{zhenghod} 
    Zheng, Z., \& Weinberg, D. H. 2007, ApJ, 659, 1
    
    \bibitem[Zheng et al.(2007)]{zheng07} 
    Zheng, Z., Coil, A. \& Zehavi, I. 2007, ApJ, 667, 760-779
    
    \bibitem[Zheng et al.(2014)]{zheng14} 
    Zheng, Z. \& Wallace, J. 2014, ApJ, 794, 116


\end{thebibliography}
\end{document}